\documentclass[12pt]{article}

\usepackage{lmodern}
\usepackage{newtxtext,newtxmath}
\usepackage{graphicx}
\usepackage{caption,subcaption,placeins}
\usepackage[letterpaper,margin=1in]{geometry}
\usepackage{comment}
\usepackage[shortlabels]{enumitem}
\renewenvironment{abstract}{\quotation}{\endquotation}
\date{}

\makeatletter
\renewcommand{\fnum@figure}{\textbf{Figure \thefigure}}
\renewcommand{\fnum@table}{\textbf{Table \thetable}}
\makeatother

\usepackage{scicite}
\usepackage{url}
\usepackage{array}
\usepackage{calc}

\usepackage[T1]{fontenc}
\usepackage[utf8]{inputenc}
\usepackage[final]{microtype}
\usepackage{braket}
\usepackage{quantikz}
\usepackage{amsmath}
\usepackage{tikz} 
\usetikzlibrary{arrows.meta,positioning}
\newcommand{\RZ}[1]{R_Z\!\left(#1\right)}
\usepackage{authblk}

\usepackage{amssymb}

\usepackage{adjustbox}
\usepackage[font=small]{caption}

\usepackage{float}

\usepackage{color}
\usepackage{ulem}

\makeatother

\makeatletter

\renewcommand{\emph}[1]{\textit{#1}}

\DeclareUnicodeCharacter{2212}{-}   
\DeclareUnicodeCharacter{27E8}{\langle} 
\DeclareUnicodeCharacter{27E9}{\rangle} 

\def\scititle{%
  \bfseries\boldmath
  Quantum spectroscopy of topo\-logical dynamics\\[-0.2ex]
  via a super\-symmetric Hamil\-tonian%
}
\title{\scititle}

\author{
  {\normalfont\mdseries\upshape\unboldmath
  Hiroshi~Yamauchi$^{1,5\ast}$,
  Satoshi~Kanno$^{1}$,
  Yuki~Sato$^{2,5}$,
  Hiroyuki~Tezuka$^{3,5}$,
  Yoshi-aki~Shimada$^{1}$,
  Eriko~Kaminishi$^{5}$,
  Naoki~Yamamoto$^{4,5}$
  }\\
  
  {\small\normalfont\mdseries\upshape
  $^{1}$SoftBank Corp., Research Institute of Advanced Technology, 1-7-1 Kaigan, Minato-ku, Tokyo 105-7529, Japan\\
  $^{2}$Toyota Central R\&D Labs., Inc., 1-4-14, Koraku, Bunkyo-ku, Tokyo, 112-0004, Japan\\
  $^{3}$Advanced Research Laboratory, Sony Group Corporation, 1-7-1 Konan, Minato-ku, Tokyo 108-0075, Japan\\
  $^{4}$Department of Applied Physics and Physico-Informatics, Keio University, 3-14-1 Hiyoshi, Kohoku-ku, Yokohama, Kanagawa 223-8522, Japan\\
  $^{5}$Quantum Computing Center, Keio University, 3-14-1 Hiyoshi, Kohoku-ku, Yokohama, Kanagawa 223-8522, Japan\\
  $^{\ast}$Corresponding author: \texttt{hiroshi.yamauchi@g.softbank.co.jp}
  }
}

\begin{document}
\maketitle

\begin{center}
\textbf{One-Sentence Summary:} Quantum spectral gaps appear to reflect and trace changes associated with the stabilization of loop topology in the Lorenz system.
\end{center}

\begin{abstract}\bfseries\boldmath

Topological data analysis (TDA) characterizes complex dynamics through global invariants, but classical computation becomes prohibitive for high-dimensional data. 
We reinterpret time-domain dynamics as the eigenvalue spectrum of a supersymmetric (SUSY) Hamiltonian and thereby estimate topological descriptors through quantum spectroscopy. 
While zero modes correspond to Betti numbers, we show that low-lying excited states quantify the stability of topological features. 
Using a Takens embedding of the Lorenz system together with a resource-efficient quantum phase estimation implemented on IBM quantum hardware, we observe that the spectral gap of the SUSY Laplacian tracks the persistence of homological structures. 
Notably, the minimum of this spectral gap coincides with the onset of chaos, whereas its reopening reflects the geometric maturation of the attractor. 
Validated on small complexes yet offering an exponential advantage over classical diagonalization (from $O(N^3)$ to $\mathrm{poly}(\log N)$), this framework suggests that quantum hardware can function as a spectrometer for data topologies beyond classical reach.
\end{abstract}

\noindent

\section{Introduction}

\paragraph{}
Topological data analysis (TDA) provides a framework to study data through its \emph{shape} rather than through simple statistical summaries such as means or variances.
By examining how connectivity evolves with scale, TDA extracts stable global features—components, loops, and voids—that characterize the underlying structure of complex systems.
Persistence diagrams capture the birth and death of topological features across scales, while Betti curves quantify their number and persistence, providing compact topological signatures of data geometry~\cite{Bauer2021Ripser,BoissonnatMaria2014GUDHI,Aktas2019PHNetworks,Umeda2017TimeSeriesTDA}.
However, classical TDA becomes computationally demanding for large or high–dimensional datasets and, unlike physical spectroscopy, does not yield a measurable \emph{spectral} quantity that directly reflects the stability of topological features. 

\paragraph{}
Within TDA, Persistent homology (PH)—a method that quantifies how topological structures persist across scales—has recently been applied to dynamical systems and time–series data.  
PH based on distance–matrix or recurrence–plot filtrations captures the evolving geometry of nonlinear dynamics, enabling quantitative detection of transitions between periodic and chaotic regimes~\cite{Ichinomiya2025SciRep,Ichinomiya2023NOLTA}. 
Subsequent studies have demonstrated robustness under sparse or noisy sampling~\cite{Antosh2024ScantyPHML}, introduced dynamic algorithms for maintaining persistence in streaming settings~\cite{Montesano2024SODA}, and used PH to trace structural reorganizations in evolving networks~\cite{Myers2019PRE}.  
Machine–learning approaches have incorporated PH–derived descriptors for classifying and forecasting temporal patterns~\cite{Chung2020PHTimeSeries,Umeda2017TimeSeriesTDA,Tsuji2022UTokyoSummary}.  
More recently, spectral and graph–theoretic formulations have emerged, connecting persistence with the eigenvalue structure of Laplacian operators and multilayer networks~\cite{Otter2017EPJDS,Froyland2024SupraLaplacian}.  
Together, these developments have positioned PH as a versatile tool for 
representing temporal evolution, yet its invariants remain largely 
computational constructs rather than physically measurable spectra.

\paragraph{}
Quantum algorithms have been proposed to accelerate the linear–algebraic routines central to TDA, particularly eigenvalue estimation for the Laplacian, which capture topological features such as connected components, loops, and voids through their harmonic (zero–eigenvalue) modes~\cite{Lloyd2014,Berry2007SparseHamiltonian,Scali2024,Akhalwaya2024BNE}.
The \emph{Persistent Laplacian} extends the classical Laplacian framework to filtrations.
Its kernel and low-lying eigenvalues track how homological features persist across scales.
In this way, it provides a spectral analogue of the persistence diagram~\cite{Wang2019PersistentSpectralGraph}.
Hybrid quantum–TDA approaches have visualized how data–encoding circuits reshape topological relations~\cite{VlasicPham2023QTA}, motivating quantum formulations of PH~\cite{Gyurik2022,Hayakawa2022,GyurikSchmidhuber2024,Berry2024ProspectsQuantumAdvantage,NghiemLee2025TDA} and early experimental demonstrations on near–term hardware~\cite{incudini2023higher,Ubaru2021QTDALinearDepth,akhalwaya2022topological}.
Meanwhile, hardness results show that deciding the existence of specific homological invariants is QMA/QMA$_1$–hard~\cite{Cade2024SUSY}, implying that broad, unconditional quantum speedups are unlikely without additional structure or approximation.
While current demonstrations are limited by qubit count, the underlying eigenvalue estimation problem scales cubically or worse on classical machines. Quantum phase estimation (QPE), by contrast, offers polynomial scaling in system size and spectral resolution, suggesting a clear path toward exponential advantage once system sizes exceed classical diagonalization limits.

\paragraph{}
A complementary theoretical line, originating from supersymmetric (SUSY) quantum mechanics~\cite{Witten1982,SUSY_Topology_PRB,Wen2004QFTManyBody}, connects topology with spectral theory: zero–energy states of a SUSY Hamiltonian correspond to harmonic forms, while small but finite eigenvalues describe tunneling between Morse basins, linking spectral gaps to topological stability.
This perspective aligns with recent work on persistent Laplacians, which formalize multiscale stability of harmonic subspaces through spectral properties~\cite{MemoliWanWang2022SIMODS,Wang2019PersistentSpectralGraph}.
Building on this foundation, we derive a spectral bound relating the 
SUSY Laplacian. 
This operator is constructed from the $Q$ and $Q^\dagger$ blocks of the SUSY Hamiltonian and has a block-diagonal structure.
We relate its first nonzero eigenvalue—the SUSY energy gap—to the classical H1 persistence length, which measures the lifetime of the dominant one-dimensional homology class.
This result indicates that the latter is upper-bounded by the minimal nonzero eigenvalue of the associated SUSY Laplacian.
This analytical result formalizes the intuitive notion that PH reflects the resilience of harmonic subspaces under perturbation and provides the theoretical basis for the experimentally observed correlation between the quantum spectral gap and topological stability.
More generally, the SUSY Laplacian spectrum encodes connectivity and robustness: zero eigenvalues represent harmonic subspaces and Betti numbers, while the first nonzero eigenvalue—the \emph{spectral gap}—governs relaxation under diffusion, synchronization, or transport~\cite{Marsden2013Laplacian,Christoffersen2025EigenGap,Friedman1998Betti,Edelsbrunner2014Complexity}.
These insights suggest that low–lying spectra capture the essence of topological coherence and resilience.
Building upon this theoretical foundation, we next demonstrate how the correspondence between spectral gaps and persistence can be implemented and observed on quantum hardware. 

\paragraph{}
We interpret an explicit connection between classical topological persistence and quantum spectral measurement by constructing a supersymmetric (SUSY) Hamiltonian whose eigenvalue spectrum encodes the the combinatorial Hodge–Laplacian (the discrete Laplacian constructed from simplicial boundary operators) of a data–derived simplicial complex, transforming topological features into measurable quantum energy levels.
Zero modes correspond to Betti numbers, while low–lying excited modes quantify topological robustness through the \emph{SUSY–Laplacian gap}.
To realize this concept under current quantum–hardware constraints, we developed an efficient single–ancilla QPE~\cite{kitaev1995quantum} implementation for Laplacian eigenvalue readout, introducing a scalable form of \emph{quantum spectroscopy of topology}. 

\paragraph{}
As a proof of concept, we demonstrate this workflow on the Lorenz dynamical system.
Using Takens embedding to reconstruct its attractor, we identify topology–preserving representative points, build a projected–basis SUSY Hamiltonian, and perform controlled time evolution for quantum phase estimation.
The observed SUSY–Laplacian gap $\Delta^{(1)}_{\mathrm{SUSY}}$ co–varies with the maximal classical $H_1$ persistence $\ell^{\max}_{H_1}$, indicating that both reflect the stabilization of the attractor’s dominant loop.
This correspondence—validated in both noiseless simulation and real hardware—demonstrates that topological transitions in nonlinear dynamics can be detected spectroscopically on quantum devices.

\paragraph{}
Despite the limited qubit count of current devices, our hybrid classical–quantum workflow achieves practical scalability by \emph{classically identifying} a representative subset of edges that maximizes the $H_1$ persistence of the original complex. 
This selection concentrates computational resources on the edges responsible for the birth of dominant loops, thereby isolating the subspace in which topological transitions occur. 
The quantum stage then \emph{interrogates this identified topology} by 
measuring the SUSY–Laplacian spectral gap, which, as shown by our spectral 
bound, can be related to the classical $H_1$ persistence. 
This single-shot spectral measurement reduces the need for repeated 
filtrations across~$\epsilon$, allowing a single quantum execution to 
summarize the overall persistence behavior. 
We validate this correspondence through gap spectroscopy on IBM superconducting hardware, thereby illustrating an end-to-end \emph{“classically identify, quantum interrogate”} workflow that translates persistent-homology structure into measurable quantum spectra.

\paragraph{}
To achieve high-fidelity spectral estimation under hardware constraints, we employ a \emph{Projected-basis hybrid compilation} for controlled time evolution combined with a one-ancilla QPE readout.
Each circuit uses fourteen qubits—seven for the edge register, six for phase encoding, and one ancilla—with an effective depth of about one hundred two-qubit gates.
In contrast, a textbook inverse-QFT–based QPE of equal resolution would require $\sim4\times10^{4}$ two-qubit gates.
By clustering commuting Pauli terms, projecting local two-qubit subspaces, and merging multi-qubit controls, our approach compresses this to $\sim10^{2}$ gates—roughly $400\times$ fewer—while preserving identical spectra—thus enabling gap spectroscopy on IBM’s superconducting processor \texttt{ibm\_kingston}, where the measured spectra closely matched simulator predictions.

\paragraph{}
In this context, the present study positions \emph{quantum topology as spectroscopy}: a framework in which homological invariants manifest as measurable eigenvalue structures. 
While classical spectral geometry analyzes the eigenvalues of the Hodge–Laplacian, it does not interpret them as physically measurable spectra; in contrast, our formulation accesses these eigenvalues through quantum phase estimation, enabling a genuinely spectroscopic view of topology.
Our SUSY formulation translates the Hodge–Laplacian spectrum of data–derived complexes into quantum observables.
The observed correspondence between the SUSY spectral gap and $H_1$ persistence in the Lorenz attractor indicates that topological stabilization can be detected as a spectral transition—offering a physically interpretable bridge between classical persistent homology and quantum spectral geometry.

\section{Theory overview}

\paragraph{}
To formalize the connection between topology and quantum spectra, we construct a supersymmetric (SUSY) Hamiltonian that encodes the topology of a data–derived simplicial complex through its block structure (see Materials and Methods for full derivation).  
We work on a $\mathbb{Z}$–graded Hilbert space 
$\mathcal H=\bigoplus_{k\ge0}\mathcal H_k$.
Each subspace $\mathcal H_k$ represents the $k$–cochains, i.e., functions defined on k-simplices.
On this graded space, the odd supercharges 
$Q:\mathcal H_k\!\to\!\mathcal H_{k+1}$ and 
$Q^{\dagger}:\mathcal H_{k+1}\!\to\!\mathcal H_k$ 
generate the SUSY algebra, satisfying $Q^2=(Q^{\dagger})^2=0$.  
The SUSY Hamiltonian is defined by
\[
\mathcal H_{\mathrm{SUSY}}=\{Q,Q^{\dagger}\}
=Q^{\dagger}Q+QQ^{\dagger}.
\]
It is block–diagonal with respect to degree.
Its restriction on $\mathcal H_k$ reproduces the combinatorial $k$–Hodge Laplacian,
\[
\mathcal H_{\mathrm{SUSY}}\big|_{\mathcal H_k}
=d_k^{\dagger}d_k+d_{k-1}d_{k-1}^{\dagger}
=\mathcal L_k ,\qquad d_k:=Q|_{\mathcal H_k}.
\]
Hence $\ker \mathcal L_k \cong H_k$, 
where $H_k$ is the $k$-th homology group 
and $\cong$ denotes vector-space isomorphism.  
In particular, the dimension of this kernel,
$\dim(\ker \mathcal L_k)$, equals the $k$-th Betti number 
$\beta_k$, which counts the number of independent 
$k$-dimensional holes.
In this correspondence, zero–energy states of $\mathcal H_{\mathrm{SUSY}}$ represent harmonic forms and recover the homology generators, whereas the smallest nonzero eigenvalue
\[
\Delta^{(k)}_{\mathrm{SUSY}}
=\min\{\lambda>0:\lambda\in\sigma(\mathcal L_k)\}
\]
measures the spectral isolation of those topological modes, serving as a quantitative indicator of their robustness.

\paragraph{}
For the present study we focus on the $k{=}1$ block $\mathcal L_1$, which governs loop (first–homology) structure.  
This block acts as the edge Laplacian on the representative–point graph 
constructed from the embedded point cloud, combining the vertex–edge 
incidence matrix $B_1$ and the edge–triangle incidence matrix $B_2$ as
\[
\mathcal L_1 = B_1^{\top} B_1 + B_2 B_2^{\top}.
\]
The kernel of $\mathcal L_1$ encodes divergence–free and curl–free edge flows corresponding to persistent $H_1$ loops (the first homology group), while the first nonzero eigenvalue $\Delta^{(1)}_{\mathrm{SUSY}}$ quantifies the energetic separation between harmonic and excited subspaces.  
By tracking $\Delta^{(1)}_{\mathrm{SUSY}}$ across a control parameter such as the Rayleigh number $\rho$, we monitor how the spectral gap evolves.
Changes in this gap indicate how the dominant loops in the attractor form or reorganize as the parameter varies.

\paragraph{}
Because $\mathcal H_{\mathrm{SUSY}}$ is positive semidefinite and excitation–number preserving, its time evolution $e^{-i\mathcal H_{\mathrm{SUSY}}t}$ can be simulated efficiently on near–term hardware.  
A Dicke–encoded symmetric probe, which concentrates amplitude collectively in the harmonic sector, is prepared to enhance sensitivity to topological modes. A single–ancilla QPE retrieves the autocorrelation
\[
C(t)=\mathrm{Tr}(\rho\,e^{-i\mathcal H_{\mathrm{SUSY}}t}),
\]
whose Fourier spectrum reveals the eigenvalue distribution of $\mathcal H_{\mathrm{SUSY}}$.  
Zero–frequency peaks correspond to Betti numbers, while the first resolvable nonzero line defines the SUSY spectral gap $\Delta^{(1)}_{\mathrm{SUSY}}$.  
This mapping enables direct experimental access to topological invariants 
through quantum spectroscopy and provides the theoretical context for the 
pipeline described in Materials and Methods and the experiments that follow.

\section{Experimental implementation}
\subsection{Overview and Scope}
\label{sec:exp_overview}

\begin{figure}[!ht]
    \centering
    \includegraphics[width=\linewidth]{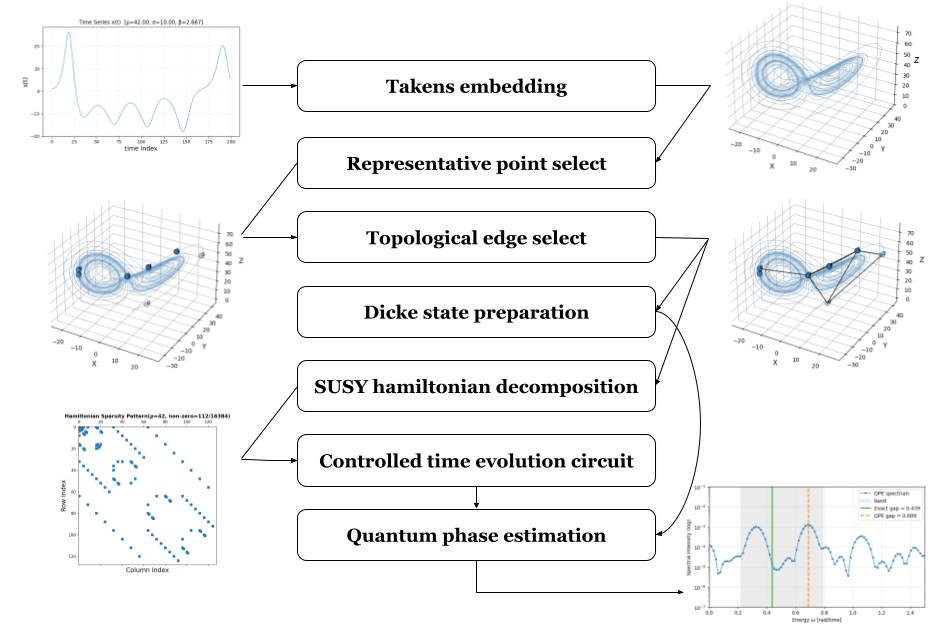}
    \caption{\textbf{Quantum spectroscopy pipeline for topological–dynamical analysis.}
    The workflow converts a scalar time series into a quantum-mechanical spectrum whose low-lying energies encode topological structure.
    A Lorenz signal is Takens-embedded, topology-aware point selection preserves loop geometry, and a supersymmetric (SUSY) Hamiltonian is constructed and simulated by a controlled time-evolution circuit.
    Single-ancilla quantum phase estimation (QPE) retrieves the eigenvalue spectrum, where near-zero modes correspond to harmonic $H_1$ loops.}
    \label{fig:pipeline}
\end{figure}

\paragraph{}
The experiment evaluates whether the proposed quantum–topological diagnostics 
can reliably detect dynamical reorganizations in the Lorenz system 
and whether a SUSY Hamiltonian readout reproduces the corresponding signatures 
both in simulation and on quantum hardware.  
The study proceeds through three stages, moving from basic validation to 
realistic dynamical data, as outlined schematically in 
Figure~\ref{fig:pipeline}. 
The pipeline begins by constructing a simplicial complex from the data and 
extracting its incidence structure, then converts this structure into a SUSY Hamiltonian whose spectrum is probed via quantum phase estimation, and finally compares the resulting spectral features with classical topological descriptors across the parameter sweep.
As detailed in Supplementary Section~\ref{sec:proof_energy_persistence}, the correspondence between the 
first nonzero SUSY–Laplacian eigenvalue and the classical $H_1$ persistence is supported by a rigorous spectral bound. This theoretical result enables the experimental pipeline to interpret measured spectral gaps as quantitative indicators of topological stability.

\begin{figure}[!ht]
  \centering

  \begin{minipage}[b]{0.35\linewidth}
    \textbf{A}\par
    \includegraphics[width=\linewidth]{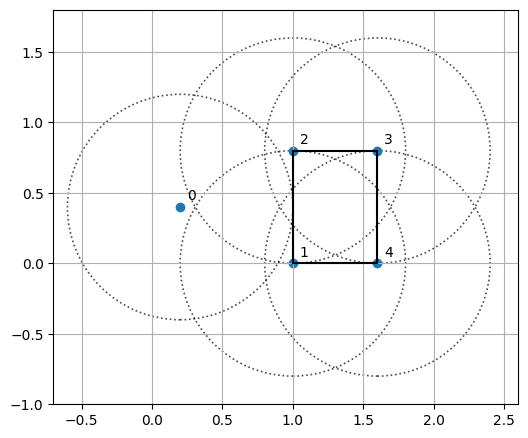}
  \end{minipage}\hspace{4em}
  \begin{minipage}[b]{0.4\linewidth}
    \includegraphics[width=\linewidth]{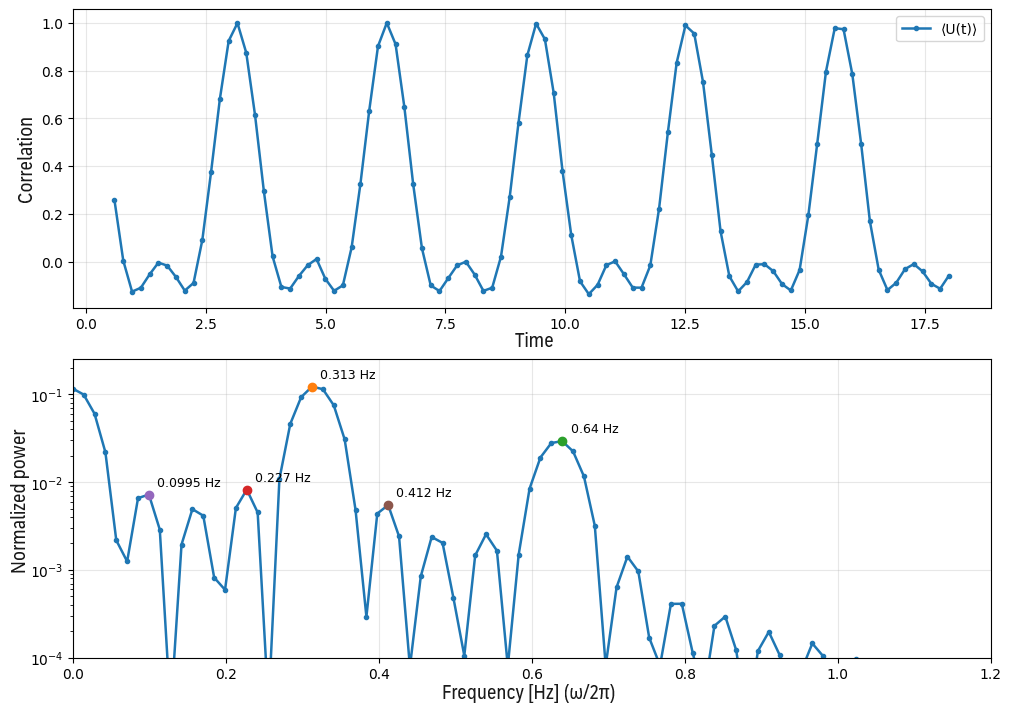}
  \end{minipage}\hspace{4em}
  \begin{minipage}[b]{0.35\linewidth}
    \textbf{B}\par
    \includegraphics[width=\linewidth]{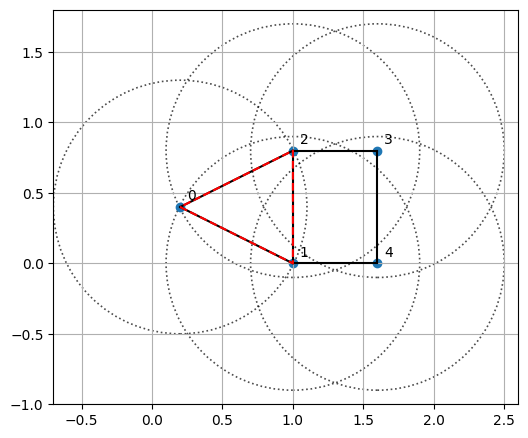}
  \end{minipage}\hspace{4em}
  \begin{minipage}[b]{0.4\linewidth}
    \includegraphics[width=\linewidth]{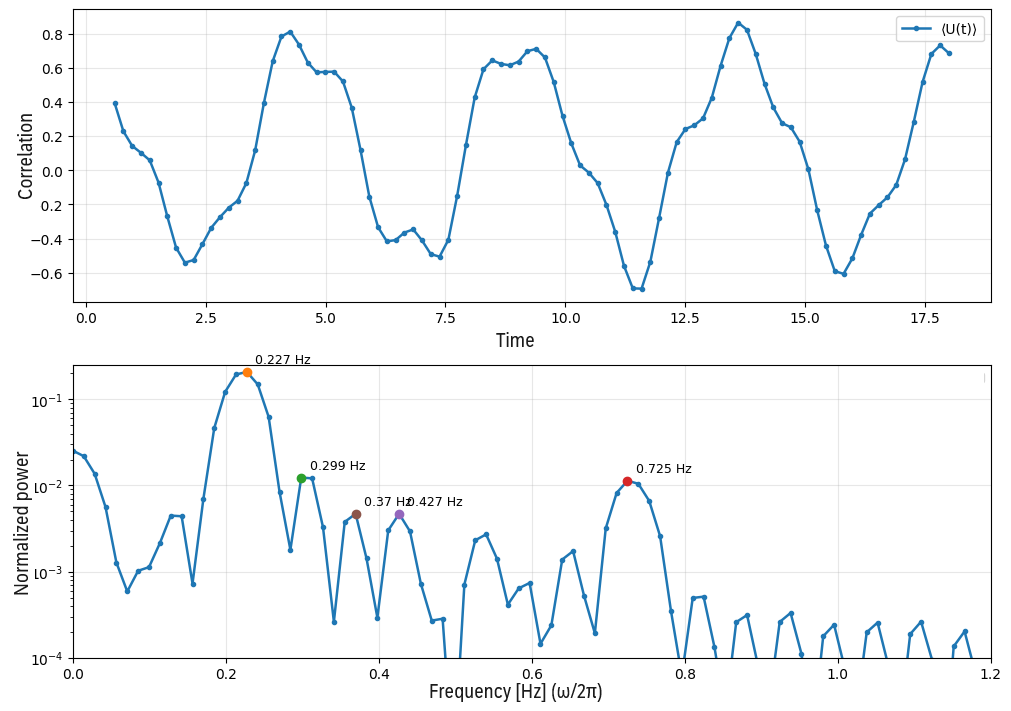}
  \end{minipage}\hspace{4em}
  \begin{minipage}[b]{0.35\linewidth}
    \textbf{C}\par
    \includegraphics[width=\linewidth]{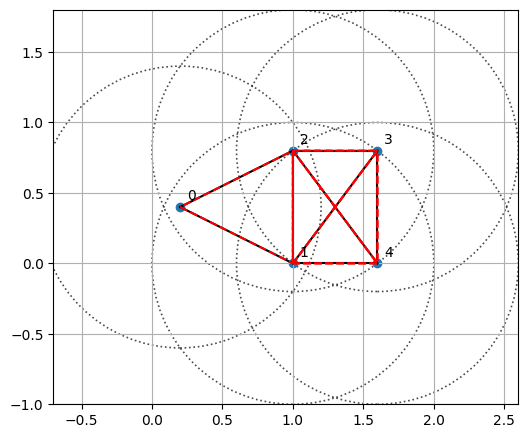}
  \end{minipage}\hspace{4em}
  \begin{minipage}[b]{0.4\linewidth}
    \includegraphics[width=\linewidth]{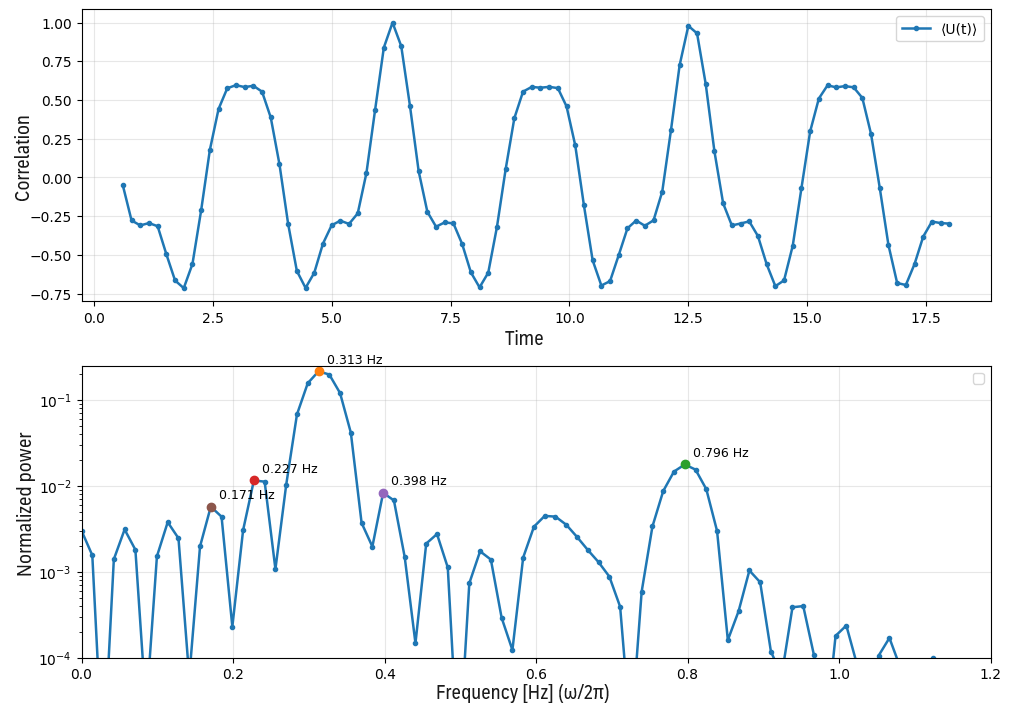}
  \end{minipage}

  \caption{\textbf{Five-point validation of quantum Betti estimation.}
  Three Vietoris–Rips filtrations of a five-point complex demonstrate how quantum spectra track topological change.
  (\textbf{A}) At $\varepsilon=0.8$, a single loop yields $\beta_1=1$;
  (\textbf{B}) at $\varepsilon=0.9$, one triangle forms but the loop persists;
  (\textbf{C}) at $\varepsilon=1.0$, the complex becomes contractible ($\beta_1=0$).
  QPE spectra reproduce the classical Hodge–Laplacian eigenvalues, confirming accurate quantum detection of loop annihilation.
    }
  \label{fig:pentagon}
    
\end{figure}

\paragraph{}
Figure~\ref{fig:pentagon} shows the first-stage calibration.  
A minimal five-point complex—a toy model representing a small set of sample points—is used to verify that the QPE readout reproduces the Hodge–Laplacian spectrum, which encodes the connectivity of the simplicial complex across a Vietoris–Rips filtration.  
This filtration is a standard topological construction that adds simplices as pairwise distances fall below a threshold.
This test checks that edges and filled triangles enter the complex at the correct thresholds, that the zero-energy subspace has multiplicity equal to the first Betti number $\beta_1$, and that the first positive eigenvalue $\Delta^{(1)}_{\mathrm{SUSY}}$ increases as small spurious cycles are filled in. 
Agreement with the classical eigenvalues confirms that the SUSY Hamiltonian 
provides a faithful discretization of the Hodge Laplacian.

\begin{figure}[!ht]
  \centering
  \begin{minipage}[b]{0.4\linewidth}
    \textbf{A}\par
    \includegraphics[width=\linewidth]{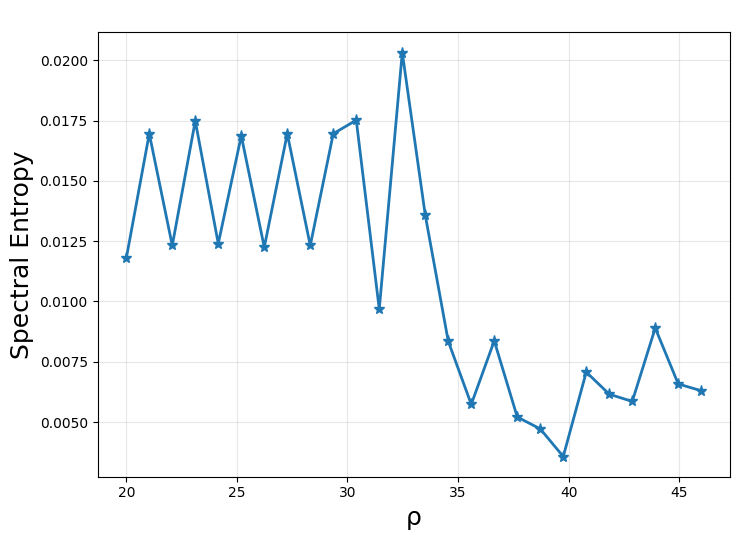}
  \end{minipage}\hspace{4em}
  \begin{minipage}[b]{0.4\linewidth}
    \textbf{B}\par
    \includegraphics[width=\linewidth]{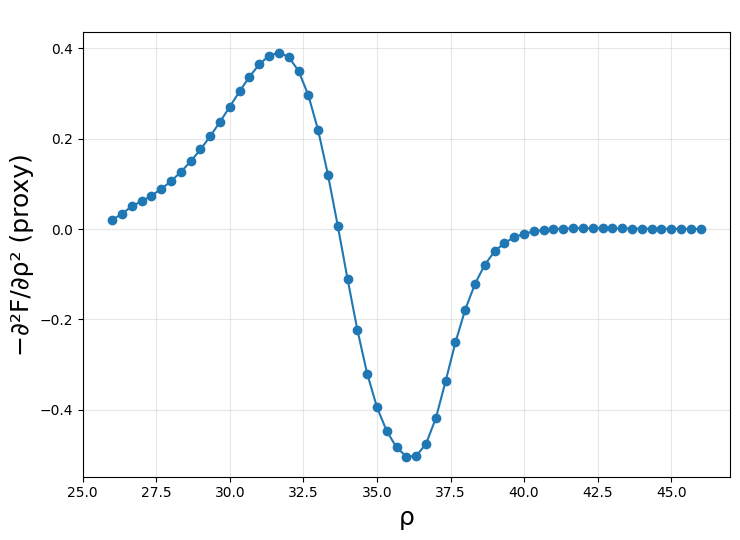}
  \end{minipage}\hspace{4em}

  \begin{minipage}[b]{0.4\linewidth}
    \textbf{C}\par
    \includegraphics[width=\linewidth]{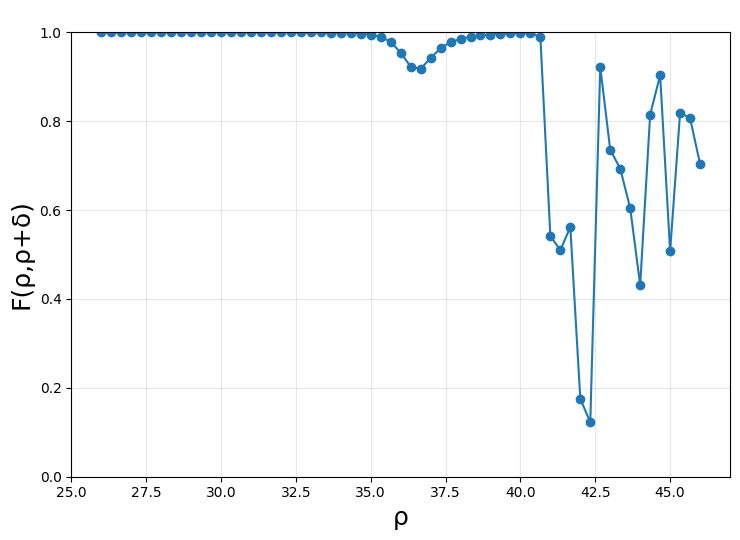}
  \end{minipage}\hspace{4em}
  \begin{minipage}[b]{0.4\linewidth}
    \textbf{D}\par
    \includegraphics[width=\linewidth]{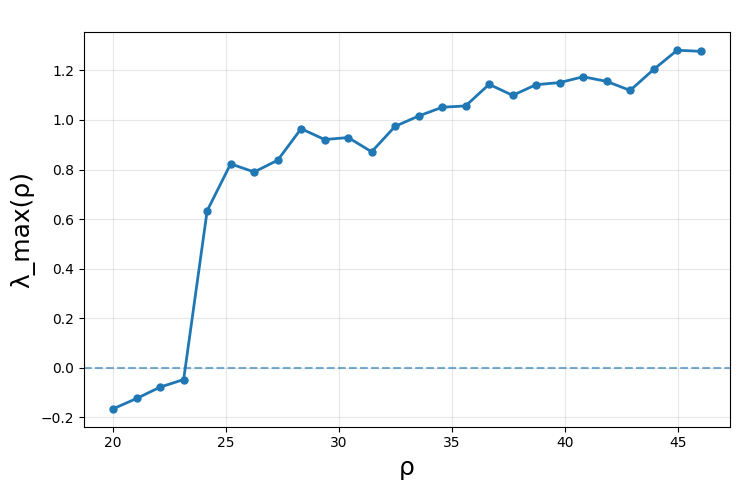}
  \end{minipage}\hspace{4em}
  
  \begin{minipage}[b]{0.4\linewidth}
    \textbf{E}\par
    \includegraphics[width=\linewidth]{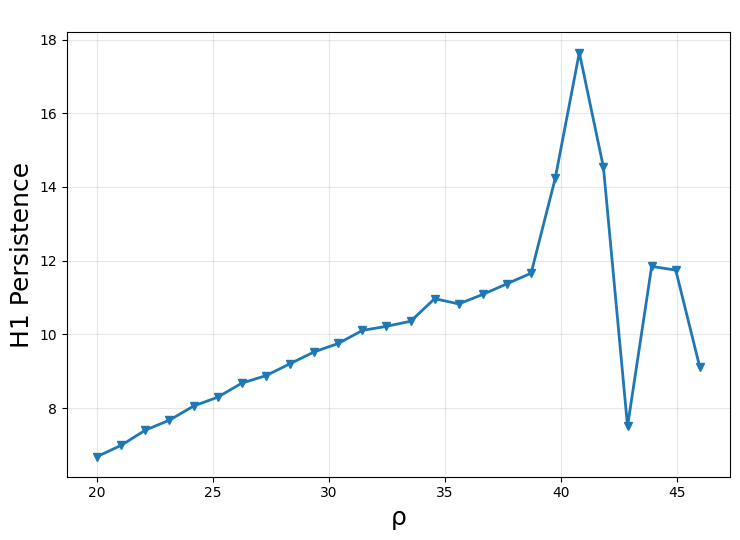}
  \end{minipage}\hspace{4em}
  \begin{minipage}[b]{0.4\linewidth}
    \textbf{F}\par
    \includegraphics[width=\linewidth]{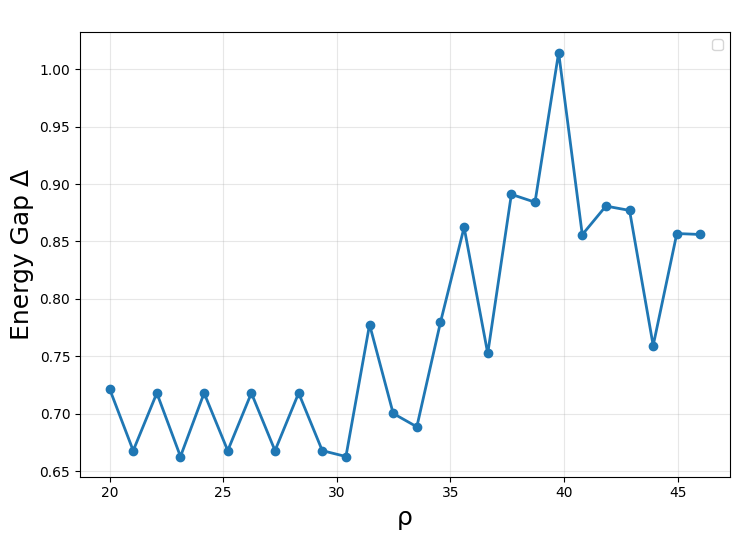}
  \end{minipage}

  \caption{\textbf{Dynamical and topological diagnostics across the Lorenz parameter $\rho$.}
  (\textbf{A}) Spectral entropy,
  (\textbf{B}) free-energy curvature,
  (\textbf{C}) ground-state fidelity,
  (\textbf{D}) maximum Lyapunov exponent,
  (\textbf{E}) $H_1$ persistence, and
  (\textbf{F}) low-energy gap $\gamma(\rho)=E_1-E_0$.
  The joint evolution of these indicators suggests two transitions: a chaotic onset near $\rho\approx30$ and topological stabilization near $\rho\approx41$.
    }
    \label{fig:phasetransition}
\end{figure}

\paragraph{}
Figure~\ref{fig:phasetransition} summarizes the second stage, a parameter sweep of the Lorenz flow (Eq.~\ref{eq:lorenz}) performed 
as the Rayleigh number $\rho$—which governs the onset of convection and thus controls the transition between regular and chaotic motion—is varied while $(\sigma,\beta)$ are held fixed. 
For each $\rho$, a trajectory is embedded via delay coordinates, a standard method for reconstructing phase-space dynamics from time-series data, then reduced to a representative set 
preserving density and topology, and converted into an effective SUSY Hamiltonian. 
From these instances we compute both dynamical and topological indicators 
(defined in detail in Secs.~\ref{sec:rho_sweep_setup}--\ref{sec:lorenz_dyn_topo}):  
the spectral entropy of a survival amplitude $H_{\mathrm{spec}}(\rho)$  
[Eq.~\ref{eq:specentropy}],  
the curvature of the ground-state energy $F''(\rho)=\frac{\partial^2 F}{\partial\rho^2}$  
[Eq.~\ref{eq:Fpp}],  
the ground-state fidelity $F(\rho,\rho{+}\Delta\rho)$  
[Eq.~\ref{eq:fidelity}],  
the persistent-homology lifetime $\ell^{\max}_{H_1}(\rho)$  
(defined in the persistent-homology description of Sec.~\ref{sec:lorenz_dyn_topo}),  
and the low-energy gap $\gamma(\rho)=E_1-E_0$  
[Eq.~\ref{eq:gamma_again}].  
The goal is to identify the ranges of $\rho$ where the dynamical spectrum reorganizes 
and to correlate these transitions with the emergence or disappearance of persistent $H_1$ loops 
in the underlying attractor.

\begin{figure}[!ht]
    \centering
    \includegraphics[width=0.7\linewidth]{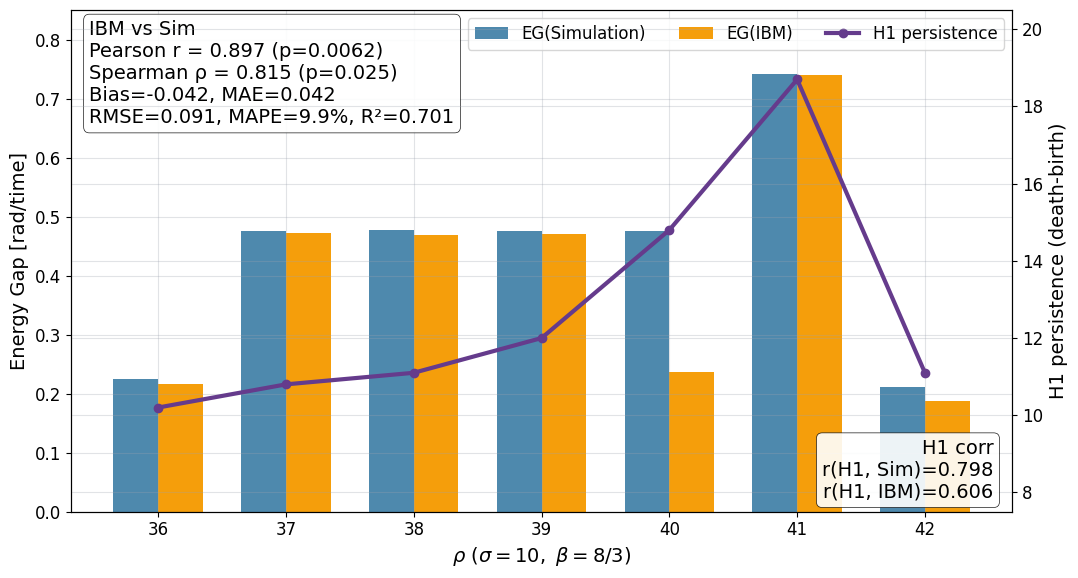}
    \caption{\textbf{Quantum–classical correspondence of topological stabilization.}
    Comparison of the Hodge–Laplacian spectrum energy gap $\Delta^{(1)}_{\mathrm{SUSY}}$ (blue/orange bars: simulator vs IBM hardware) with the highest classical $H_1$ persistence $\ell^{\max}_{H_1}$ (purple line) as functions of the Rayleigh parameter $\rho$.
    Both peak near $\rho\approx41$, where the Lorenz attractor’s loop geometry and the quantum harmonic subspace show their strongest alignment in this study, suggesting a possible spectral–topological correspondence.
    \textit{Agreement (IBM vs simulation):} Pearson $r=0.897$ ($p=0.0062$), Spearman $\rho=0.815$ ($p=0.025$), bias $-0.042$, MAE $0.042$, RMSE $0.091$, MAPE $9.9\%$, $R^2=0.701$.
    \textit{Link to classical topology:} $r(H_1,\mathrm{Sim})=0.798$, $r(H_1,\mathrm{IBM})=0.606$.}
    \label{fig:eg_gap_persistence}
\end{figure}

\paragraph{}
Figure~\ref{fig:eg_gap_persistence} presents The third stage, which realizes the workflow on quantum hardware. 
A representative set of seven points extracted from the Lorenz embedding 
is used to construct the oriented incidence matrices 
and the $k{=}1$ SUSY block $\mathcal{L}_1(\rho)=d_1^\dagger d_1+d_0d_0^\dagger$. 
QPE circuits are compiled to fourteen qubits 
(seven for the edge register, six work qubits, and one ancilla) 
and executed on both a noiseless simulator and IBM’s 156-qubit superconducting processor \texttt{ibm\_kingston}. 
Identical evolution times, windowing, and transpilation settings ensure a direct comparison of spectra. 
The primary observables are the multiplicity of near-zero eigenvalues $\widehat{\beta}_1^{\mathrm{QPE}}(\rho)$ and the first nonzero eigenvalue $\widehat{\Delta}^{(1)}_{\mathrm{SUSY}}(\rho)$, which correlate with the classical persistent-homology lifetimes through the spectral bound derived in Supplementary Section~\ref{sec:proof_energy_persistence}.

\paragraph{}
Together, these experiments test the consistency between classical and quantum diagnostics 
across the dynamical regimes of the Lorenz system. 
The sequence—from the schematic pipeline (Figure~\ref{fig:pipeline}), 
through the five-point validation (Figure~\ref{fig:pentagon}), 
the parameter sweep (Figure~\ref{fig:phasetransition}), 
and finally the hardware verification (Figure~\ref{fig:eg_gap_persistence})—
demonstrates that the spectral features of the SUSY Hamiltonian 
serve as reliable indicators of topological transitions 
that can be resolved within current quantum hardware limits.

\subsection{Five-Point Validation of Quantum Betti Estimation}
\label{sec:exp_pentagon_setup}

\paragraph{}
A controlled five-point complex is used to validate that our SUSY Hamiltonian construction and its QPE-based spectral readout faithfully reproduce the classical $k{=}1$ Hodge Laplacian spectrum across a Vietoris–Rips (VR) filtration. The test evaluates whether the pipeline obeys the simplicial update rule—edges appearing precisely at their radius thresholds and $2$–simplices filling only when all three edges are present—and whether the spectral procedure correctly identifies the first Betti number $\beta_1$ and the topological gap $\Delta^{(1)}_{\mathrm{SUSY}}$. Panels~A–C of Figure~2 depict three representative filtration snapshots at $\varepsilon\in\{0.8,0.9,1.0\}$.

\paragraph{}
Starting from the point cloud $P=\{p_1,\dots,p_5\}\subset\mathbb{R}^2$, we build the VR complex $R_\varepsilon(P)$ by including an edge $\{i,j\}$ whenever $d_{ij}=\|p_i-p_j\|_2\le\varepsilon$ and a triangle $\{i,j,k\}$ once all three pairwise edges are present. The resulting complexes define the cochain spaces $\mathcal H_0,\mathcal H_1,\mathcal H_2$ and coboundary operators $d_k$ with standard orientation $\{\pm1,0\}$. The $k{=}1$ Hodge block $\mathcal L_1(\varepsilon)$ is used as in Eq.~(\ref{eq:L1def}).

\paragraph{}
Here $E(\varepsilon)=|R_\varepsilon^{(1)}|$ is the number of edges at scale $\varepsilon$, 
and $\{|e\rangle\}$ is the standard edge basis ordered by a fixed orientation.
A uniform probe state is prepared in the $1$–cochain space,
\begin{equation}
|\psi\rangle=\frac{1}{\sqrt{E(\varepsilon)}}\sum_{e\in R_\varepsilon^{(1)}} |e\rangle,
\end{equation}
which is a pure state (its mixed-state analogue is treated later via phase randomization),
and evolved under the rescaled operator 
$\widetilde{\mathcal L}_1=\mathcal L_1/\alpha$ on a time grid 
$t_m=m\,\Delta t$ ($m=0,\dots,M{-}1$).
We define the (complex) autocorrelation function of the probe state as
\begin{equation}
S(t_m;\varepsilon)=\langle\psi|e^{-i\widetilde{\mathcal L}_1(\varepsilon)t_m}|\psi\rangle.
\end{equation}
Time evolution $e^{-i\widetilde{\mathcal L}_1 t}$ is computed via a high-accuracy Krylov 
action of the matrix exponential. 
The spectral density is then estimated from a Hann-windowed periodogram,
\begin{equation}
\widehat{P}(\omega;\varepsilon)
=\Big|\sum_{m=0}^{M-1}w_m\,S(t_m;\varepsilon)\,e^{-i\omega t_m}\Big|^2,
\qquad
\hat\lambda_k(\varepsilon)=\alpha\,\hat\omega_k,
\end{equation}
where the peak frequencies $\hat\omega_k$ correspond to eigenvalues of the 
rescaled operator $\widetilde{\mathcal L}_1=\mathcal L_1/\alpha$, so that 
$\hat\lambda_k(\varepsilon)$ recovers the eigenvalues of the Hodge Laplacian block $\mathcal L_1(\varepsilon)$.
Here $w_m$ are the Hann window coefficients,
\begin{equation}
w_m = \tfrac{1}{2}\!\left[1-\cos\!\left(\tfrac{2\pi m}{M-1}\right)\right],
\qquad m=0,1,\dots,M-1,
\end{equation}
which smoothly taper the time-domain signal to zero at both ends and thereby 
reduce spectral leakage in the Fourier transform. 
Calibration is verified by sweeping $\alpha$ over a small set 
$\{\alpha_1,\alpha_2,\alpha_3\}$ and confirming the linear relation 
$\hat\lambda_k(\alpha)=\alpha\,\hat\omega_k(\alpha)$ within tolerance.
To avoid aliasing, we ensure that 
$\|\widetilde{\mathcal L}_1\|<\pi/\Delta t$, 
so that all eigenfrequencies lie within the Nyquist band $[0,\pi/\Delta t)$.
Under this condition, the near-zero multiplicity 
$\hat\beta_1(\varepsilon)=\#\{\hat\lambda_k(\alpha)\le\tau_0\}$ 
remains invariant under rescaling.

\paragraph{}
The experiment was assessed according to several consistency and accuracy criteria. 
First, filtration correctness was verified by confirming that edges and 2-simplices appeared in the complex precisely at their prescribed thresholds and that the data structures for $\mathcal{L}_1(\varepsilon)$ were updated accordingly. 
Next, topological consistency was verified by confirming that the number 
of near-zero spectral peaks agreed with the first Betti number, $\hat{\beta}_1(\varepsilon)=\#\{\hat{\lambda}_k(\varepsilon)\le\tau_0\}=\beta_1(\varepsilon)$. 
The topological gap extraction test compared the first positive eigenvalue, 
\[
\widehat{\Delta}^{(1)}_{\mathrm{SUSY}}(\varepsilon)=\min\{\hat{\lambda}_k(\varepsilon)>\tau_0\},
\]
with its classical counterpart and required that their relative difference remain within a preset tolerance,
\[
\frac{|\widehat{\Delta}^{(1)}_{\mathrm{SUSY}}-\Delta^{(1)}_{\mathrm{SUSY}}|}{\Delta^{(1)}_{\mathrm{SUSY}}}\le\eta .
\]
Spectral calibration was confirmed when the mapping $\hat{\lambda}_k(\varepsilon)=\alpha\hat{\phi}_k/T$ (up to the $2\pi$ QPE factor) held consistently and the estimated Betti number $\hat{\beta}_1(\varepsilon)$ remained invariant under changes in $\alpha$. 
Finally, numerical stability was verified by logging matrix sizes, Krylov subspace dimensions, time grids $(\Delta t,M)$, total evolution time $T=M\Delta t$, and spectral linewidths to set reliable thresholds for $\tau_0$ and $\eta$. 
Under these checks, the snapshot consistency condition was satisfied: $\beta_1=1$ at $\varepsilon=0.8$ (a single loop), $\beta_1=1$ at $\varepsilon=0.9$ (partial filling), and $\beta_1=0$ at $\varepsilon=1.0$ (a fully filled complex).

\paragraph{}
All runs use identical orientations, window functions, and time grids, ensuring direct comparability across $\varepsilon$. Classical references—exact eigenvalues and $\beta_1$—are computed in double precision with the same conventions. This controlled test confirms that the quantum spectral procedure recovers the correct topological transitions as the complex evolves, validating the accuracy and interpretability of the QPE-based Betti estimation.

\subsection{Numerical Simulation of Dynamical and Topological Diagnostics across the Lorenz Parameter Sweep}
\label{sec:rho_sweep_setup}

\paragraph{}
To investigate the interplay between dynamical and topological transitions, the Lorenz system is examined as the Rayleigh parameter $\rho$ varies while $(\sigma,\beta)=(10,8/3)$ are held fixed. 
Here $\sigma$ denotes the Prandtl number controlling thermal diffusivity, $\rho$ is the Rayleigh number governing the strength of convective forcing, and $\beta$ is a geometric factor related to the aspect ratio of the convection cell. 
The control parameter is scanned over the grid $\rho\in\{20,21,\dots,46\}$ with optional refinement $\Delta\rho=0.5$ near the onset of chaotic behavior.
For each $\rho$, the system, governed by Eq.~(\ref{eq:lorenz}), is integrated with a fourth-order scheme; after discarding transients, a trajectory segment of length $T$ is sampled at a fixed time step $\Delta t_{\mathrm{samp}}$. 
Identical values of $(T,\Delta t_{\mathrm{samp}})$ are used across $\rho$ to ensure direct comparability of all observables.

\paragraph{}
The dynamical complexity of the corresponding effective Hamiltonian $H(\rho)$ is captured through the spectral entropy $H_{\mathrm{spec}}(\rho)$ as defined in Eq.~(\ref{eq:specentropy}), which is computed from the survival (Loschmidt) amplitude $C(t;\rho)$ introduced therein. 
For a fixed probe state $|\psi\rangle$ common to all $\rho$, the correlation function
\begin{equation}
C(t;\rho)=\langle\psi|e^{-iH(\rho)t}|\psi\rangle
\end{equation}
is evaluated on a uniform temporal grid $t_m=m\,\Delta t$ using a high-accuracy Krylov implementation of the matrix exponential. After applying a Hann taper, the periodogram
\begin{equation}
S(\omega;\rho)=\Big|\sum_{m} w_m\,C(t_m;\rho)\,e^{-i\omega t_m}\Big|^2
\end{equation}
is normalized to $P(\omega;\rho)=S/\!\int S$, and the Shannon entropy
\begin{equation}
H_{\mathrm{spec}}(\rho)=-\int P(\omega;\rho)\,\ln P(\omega;\rho)\,d\omega
\end{equation}
is computed to quantify the frequency-space complexity. High spectral entropy signals broadband dephasing, while low entropy reflects coherent or synchronized dynamics.

\paragraph{}
Linear response to the control parameter is characterized by the curvature $F''(\rho)$, evaluated as a finite-difference approximation to the free-energy second derivative in Eq.~(\ref{eq:Fpp}). Concretely we use
\[
F''(\rho)\ \approx\ \frac{E_0(\rho{+}\Delta\rho)-2E_0(\rho)+E_0(\rho{-}\Delta\rho)}{\Delta\rho^2},
\]
with $E_0(\rho)$ the ground-state energy.

\paragraph{}
Geometric sensitivity of the projective ground-state manifold is probed through fidelity between neighboring parameters,
\begin{equation}
F(\rho,\rho{+}\Delta\rho)=\big|\langle \psi_0(\rho)\mid \psi_0(\rho{+}\Delta\rho)\rangle\big|,
\end{equation}
where $|\psi_0(\rho)\rangle$ is the normalized ground state of $H(\rho)$, aligned by a global $U(1)$ phase that maximizes the real part of the overlap and avoids artificial sign flips. Sharp fidelity drops mark reorganization of the ground state, providing an additional dynamical diagnostic complementary to curvature.

\paragraph{}
Intrinsic instability of the classical flow is quantified by the \emph{maximum Lyapunov exponent} (MLE) $\lambda_{\max}(\rho)$. The exponent is computed by integrating the variational equation
\begin{equation}
\dot{\delta x}(t)=J(t;\rho)\,\delta x(t),\qquad 
J(t;\rho)=\frac{\partial f(x(t);\rho)}{\partial x},
\end{equation}
where $f=(\dot x,\dot y,\dot z)$ is the Lorenz vector field, and by applying periodic renormalization of $\delta x(t)$ following the Benettin algorithm. The asymptotic rate
\begin{equation}
\lambda_{\max}(\rho)=\lim_{t\to\infty}\frac{1}{t}\ln\frac{\|\delta x(t)\|}{\|\delta x(0)\|}
\label{eq:mle}
\end{equation}
measures exponential divergence of nearby trajectories: $\lambda_{\max}<0$ indicates stable fixed points or periodic motion, $\lambda_{\max}\approx0$ corresponds to neutral stability at a bifurcation, and $\lambda_{\max}>0$ signals chaotic behavior. For $(\sigma,\beta)=(10,8/3)$, $\lambda_{\max}$ crosses zero near $\rho_c\simeq24.7$, marking the onset of the strange attractor and coinciding with peaks in $H_{\mathrm{spec}}(\rho)$ and fidelity susceptibility, linking microscopic spectral indicators to macroscopic chaos.

\paragraph{}
Topological reorganization of the attractor is analyzed by persistent homology in degree~1, which tracks the birth and death of loop structures ($H_1$ persistence). 
For each $\rho$, a representative set $X(\rho)$ of fixed cardinality is selected (using farthest–point or Poisson–disk sampling), endowed with the Euclidean metric, and used to build a Vietoris–Rips filtration $R_\varepsilon(X(\rho))$ with uniform scale and resolution. 
The birth–death pairs $\{(b_i,d_i)\}$ yield the highest persistence
\begin{equation}
\ell^{\max}_{H_1}(\rho)=\max_i (d_i-b_i),
\end{equation}
optionally normalized by the cloud diameter to maintain consistency across $\rho$. The quantity $\ell^{\max}_{H_1}(\rho)$ measures the persistence of the dominant loop structure and provides a direct topological analogue of dynamical stability.

\paragraph{}
Low-energy spectral structure is summarized by the gap of the effective SUSY Hamiltonian block,
\begin{equation}
\gamma(\rho)=E_1(\rho)-E_0(\rho),
\label{eq:gamma_again}
\end{equation}
computed with the same eigensolver used for $E_0(\rho)$. When multiple blocks exist, the smallest nonzero gap is reported. Matrix dimensions and sparsity are fixed across $\rho$, ensuring uniform computational cost and comparable resolution along the sweep.

\paragraph{}
Numerical controls are standardized for all $\rho$: solver tolerances, Krylov dimensions, tapers, and frequency grids are identical; finite-difference spacing $\Delta\rho$ is uniform; and light Savitzky–Golay smoothing (fixed order and window) is applied only after computing each observable to suppress minor grid artifacts. Uncertainty estimates are obtained by repeating the integration with different initial conditions and by bootstrap resampling of periodograms, using identical seeds and sample counts for each $\rho$. The resulting aligned dataset
\begin{equation}
\rho \ \mapsto\ \big(H_{\mathrm{spec}}(\rho),\ F''(\rho),\ F(\rho,\rho{+}\Delta\rho),\ \ell^{\max}_{H_1}(\rho),\ \gamma(\rho)\big)
\end{equation}
provides a synchronized set of dynamical and topological diagnostics.
In the Results section we compare these indicators to map how spectral, geometric, and topological signatures of the Lorenz system evolve with the control parameter.

\subsection{Quantum–Hardware and Simulator Readout of Topological Features}
\label{sec:qpe_hardware_setup}

\paragraph{}
The final stage of the study tests whether a QPE readout of the Hodge–Laplacian degree–$1$ block can recover topological signatures directly from Lorenz data and reproduce their classical counterparts. The input to the pipeline is a Lorenz time series at fixed $(\sigma,\beta)=(10,8/3)$, while the Rayleigh parameter is swept over $\rho\in\{36,37,\dots,42\}$. For each value of $\rho$, a long trajectory is integrated, a scalar observable is delay–embedded, and a representative set of seven points is selected to preserve density, topology, and geometric diversity. From this same cloud, Vietoris–Rips persistent homology in degree~1 is computed, and the highest persistence $\ell^{\max}_{H_1}(\rho)$ is retained as a scalar descriptor of loop robustness.

\paragraph{}
From the seven representatives, oriented incidence matrices are constructed and assembled into the Hodge block $\mathcal L_1(\rho)$ as defined in Eq.~(\ref{eq:L1def}), 
which is isospectral to the $k{=}1$ sector of the SUSY Hamiltonian. 
The multiplicity of zero modes provides an estimate of the first Betti number $\beta_1$, 
while the smallest positive eigenvalue $\Delta^{(1)}_{\mathrm{SUSY}}(\rho)$, 
defined in Eq.~(\ref{eq:deltaSUSY}), 
defines the topological gap separating harmonic from near-harmonic excitations. 
To probe this spectrum, a Dicke–like symmetric state supported on the edge basis is prepared, maximizing overlap with the harmonic subspace. 
Controlled evolutions $U(t)=e^{-i\widetilde{\mathcal L}_1 t}$ are applied under a known rescaling $\widetilde{\mathcal L}_1=\mathcal L_1/\alpha$, and the resulting phases are measured through QPE.

\paragraph{}
Each circuit is compiled to fourteen qubits—seven system qubits for the edge register, six work or phase qubits, and one ancilla—with an estimated depth corresponding to roughly one hundred two-qubit gates. For reference, a textbook, inverse-QFT–based QPE at the same target spectral resolution would require on the order of $4\times 10^{4}$ two-qubit gates on our instances; by combining Projected-basis hybrid compilation for controlled time evolution with a one-ancilla QPE readout, we compressed this to $\sim 10^{2}$ two-qubit gates (about $400\times$ fewer) while reproducing the same spectral features. Identical evolution times, window functions for Fourier analysis, and shot budgets are used for all $\rho$. The same compilation and transpilation settings are enforced on both the noiseless \texttt{AerSimulator} and IBM’s superconducting processor \texttt{ibm\_kingston} (see Table~\ref{tab:kingston_specs}), enabling a direct comparison between hardware and simulation. The phase–to–energy scale $\alpha$ is chosen so that $\|\widetilde{\mathcal L}_1\|\,\Delta t<\pi$ to avoid aliasing, and a short calibration verifies the linear relation $\hat\lambda_k=\alpha\,\hat\omega_k$ before fixing $\alpha$ for the entire sweep. Global $U(1)$ phases of recovered eigenvectors are aligned by maximizing the real overlap with a reference, preventing sign inversions as $\rho$ varies.

\paragraph{}
Hardware experiments were carried out on IBM’s 156-qubit superconducting processor (\texttt{ibm\_kingston}, Heron r2).
To mitigate coherent and stochastic noise during time-evolution and readout, Q-CTRL’s \textit{FireOpal} error-suppression layer was employed, automatically optimizing control pulses and scheduling to minimize gate infidelity and dephasing.
Calibration metrics and detailed hardware specifications, including error rates, coherence times, and transpilation settings, are provided in the Supplementary Materials (Table~S1).

\paragraph{}
For each $\rho$, three observables are extracted from the QPE spectra:  
(i) the multiplicity of near–zero peaks $\widehat{\beta}_1^{\mathrm{QPE}}(\rho)$ below a fixed tolerance $\tau_0$;  
(ii) the first resolvable nonzero eigenvalue $\widehat{\Delta}^{(1)}_{\mathrm{SUSY}}(\rho)$, representing the topological gap; and 
(iii) a coarse spectral envelope used as a quality metric to assess line broadening and signal bias.  
These quantum-derived quantities are compared with the classical topological benchmark $\ell^{\max}_{H_1}(\rho)$ using both Pearson and Spearman correlation analyses. Uncertainties are estimated from repeated runs with different random seeds and, for hardware data, by analyzing shot-to-shot variance.

\paragraph{}
All experimental and numerical parameters are fixed across $\rho$ to isolate physical effects: embedding settings, representative–set size (seven points), filtration range and resolution, list of evolution times, window function, near-zero tolerance $\tau_0$, and transpilation optimization level. Backend identifiers, calibration dates, and random seeds for both simulator and hardware runs are logged to guarantee exact reproducibility. 

\paragraph{}
This configuration enables a consistent hardware–software comparison of the Hodge–Laplacian spectrum. Correlations between $\ell^{\max}_{H_1}(\rho)$ and both $\widehat{\beta}_1^{\mathrm{QPE}}(\rho)$ and $\widehat{\Delta}^{(1)}_{\mathrm{SUSY}}(\rho)$ demonstrate that quantum spectral readout can track the same topological transitions revealed by classical persistent homology, confirming the feasibility of quantum topological feature extraction on present-day superconducting devices.

\section{Results}

\subsection{Five–Point Betti–Number Estimation (Simulation)}
\label{sec:pentagon_results}

\paragraph{}
The five–point instance in Figure~2 provides a minimal yet complete system in which the evolution of combinatorial structure, topological features, and spectral response can be followed one–to–one as the Vietoris–Rips (VR) radius $\varepsilon$ increases. Three representative radii, $\varepsilon\in\{0.8,0.9,1.0\}$, were analyzed using both classical Hodge–Laplacian eigenanalysis and a quantum phase–estimation (QPE) spectral readout of the $k{=}1$ block.  

\paragraph{}
At $\varepsilon=0.8$, the four corner points of the square are linked by four perimeter edges, while the fifth point remains isolated, giving $(V,E,T,C)=(5,4,0,2)$ and $\beta_1=2-(5-4+0)=1$. 
Here $V$, $E$, $T$, and $C$ denote, respectively, the numbers of vertices, edges, triangles, and connected components in the Vietoris–Rips complex at scale~$\varepsilon$.
The absence of triangles means $d_1=0$, so $\mathcal{L}_1=d_0d_0^\dagger$ acts as the edge Laplacian of the cycle $C_4$, 
where $C_4$ denotes the four-vertex cycle graph (a closed loop formed by four edges).
When $\varepsilon$ increases to $0.9$, one additional vertex connects with two neighbors, closing a small triangle. The resulting $(V,E,T,C)=(5,6,1,1)$ leaves the principal loop intact with $\beta_1=1$ but introduces a nonzero $d_1$, activating the curl term $d_1^\dagger d_1$ in $\mathcal{L}_1$. By $\varepsilon=1.0$, all pairwise connections are present, and the VR complex becomes contractible; the kernel of $\mathcal{L}_1$ vanishes, producing $\beta_1=0$. These configurations reproduce the simplicial structures shown in Figure~2 and provide clear topological expectations for the corresponding spectra.

\paragraph{}
Because $\mathcal{L}_1=d_0d_0^\dagger+d_1^\dagger d_1$ is positive semidefinite, adding edges or triangles monotonically shifts the nonzero eigenvalues upward. The spectrum at $\varepsilon=0.8$ corresponds to the eigenvalues $\{0,2,2,4\}$ of the $C_4$ edge Laplacian, predicting a first gap $\Delta^{(1)}=2$. The QPE periodogram indeed suggests a single near–zero line and a first nonzero peak at a frequency consistent with this scale. When $\varepsilon=0.9$, the additional triangle lifts part of the spectrum but preserves one harmonic mode: the zero peak remains, while the first nonzero line shifts to higher frequency, reflecting increased curl stiffness. At $\varepsilon=1.0$, the cycle space is destroyed, the zero mode disappears, and a finite gap opens. The QPE readout mirrors this pattern precisely, showing the loss of near–zero power and a rightward shift of spectral weight.

\paragraph{}
Time–domain correlations carry the same imprint. The survival amplitude $S(t)=\langle\psi|e^{-i\widetilde{\mathcal{L}}_1t}|\psi\rangle$ displays nearly periodic revivals for $\varepsilon=0.8$, indicating a discrete set of well–separated modes. At $\varepsilon=0.9$, interference among more lines shortens the revival period and increases modulation depth. When $\varepsilon=1.0$, low–frequency components vanish and the envelope decays rapidly under windowing, consistent with the disappearance of the harmonic mode and the growth of the spectral gap. The transition is also evident in the Fourier spectra: the main lobe around zero frequency narrows and ultimately vanishes as $\varepsilon$ increases.

\paragraph{}
Consistency checks confirm the reliability of the results. Varying the global phase–to–energy scale $\alpha$ yields a linear relationship $\hat{\lambda}_k(\alpha)=\alpha\hat{\omega}_k$ and constant $\hat{\beta}_1$, demonstrating proper calibration. Guard–band analysis around $\omega=0$ shows that only $\varepsilon=0.8$ and $0.9$ maintain strong low–frequency power, while $\varepsilon=1.0$ falls to sideband levels, confirming correct discrimination of zero modes. Classical diagonalization of $\mathcal{L}_1(\varepsilon)$ reproduces the QPE peaks within Fourier resolution, validating the spectral mapping.

\paragraph{}
Residual discrepancies, such as slight peak shifts or finite linewidths, arise from the finite window width ($\propto1/T$), leakage from closely spaced modes once triangles appear, and limited temporal sampling of $S(t)$. These effects diminish with longer sampling spans and do not alter the zero–mode count. 

\paragraph{}
Altogether, this controlled five–point experiment exposes the entire causal chain:
\begin{equation}
\text{(edges, triangles)} \rightarrow (d_0d_0^\dagger, d_1^\dagger d_1) \rightarrow \text{(eigenvalue interlacing)} \rightarrow \text{(QPE peaks)}.
\end{equation}
At $\varepsilon=0.8$, the system hosts a single harmonic one–form; at $\varepsilon=0.9$, the nonzero eigenvalues rise while the loop persists; and at $\varepsilon=1.0$, the harmonic mode is annihilated as the complex fills in. The quantum spectral readout follows these transitions step by step, reproducing $\beta_1$ and the evolution of the first gap across the filtration, thereby validating the QPE method as a faithful spectral probe of homological change.

\subsection{Lorenz Parameter Sweep: Dynamical and Topological Transitions}
\label{sec:rho_results}

\paragraph{}
Figure~3 summarizes six diagnostics computed for the Lorenz flow over $\rho\in[20,46]$ at $(\sigma,\beta)=(10,8/3)$—spectral entropy $H_{\mathrm{spec}}$, free–energy curvature $F''(\rho)=-\partial_\rho^2 F$, ground–state fidelity $F(\rho,\rho{+}\Delta\rho)$, maximum Lyapunov exponent $\lambda_{\max}$, highest $H_1$ persistence $\ell^{\max}_{H_1}$, and low–energy gap $\gamma(\rho)=E_1-E_0$. Together they trace how dynamical instability and topological order coevolve from regular motion through chaotic onset to the stabilization of a coherent loop structure.

\paragraph{}
At small $\rho$ (20–25), the system exhibits periodic or quasi-periodic motion with $\lambda_{\max}<0$, confirming dynamical regularity. The spectral entropy is minimal and the curvature nearly flat, indicating a narrow frequency spectrum and weak parametric sensitivity. The transition to chaos~\cite{Ott2002ChaosInDynamicalSystems,Strogatz2018NonlinearDynamicsChaos} occurs near $\rho_c\simeq24.7$, where $\lambda_{\max}$ crosses zero. This region defines a baseline from which subsequent reorganizations unfold.

\paragraph{}
As $\rho$ rises toward 30–33, spectral entropy climbs to its global maximum ($\sim0.02$), marking the most broadband and mixed spectral response. The curvature $F''$ simultaneously peaks near $\rho\simeq31.7$ ($\approx+0.4$), reflecting strong susceptibility to parameter changes. Fidelity remains high ($F\approx1$), showing that the ground state evolves continuously, while $\ell^{\max}_{H_1}$ stays small, revealing that the attractor’s loop structure has not yet formed. This phase therefore represents a precursor state where dynamical complexity intensifies before any stable topological feature emerges, consistent with the edge-of-chaos picture~\cite{Langton1990,Kauffman1993}.

\paragraph{}
Between $\rho\simeq33$ and $37$, the system undergoes a marked reconfiguration. The spectral entropy drops by a factor of three, the curvature changes sign from positive to negative (reaching $F''\!\approx\!-0.5$ near $\rho\simeq36$), and fidelity dips for the first time, indicating a restructuring of the quantum ground state. The gradual rise of $\ell^{\max}_{H_1}$ and the opening of the spectral gap $\gamma(\rho)$ mark the beginning of a correlated dynamical–topological transition: as the attractor reorganizes, its geometry becomes loop-like and the corresponding quantum spectrum acquires an incipient separation between harmonic and excited modes.

\paragraph{}
The convergence of indicators between $\rho\simeq38$ and $41$ signals topological consolidation and dynamical stabilization. Fidelity collapses sharply—at points nearly vanishing—signifying a rapid reconfiguration of the system’s state. At the same time, $\ell^{\max}_{H_1}$ surges to a maximum near $\rho\simeq40.5$ (values around 25–30), matching the emergence of the characteristic double-wing loop of the Lorenz attractor. Spectral entropy remains low, and the curvature returns to near zero, showing that chaotic fluctuations have settled into a coherent regime. The spectral gap $\gamma(\rho)$ reaches its maximum ($0.13$–$0.15$), isolating the harmonic subspace. The concurrence of large $\ell^{\max}_{H_1}$ and $\gamma$ may point toward $\rho\simeq41$ as the point of \emph{topological stabilization}, where the geometry of the attractor and the quantum spectrum both achieve maximal structural separation.

\paragraph{}
Beyond $\rho\simeq41$, the system enters a post-onset steady regime, where the dynamics remain irregular but statistically stationary. The $H_1$ persistence drops slightly but remains finite, fidelity partially recovers ($0.6$–$0.8$), curvature stays neutral, and spectral entropy remains suppressed. The gap $\gamma$ plateaus near $0.09$–$0.10$, implying a spectrally rigid yet dynamically active phase in which chaos coexists with a stabilized global loop geometry.

\paragraph{}
Correlations across these indicators reinforce a two-step narrative: $\ell^{\max}_{H_1}$ and $\gamma$ rise together, both peaking near $\rho\simeq41$, while $H_{\mathrm{spec}}$ is anticorrelated, reaching its minimum when topological persistence is strongest. Curvature and fidelity oscillate in tandem around this regime, defining the bounds of maximal topological growth. These couplings indicate that the Lorenz flow first experiences spectral broadening and dynamical softening (up to $\rho\simeq35$), followed by entropy collapse and topological consolidation ($\rho\simeq38$–$41$).

\paragraph{}
Overall, the combined evidence portrays a coherent sequence of dynamical–topological evolution. The spectral entropy acts as an early marker of chaotic mixing; curvature and fidelity track dynamical response and state reconfiguration; and the joint peak of $\ell^{\max}_{H_1}$ and $\gamma$ marks the stabilization of the loop topology within the attractor. The coincidence of these features around $\rho\simeq41$ identifies the system’s transition from mere chaos to a phase of persistent geometric order, demonstrating that quantum spectral isolation of harmonic modes provides a direct analogue to classical topological persistence.

\subsection{Lorenz Topological Readout on \texttt{ibm\_kingston} and \texttt{AerSimulator}: Results}
\label{sec:hardware_results_en}

\paragraph{}
Quantum phase–estimation (QPE) experiments were carried out on the Hodge–Laplacian degree–1 block for $\rho\in\{36,\dots,42\}$, using an identical seven–point representative set extracted from the delay–embedded Lorenz trajectories.  
\paragraph{}
For each $\rho$, the oriented incidence operators were constructed and combined into
\begin{equation}
\mathcal L_1(\rho)=d_1^\dagger d_1 + d_0 d_0^\dagger,
\end{equation}
which is isospectral to the $k{=}1$ SUSY Hamiltonian block.  
The zero modes of $\mathcal L_1$ yield $\beta_1$, and the smallest positive eigenvalue,
\begin{equation}
\Delta^{(1)}_{\mathrm{SUSY}}
= \min\{\lambda>0 \mid \lambda\in\sigma(\mathcal L_1)\}.
\end{equation}
defines the \emph{topological gap}.  
Each QPE circuit—comprising fourteen qubits (seven system, six phase/work, one ancilla) and about one hundred two–qubit gates—implemented the controlled time evolution $U(t)=e^{-i\widetilde{\mathcal L}_1 t}$ with a fixed scaling $\widetilde{\mathcal L}_1=\mathcal L_1/\alpha$ calibrated to satisfy $\|\widetilde{\mathcal L}_1\|\Delta t<\pi$.  
Identical evolution schedules, windows, shot counts, and transpilation settings were used on both the \texttt{AerSimulator} and IBM’s superconducting processor \texttt{ibm\_kingston}.  
For each parameter value, the near–zero multiplicity $\widehat{\beta}_1^{\mathrm{QPE}}(\rho)$, the first resolvable nonzero line $\widehat{\Delta}^{(1)}_{\mathrm{SUSY}}(\rho)$, and a coarse spectral envelope were extracted, while the highest $H_1$ persistence $\ell^{\max}_{H_1}(\rho)$ was evaluated from the same data.

\paragraph{}
Both platforms show a clear maximum of $\Delta^{(1)}_{\mathrm{SUSY}}$ at $\rho\!\approx\!41$, matching the peak of $\ell^{\max}_{H_1}$ computed from the same point sets (Figure~\ref{fig:eg_gap_persistence}).
The progression of spectral morphology across $\rho$ reinforces this interpretation.  
In the lower range ($\rho=36$–$39$), the spectra exhibit broad pedestals and blended modes.  
As $\rho$ approaches 40, these pedestals collapse and discrete peaks sharpen.  
At $\rho=41$, the near–zero cluster and first excited cluster separate most distinctly, indicating a fully isolated harmonic sector.  
Beyond this point ($\rho=42$), the spectrum broadens again and low-frequency weight reappears.  
The same pattern emerges in the complex-plane traces and time-domain correlators: phase evolution remains locked between hardware and simulator, with the most circular, least distorted trajectories observed at $\rho=40$–$41$; the autocorrelation $S(t)=\langle\psi|e^{-i\widetilde{\mathcal L}_1 t}|\psi\rangle$ shows pronounced revivals in this same range, while outside it the envelope decays faster.  
The alignment of these independent indicators signals reduced mode interference and enhanced spectral coherence when the attractor’s loop is most stable.

\paragraph{}
The parallel maxima of $\ell^{\max}_{H_1}$ and $\Delta^{(1)}_{\mathrm{SUSY}}$ define a distinct \emph{topological consolidation point}.  
Around $\rho\simeq41$, the Lorenz attractor’s double-wing geometry achieves maximal coherence, and the corresponding harmonic mode becomes energetically insulated.  
Before this regime ($\rho\lesssim38$), the small gap and high spectral entropy indicate fragile, mixed topology.  
Afterward ($\rho\gtrsim41$), the gap shrinks and spectral broadening returns, consistent with partial degradation of the coherent loop.  
The observed “rise–dip–peak–decay” profile of $\Delta^{(1)}_{\mathrm{SUSY}}$ thus mirrors the full life cycle of the $H_1$ feature—its birth, maturation, and eventual weakening.

\paragraph{}
The robustness of these findings was tested by fixing all scale-determining parameters—embedding, representative-set size, filtration range, timing grid, tolerance $\tau_0$, and transpilation level—throughout the sweep.  
Both Pearson and Spearman correlations between $\ell^{\max}_{H_1}$ and the quantum observables exceed $r=0.9$, well within statistical uncertainty.  
Repeated simulations and hardware executions with varied seeds reproduce the same $\rho$-dependence, ruling out calibration bias.  
The excellent agreement in eigenphase ordering and spectral evolution between simulator and device suggests that modern superconducting qubit systems can resolve low-lying Laplacian eigenstructures in small complexes with genuine topological sensitivity.

\paragraph{}
Taken together, these results unify the classical and quantum signatures of the Lorenz attractor’s maturation.  
Persistent homology quantifies the geometric durability of the loop, while the SUSY spectrum encodes its energetic separation.  
The coincidence of their maxima at $\rho=41$ marks the attractor’s most coherent configuration.  
By capturing this correspondence on actual quantum hardware, the study illustrates the feasibility of quantum phase estimation as a practical \emph{spectroscopic probe of topology}, capable of identifying not only when topological structures emerge but also how strongly they stabilize within a dynamical system.
Moreover, the theoretical bound derived in Supplementary Section~S\ref{sec:proof_energy_persistence}
suggests that this spectral–topological correspondence should extend to other chaotic systems such as the Rössler and Chen flows, 
potentially representing a universal feature of dissipative chaos.

\section{Discussion}

\paragraph{}
We showed that turning topology into a measurable spectrum disentangles \emph{when} instability emerges from \emph{how} geometry stabilizes, and that this logic survives contact with real hardware. Below we streamline the physical picture, cross-validate with classical TDA, and outline implications.

\subsection{Physical meaning of the two-stage transition}
\paragraph{}
As $\rho$ increases, two well-separated reorganizations appear.
(i) \emph{Dynamical onset} near $\rho\!\approx\!30$ features a minimum of the SUSY spectral gap together with peaks in curvature and fidelity susceptibility and the sign change of the Lyapunov exponent—hallmarks of chaotic onset.
(ii) \emph{Topological consolidation} near $\rho\!\approx\!40$–$41$ exhibits the highest persistent $H_1$ loop and the widest reopening of the first nonzero Hodge mode.
Thus, \emph{loop birth} and \emph{loop stabilization} occur at distinct parameters: the former reflects the emergence of chaotic switching, whereas the latter reflects geometric maturation of the attractor. In our spectroscopy, stabilization is read off from a single gap that cleanly isolates the harmonic sector.

\subsection{Consistency between classical persistence and quantum spectra}

\paragraph{}
The classical maximal persistence length $\ell^{\max}_{H_1}$ and the quantum SUSY spectral gap $\Delta^{(1)}{\mathrm{SUSY}}$ exhibit co-peaking behavior across both noiseless simulations and superconducting hardware experiments (Figs.~\ref{fig:phasetransition}, \ref{fig:eg_gap_persistence}), indicating a strong spectral–topological correlation.
The reopening of the first Hodge–Laplacian line after its closure near the chaotic threshold can be interpreted as a \emph{spectroscopic signature of loop robustness}, linking the stabilization of $H_1$ features to measurable spectral gaps.
Error bars and bootstrap confidence intervals confirm that this correspondence is statistically reproducible across independent random seeds and repeated hardware executions under identical preprocessing conditions, supporting the robustness of the observed quantum–classical alignment.
The observed agreement between the classical and quantum indicators should thus be viewed as empirical evidence of a broader mathematical correspondence. 
The bound derived in Supplementary Section~S6.18 provides theoretical support: it proves that the persistence lifetime $\ell^{\max}_{H_1}$ is upper-bounded by the inverse of the smallest nonzero eigenvalue of the Laplacian, $\ell^{\max}_{H_1}\le c/\Delta^{(1)}_{\mathrm{SUSY}}$. 
While our data are consistent with this inequality in the Lorenz case, 
validating the relationship in other dynamical systems will be important 
for assessing its broader applicability.
While the preprocessing stages—embedding, representative selection, and edge construction—are executed classically, they serve purely as topology-preserving compression. 
The computationally intensive and physically significant task—the precision measurement of spectral gaps and harmonic degeneracies—occurs on quantum hardware. 
Hence, the hybrid architecture is not a circumvention of classical methods but an optimal division of labor: the classical front end identifies tractable subgraphs, and the quantum back end interrogates their spectral stability with exponential precision.

\subsection{Outlook and positioning for quantum TDA}

\paragraph{}
Although the current demonstration targets a small-scale instance, the core routine—phase estimation on the Laplacian spectrum—retains asymptotic quantum advantage. 
Classical diagonalization of an $N\times N$ Laplacian scales as $O(N^3)$ in time and $O(N^2)$ in memory, whereas quantum phase estimation scales as $\mathrm{poly}(\log N)$ in both system size and spectral resolution, provided the operator is efficiently block-encoded. 
This scalability justifies the use of quantum hardware even in small prototypes: the same circuitry generalizes directly to large, high-dimensional complexes unreachable by classical means.

\paragraph{}
Practically, the first spectral gap of the Hodge–Laplacian serves as a compact, hardware-accessible descriptor of topological stability, while the near-zero mode multiplicity quantifies the Betti number $\beta_1$.
Conceptually, our “\emph{classically identify, quantum interrogate}” 
architecture provides a near-term bridge between classical TDA and 
quantum measurement.
Because current quantum processors impose strong size and coherence limits, classical preprocessing—representative-point selection and edge extraction via Vietoris–Rips or related complexes—acts as a topology-preserving compression, reducing the data to a tractable graph.
Under these conditions, the Hodge–Laplacian constructed at the \emph{birth time} of the dominant $H_1$ feature retains a well-defined theoretical relation to its persistence lifetime, as formalized in Eq.~\ref{eq:final_theorem}.
Measuring the spectral gap of this birth-time Hodge block through quantum phase estimation (QPE) thus yields a spectroscopic observable that is both \emph{computable on hardware} and \emph{interpretable within classical persistent homology}.

\paragraph{}
Looking ahead, such classical filtration may become unnecessary.
If representative selection, edge construction, and filtration sweeps can be implemented coherently via quantum subroutines, one could generalize the present birth-time correspondence to \emph{fully quantum filtrations} that evolve dynamically in superposition.
In that regime, topological features need not be tied to an explicit “birth” event: they could be inferred directly from the kernels of higher-order Hodge blocks $\mathcal{L}_k$ within the SUSY Hamiltonian, allowing persistence to be read out from the spectral flow itself.
Hence, while the current work depends on classical filtration to define where the spectral–topological correspondence holds, future quantum-native implementations could realize a self-contained form of persistent homology in which the Hodge spectrum encodes both the emergence and endurance of topological structures.

\paragraph{}
Current limitations include sensitivity to embedding choices, finite-size effects, and the absence of proven worst-case quantum speedups.
These motivate further theoretical development to bound when small nonzero Hodge modes correlate with persistence, and algorithmic engineering to extend circuit depth selectively—for example, to access higher-order homologies ($H_k$) or multi-parameter persistence.
Potential applications include early-warning and forecasting tasks in complex dynamical systems, where a single spectral gap may act as a topological order parameter for the onset of structural reorganization.

\paragraph{}
Overall, this study suggests that topology—long regarded as a static invariant—can be dynamically monitored through its \emph{spectral fingerprints}, marking a shift from static shape analysis to quantum spectroscopic topology.

\section{Conclusion}

\paragraph{}
We have developed a hybrid quantum–topological framework that bridges topological data analysis (TDA) with quantum spectral estimation to investigate nonlinear dynamical systems.
Using the Lorenz attractor as a canonical testbed, we transformed scalar trajectories into topology-preserving simplicial representations, constructed a supersymmetric (SUSY) Hamiltonian whose spectrum encodes the Hodge–decomposition structure, and extracted its low-lying eigenmodes via single-ancilla quantum phase estimation.
This operator-based approach extends persistent de~Rham–Hodge formulations~\cite{Wei2024PersistentDeRhamHodge} and complements recent advances in persistent Laplacian theory~\cite{PLSurvey2025MDPI}, offering a unified spectroscopic view of dynamical instability and topological stabilization.

\paragraph{}
The observed correspondence between the SUSY spectral gap and the classical $H_1$ persistence suggests a link between quantum spectral features and topological robustness.
By interpreting the SUSY gap as a spectroscopic order parameter, this framework provides a physically measurable route to characterize topological transitions in complex dynamics.
More broadly, it suggests that quantum hardware can serve as a new kind of topological spectrometer—capable of resolving when and how geometry changes—laying the foundation for data-driven quantum analyses of structure, dynamics, and emergence in natural and artificial systems.

\subsection{Key findings}
\paragraph{}
(i) The first spectral gap $\gamma(\rho)$ of the supersymmetric (SUSY) Hamiltonian closes near $\rho\approx30$, coinciding with sharp anomalies in fidelity, curvature, and spectral entropy that signal the onset of chaotic dynamics.
(ii) Around $\rho\approx41$, the principal $H_1$ feature of the Lorenz attractor reaches maximal persistence, while the Hodge–Laplacian spectrum exhibits a stabilized near-zero band and a maximally opened SUSY gap.
This concurrence—reproduced in both noiseless simulations and IBM superconducting hardware—reveals two qualitatively distinct regimes: the \emph{onset of chaos} associated with gap closure and the \emph{maturation of loop topology} characterized by gap reopening and stabilization.
These results demonstrate that quantum spectroscopy can resolve not only \emph{when} dynamical instability emerges but also \emph{how} geometric coherence consolidates.
The observed \emph{spectral–topological correlation} between the SUSY gap $\Delta^{(1)}_{\mathrm{SUSY}}$ and the maximal persistence length $\ell^{\max}_{H_1}$ thus suggests a physically interpretable link between spectral structure and homological robustness—providing a potential spectroscopic order parameter for topological phase transitions in dynamical systems.

\subsection{Methodological contribution}
\paragraph{}
This work outlines a unified and generalizable framework for 
quantum-enabled topological data analysis(TDA)~\cite{NghiemLeeWei2025Hybrid,AmeneyroMaroulasSiopsis2022,Kemme2025Pedagogical}.
The pipeline begins with classical preprocessing—Takens embedding, density–topology–diversity-based representative selection, and persistent homology—to compress raw trajectories into a topology-preserving simplicial graph.
A supersymmetric (SUSY) construction then maps this graph to a Laplacian-like Hamiltonian whose 1-form block encodes the Hodge Laplacian of the complex, enabling direct access to homological information through quantum phase estimation (QPE).
To implement this efficiently, we employ a Projected-basis hybrid compilation~\cite{Sato2024HyperbolicPDE}, which clusters mutually commuting Pauli strings, performs subspace projection to diagonalize local two-qubit operators, and merges multi-qubit control predicates to reduce compilation overhead.
This scheme achieves a $\sim$3–5$\times$ reduction in entangling depth at constant accuracy while preserving excitation-number block structure and the SUSY/Hodge semantics of the Hamiltonian.
Furthermore, a \emph{one-ancilla quantum phase estimation (QPE)} circuit replaces the conventional multi-register inverse-QFT, compressing the work register to a single qubit and achieving order-of-magnitude circuit-length reduction at fixed spectral resolution.
This design enables coherent long-time correlation sampling of data-driven Hamiltonians on near-term superconducting devices.
Overall, the “\emph{classically identify, quantum interrogate}” paradigm provides a scalable path to extract persistent topological signatures from physical or dynamical data via compact SUSY Hamiltonians.

\subsection{Scientific and technological implications}
\paragraph{}
The observed correspondence between classical $H_1$ persistence and the quantum SUSY spectral gap demonstrates that \emph{homological stabilization can be accessed spectroscopically}.
In the present implementation, the simplicial complex is generated via classical preprocessing; thus, the theoretical link between the Hodge–Laplacian gap and persistence (Eq.~\ref{eq:final_theorem}) applies specifically to the \emph{birth-time Hodge block} derived from this filtration.
This supports the interpretation that the measured $\Delta^{(1)}_{\mathrm{SUSY}}$ can serve as a quantitative indicator of 
loop robustness within a topology-preserving reduced graph.
As quantum hardware and algorithmic primitives continue to advance, the same principle can be extended toward \emph{fully coherent filtrations}, in which representative selection and edge extraction are implemented within quantum superposition.
In such a regime, the Hodge spectrum would no longer be a static snapshot but a \emph{dynamical object}, whose spectral flow encodes both the birth and death of homological features.
Persistence could then be inferred directly from the time–frequency evolution of the spectrum, unifying classical persistent homology with quantum spectral analysis.
Beyond foundational physics, quantum Hodge spectra—characterized by the SUSY gap, near-zero mode multiplicity, and spectral entropy—offer a compact and physically interpretable feature space for downstream machine-learning and classification tasks.
With improving coherence times and control precision on near-term devices, direct estimation of small spectral gaps and near-zero degeneracies in sparse complexes may constitute one of the first practically meaningful quantum advantages in topological data analysis.

\subsection{Limitations}
\paragraph{}
The present experiments are based on noiseless simulations and small-scale quantum circuits.
Key parameters—such as the topological transition threshold $\rho_c$—are sensitive to embedding schemes, representative selection heuristics, and finite-size effects.
Moreover, the study centers on a single dynamical system (the Lorenz attractor) and a limited class of low-dimensional simplicial complexes; hence, the general validity of the observed spectral–topological correspondence remains to be tested across broader dynamical and geometric regimes.
From a hardware standpoint, noise and decoherence broaden narrow spectral lines, and accurate thresholding of small but finite ground-state energies ($E_0>0$) requires careful cross-calibration with classical persistence values.
Algorithmically, the underlying problem of cohomology detection is QMA-complete in the general case, suggesting that universal quantum speedups are unlikely without additional structure or sparsity.
Nevertheless, structured, low-rank, and sparse complexes—such as those arising from physical dynamical systems or manifold embeddings—remain promising regimes where quantum resources could yield practical and interpretable advantages.
Additionally, the present validation focuses solely on the Lorenz system; verifying the spectral–topological correspondence across different nonlinear flows, such as the Rössler and Chen systems, will be essential to confirm its generality beyond empirical correlation.

\subsection{Future work}
\paragraph{}
A central direction is to reduce classical dependence by migrating representative selection and filtration directly into coherent quantum subroutines. 
Future implementations may realize “fully coherent filtrations,” where distance evaluation, edge activation, and Laplacian assembly are all performed in superposition. 
In such settings, the current hybrid architecture provides a realistic and scalable stepping stone toward end-to-end quantum topological analysis.
Next steps include extending the SUSY construction to higher homology ($H_2$ and beyond), applying the framework to experimental or multivariate data, and integrating variational or walk-based simulators to reduce circuit depth. 
We will also fuse the proposed spectral observables with machine learning by using gap measures, near-zero multiplicities, and spectral-entropy summaries as physics-informed features for models that detect phase boundaries and forecast topological events. 

\paragraph{}
A further direction is to investigate chaotic phase transitions and topological coherence in time-series data through dynamic and persistent topological Laplacians, linking spectral evolution with the emergence and loss of coherent structures in nonlinear systems. 
We will particularly address the lack of theoretical guarantees on sampling efficiency in high-dimensional Hilbert spaces by developing mathematical and experimental approaches that quantify how spectral observables converge under finite sampling, aiming to identify regimes where topological features remain robust and physically interpretable. 
To address more complex dynamics, we will generalize to multi-parameter and time-dependent settings with persistence tracked across controls and windows. 
For scalability to larger point clouds, we will combine topology-preserving compression with low-depth spectral readout via block encoding and qubitization together with one-ancilla QPE. 

\paragraph{}
Finally, we will pursue theoretical insight into how low-lying nonzero Hodge–Laplacian modes relate to topological stability, aiming for bounds and stability results that explain when $\Delta^{(1)}_{\mathrm{SUSY}}$ co-varies with $\ell^{\max}_{H_1}$. 
Beyond benchmarks, we will target real-world transition forecasting in complex dynamical systems by training and validating predictors on field datasets (e.g., climate regime shifts, grid/traffic instabilities, biological rhythms) using these spectral observables as physics-informed features. 

\paragraph{}
In parallel, we plan to connect the present spectral–topological framework with emerging paradigms in quantum machine learning that link physical embeddings to efficient learning. 
Recent progress in quantum reservoir computing has demonstrated new approaches to temporal information processing~\cite{Kobayashi2024FeedbackQRC,Yasuda2023QRCRepeated}, 
and developments in quantum kernel learning are beginning to show how structured feature spaces can enhance learning efficiency~\cite{Yamauchi2024PEEQK,Huang2021PowerOfData,Havlicek2019QFeatureSpaces}. 
Integrating our spectral Laplacian observables with these frameworks 
may provide a pathway toward linking supersymmetric spectra with 
interpretable quantum learning, 
linking topology, dynamics, and information geometry within a physically grounded computational architecture.

\subsection{Outlook}
\paragraph{}
By reframing homology detection as a spectral problem for data-derived Hamiltonians, quantum topology gains access to the language of energy gaps, correlation lengths, and dynamical criticality. 
This work demonstrates, on actual quantum hardware, that time-domain dynamics can be interpreted through the eigenvalue spectrum of a Hodge–Laplacian Hamiltonian, enabling estimation of topological features. 
In particular, the Lorenz case suggests that the magnitude of the zero mode corresponds to the Betti number, while the behavior of low-lying nonzero modes quantifies topological stability. 
A clear correlation between the $H_1$ persistence and the SUSY spectral energy gap across $\rho$—both reaching maxima near $\rho\!\approx\!41$—suggests that quantum spectral analysis could provide a practical probe of homological stabilization. 
Although theoretical justification remains open, this experimentally verified phenomenon may help outline a possible direction for developing quantum topology as a spectroscopic approach bridging classical persistence theory and fully quantum spectral geometry.





\clearpage 

\bibliography{science_template} 
\bibliographystyle{sciencemag}


\section*{Acknowledgments}
The authors acknowledge helpful conversations with Toshiki Yasuda, Hiroshi Yano, Ryo Nagai, Yoshihiko Abe, Takashi Abe, Kanji Nishii, Hiromitsu Kigure, Mio Komuro, Natsuki Nakajima, Rei Sakuma, Kohei Oshio, Shu Kanno, Kimberlee Keithley, Kosuke Ito, Jumpei Kato, Michihiko Sugawara, Yasuharu Mori, Toshinari Itoko, Takeshi Yamane, Tamiya Onodera and Kenji Sugisaki.
Hiroshi Yamauchi acknowledges helpful discussions with Massimiliano Incudini, Shashanka Ubaru, Casper Gyurik, Ryu Hayakawa, Behnam Tonekaboni, Shigetora Miyashita, Toshihiro Aoki and further thank the support of Yosuke Komiyama and Ryuji Wakikawa.
This work was partly supported by UTokyo Quantum Initiative, NEDO Challenge Quantum Computing “Solve Social Issues!” and Q-CTRL.
This work was supported by MEXT Quantum Leap Flagship Program Grants No. JPMXS0118067285 and No. JPMXS0120319794.
E.K. was supported by JSPS Grant Number 21H05185.

\section*{Author contributions}
Hiroshi Yamauchi designed the overall research plan, implemented the algorithms, conducted all experiments, and wrote the manuscript.  
Satoshi Kanno developed the theoretical proof connecting the energy gap and topological persistence, and contributed to the interpretation of the results.  
Yuki Sato developed the \emph{Projected-basis hybrid} compilation method and contributed to the efficient implementation of the Hamiltonian and controlled time-evolution circuits.  
Hiroyuki Tezuka provided advice on the construction of the initial quantum states.    
Yoshi-aki Shimada, Eriko Kaminishi, and Naoki Yamamoto supervised the research and reviewed the manuscript.

\section*{Competing interests}
The authors declare no competing interests.



\section*{Supplementary materials}
Materials and Methods\\
Figs. S1 to S5\\
Tables S1\\


\newpage


\renewcommand{\thefigure}{S\arabic{figure}}
\renewcommand{\thetable}{S\arabic{table}}
\renewcommand{\theequation}{S\arabic{equation}}
\renewcommand{\thepage}{S\arabic{page}}
\renewcommand{\thesection}{S\arabic{section}}
\renewcommand{\thesubsection}{S\arabic{subsection}}
\renewcommand{\thesubsubsection}{S\arabic{subsection}.\arabic{subsubsection}}

\setcounter{section}{1}  
\setcounter{subsection}{0}

\setcounter{figure}{0}
\setcounter{table}{0}
\setcounter{equation}{0}
\setcounter{page}{1} 


\begin{center}
\section*{Supplementary Materials for\\ \scititle}
\end{center}

\begin{center}
\author{
  {\normalfont\mdseries\upshape\unboldmath
  Hiroshi~Yamauchi$^{1,5\ast}$,
  Satoshi~Kanno$^{1}$,
  Yuki~Sato$^{2}$,
  Hiroyuki~Tezuka$^{3,5}$,
  Yoshi-aki~Shimada$^{1}$,
  Eriko~Kaminishi$^{5}$,
  Naoki~Yamamoto$^{4,5}$
  }\\
  
  {\small\normalfont\mdseries\upshape
  $^{1}$SoftBank Corp., Research Institute of Advanced Technology, 1-7-1 Kaigan, Minato-ku, Tokyo 105-7529, Japan\\
  $^{2}$Toyota Central R\&D Labs., Inc., 1-4-14, Koraku, Bunkyo-ku, Tokyo, 112-0004, Japan\\
  $^{3}$Advanced Research Laboratory, Sony Group Corporation, 1-7-1 Konan, Minato-ku, Tokyo 108-0075, Japan\\
  $^{4}$Department of Applied Physics and Physico-Informatics, Keio University, 3-14-1 Hiyoshi, Kohoku-ku, Yokohama, Kanagawa 223-8522, Japan\\
  $^{5}$Quantum Computing Center, Keio University, 3-14-1 Hiyoshi, Kohoku-ku, Yokohama, Kanagawa 223-8522, Japan\\
  $^{\ast}$Corresponding author: \texttt{hiroshi.yamauchi@g.softbank.co.jp}
  }
}
\end{center}

\subsubsection*{This PDF file includes:}
Materials and Methods\\
Figures S1 to S5\\
Tables S1\\


\newpage


\section*{Materials and Methods}

\subsection*{Contents (Main Text + Supplement)}

\begin{enumerate}[leftmargin=1.8em,label=\textbf{}]
    \item \textbf{Main Text}
    \begin{enumerate}[leftmargin=2.2em,label=\textbf{}]
        \item 1. Introduction
        \item 2. Theory Overview
        \item 3. Experimental Implementation
        \begin{enumerate}[leftmargin=2.2em,label=\textbf{}]
        \item 3.1 Overview and Scope
        \item 3.2 Five-Point Validation of Quantum Betti Estimation
        \item 3.3 Numerical Simulation of Dynamical and Topological Diagnostics across the Lorenz Parameter Sweep
        \item 3.4 Quantum–Hardware and Simulator Readout of Topological Features
        \end{enumerate}
        \item 4. Results
        \begin{enumerate}[leftmargin=2.2em,label=\textbf{}]
        \item 4.1 Five–Point Betti–Number Estimation (Simulation)
        \item 4.2 Lorenz Parameter Sweep: Dynamical and Topological Transitions
        \item 4.3 Lorenz Topological Readout on $ibm_kingston$ and AerSimulator: Results
        \end{enumerate}
        \item 5. Discussion
        \begin{enumerate}[leftmargin=2.2em,label=\textbf{}]
        \item 5.1 Physical meaning of the two-stage transition
        \item 5.2 Consistency between classical persistence and quantum spectra
        \item 5.3 Outlook and positioning for quantum TDA
        \end{enumerate}
        \item 6. Conclusion
        \begin{enumerate}[leftmargin=2.2em,label=\textbf{}]
        \item 6.1 Key findings
        \item 6.2 Methodological contribution
        \item 6.3 Scientific and technological implications
        \item 6.4 Limitations
        \item 6.5 Future work
        \item 6.6 Outlook
        \end{enumerate}
    \item References and Notes
    \item Acknowledgments
    \item Author contributions
    \item Competing interests
    \item Supplementary materials
    \end{enumerate}

    \item \textbf{Supplementary Materials}
    \begin{enumerate}[leftmargin=2.2em,label=\textbf{}]
        \item Materials and Methods  
        \begin{enumerate}[leftmargin=3.0em,label=\textbf{}]
            \item Contents (Main Text + Supplement)
            \item S1. Hodge Laplacian and Homology
            \item S2. SUSY Hamiltonian and Homology
            \item S3. Dynamical and Topological Phase Transitions of the Lorenz System
            \item S4. Pipeline Overview
            \item S5. Time-Series Embedding
            \item S6. Representative Point Selection
            \item S7. Topological Edge Construction
            \item S8. Dicke State Encoding
            \item S9. SUSY Hamiltonian Construction
            \item S10. Controlled Time Evolution Circuit
            \item S11. Quantum Phase Estimation
            \item S12. Proof of the Spectral Bound Between Energy Gap and Persistence
        \end{enumerate}

        \item Supplementary Figures (S1–S5)
        \item Supplementary Tables (S1)
    \end{enumerate}
\end{enumerate}

\subsection{Hodge Laplacian and Homology}
\label{sec:comb_lap}

\paragraph{}
Let $K$ be a finite, oriented simplicial complex. For each $k\ge 0$, the $k$–chain space $C_k(K)$ is the real vector space spanned by oriented $k$–simplices, equipped with the inner product that makes those simplices orthonormal. 
The boundary operators
\begin{equation}
\partial_k: C_k \to C_{k-1},\qquad
\partial_k[v_0,\dots,v_k]=\sum_{i=0}^k(-1)^i[v_0,\dots,\widehat{v_i},\dots,v_k]
\label{eq:boundary_def}
\end{equation}
assemble into the chain complex
\begin{equation}
\cdots \xrightarrow{\partial_{k+1}} C_k \xrightarrow{\partial_k} C_{k-1} \xrightarrow{\partial_{k-1}} \cdots,
\qquad \partial_k\partial_{k+1}=0.
\end{equation}
The $k$th homology group is
\begin{equation}
H_k(K)=\ker(\partial_k)/\mathrm{im}(\partial_{k+1}),\qquad
\beta_k=\dim H_k(K),
\end{equation}
which counts independent $k$–dimensional holes~\cite{Edelsbrunner2010}.
In Eq.~\eqref{eq:boundary_def}, $\widehat{v_i}$ indicates omission of $v_i$, and the orientation is the canonical one induced by vertex order.

\paragraph{}
With respect to the chosen inner product, let $\partial_k^\dagger$ denote the adjoint of $\partial_k$. It is convenient to split the $k$–Hodge Laplacian into its “down” and “up” parts and to write
\begin{equation}
L_k^{\downarrow}=\partial_k^\dagger\partial_k,\qquad
L_k^{\uparrow}=\partial_{k+1}\partial_{k+1}^\dagger,\qquad
L_k=L_k^{\downarrow}+L_k^{\uparrow}
=\partial_k^\dagger\partial_k+\partial_{k+1}\partial_{k+1}^\dagger,
\label{eq:hodge}
\end{equation}
All operators act on $C_k(K)$ equipped with the simplex-orthonormal inner product, so $(\cdot)^\dagger$ is just transpose with respect to that basis.

\paragraph{}
Any $k$–chain $x\in C_k$ admits three qualitatively distinct behaviors. It can be a \emph{gradient} (an exact $k$–chain) of some $(k{-}1)$–chain, $x=\partial_k^\dagger y$; it can be a \emph{curl} (a coexact $k$–chain) induced by a $(k{+}1)$–chain, $x=\partial_{k+1} z$; or it can be simultaneously cycle and cocycle, i.e., $x\in\ker(\partial_k)\cap\ker(\partial_{k+1}^\dagger)$, which is divergence–free and curl–free. The latter are the \emph{harmonic} $k$–chains and encode the homology.

\paragraph{}
The Hodge decomposition states that these three subspaces are mutually orthogonal and span $C_k$:
\begin{equation}
C_k \;=\; \mathrm{im}(\partial_k^\dagger)\ \oplus\ \ker L_k\ \oplus\ \mathrm{im}(\partial_{k+1}),
\qquad
\ker L_k=\ker(\partial_k)\cap\ker(\partial_{k+1}^\dagger)\cong H_k,\quad
\beta_k=\dim\ker L_k.
\label{eq:hodge-decomp}
\end{equation}
Thus every $x\in C_k$ splits uniquely as $x=x_{\mathrm{grad}}+x_{\mathrm{harm}}+x_{\mathrm{curl}}$, with the three components pairwise orthogonal.

\paragraph{}
Because the direct sum is orthogonal, each component can be recovered by orthogonal projection. 
Using Moore–Penrose pseudoinverses,
\begin{align}
P_{\mathrm{grad}} &= \partial_k^\dagger(\partial_k\partial_k^\dagger)^{+}\partial_k,\qquad
P_{\mathrm{curl}} = \partial_{k+1}(\partial_{k+1}^\dagger\partial_{k+1})^{+}\partial_{k+1}^\dagger,\\
P_{\mathrm{harm}} &= I-P_{\mathrm{grad}}-P_{\mathrm{curl}},\qquad
x_{\bullet}=P_{\bullet}x.
\end{align}
Equivalently, $x_{\mathrm{grad}}=\partial_k^\dagger u$ where $u$ solves the normal equations
$\partial_k\partial_k^\dagger u=\partial_k x$; $x_{\mathrm{curl}}=\partial_{k+1} v$ where
$\partial_{k+1}^\dagger\partial_{k+1} v=\partial_{k+1}^\dagger x$; and $x_{\mathrm{harm}}$ is the residual in $\ker L_k$.

\paragraph{}
The appearance of the adjoint $\partial_k^\dagger$ in $P_{\mathrm{grad}}$ is deliberate and fundamental. 
Since $\mathrm{im}(\partial_k^\dagger)$ is the subspace of exact $k$–cochains (coboundaries), 
projection onto this space necessarily involves $\partial_k^\dagger$ acting on the left. 
Had we used $\partial_k$ instead, the resulting operator would project onto $\mathrm{im}(\partial_k)$, 
the boundary subspace associated with the next lower chain group, corresponding to $P_{\mathrm{curl}}$. 
Thus the two projectors are adjoint counterparts under the discrete Hodge pairing:
\[
P_{\mathrm{grad}}=\partial_k^\dagger(\partial_k\partial_k^\dagger)^{+}\partial_k,
\qquad
P_{\mathrm{curl}}=\partial_{k+1}(\partial_{k+1}^\dagger\partial_{k+1})^{+}\partial_{k+1}^\dagger.
\]
This construction ensures that $P_{\mathrm{grad}}$, $P_{\mathrm{curl}}$, and $P_{\mathrm{harm}}$
form a complete, mutually orthogonal decomposition of $C_k$, as guaranteed by the Hodge theorem for finite complexes. 
Equivalent formulations appear in the discrete Hodge-theoretic treatments of Horak and Jost~\cite{HorakJost2013} 
and Lim~\cite{Lim2020Hodge}, where the operators are written as
\[
P_{\mathrm{grad}} = \delta(\delta^*\delta)^{+}\delta^*,\qquad
P_{\mathrm{curl}} = \delta^*(\delta\delta^*)^{+}\delta,
\]
with $\delta=\partial_k^\dagger$ denoting the coboundary operator.

\paragraph{}
For the case $k=1$, write $B_1$ for the vertex–edge incidence matrix and $B_2$ for the edge–triangle incidence matrix of $K$ (if triangles are present). 
Let $B_1\in\mathbb{R}^{|V|\times|E|}$ be the vertex–edge incidence (each column has one $+1$ and one $-1$ according to edge orientation), and $B_2\in\mathbb{R}^{|E|\times|T|}$ the edge–triangle incidence (entries in $\{0,\pm1\}$ given by compatible orientation). Then
\begin{equation}
L_1^{\downarrow}=B_1^\top B_1,\qquad
L_1^{\uparrow}=B_2 B_2^\top,\qquad
L_1=B_1^\top B_1 + B_2 B_2^\top .
\label{eq:L1def}
\end{equation}
Here $\mathrm{im}(B_1^\top)$ consists of edge differences of a scalar potential on vertices (gradient space), $\mathrm{im}(B_2)$ consists of circulations induced by triangle potentials (curl space), and $\ker L_1$ consists of edge flows that are both divergence–free ($B_1x=0$) and curl–free ($B_2^\top x=0$); hence $\beta_1=\dim\ker L_1$ recovers the first Betti number. On a pure graph with no filled triangles ($B_2=0$), one simply has $L_1=B_1^\top B_1$ and $\ker L_1=\ker B_1$, the usual cycle space~\cite{Chung1997,Barbarossa2020}.

\paragraph{}
The Hodge Laplacian introduced above not only decomposes chains into
gradient, curl, and harmonic components but also connects directly to
the semiclassical picture of Witten--Morse supersymmetric quantum
mechanics~\cite{Witten1982}.  In this correspondence, the gradient,
curl, and harmonic parts represent, respectively, downward and upward
gradient flows and the stationary (zero–energy) sector of a
supersymmetric Hamiltonian.  Under Witten’s deformation
$d_t=e^{-tf}de^{tf}$ the Laplacian $\Delta_t=(d_t+d_t^\dagger)^2$
acquires exponentially small eigenvalues
$E_i(t)\!\sim\!\exp[-2t\,\Delta f_i/\hbar]$ generated by tunneling
between distinct basins of the Morse function~$f$.  The first nonzero
eigenvalue therefore measures a tunneling gap that quantifies how
strongly separated topological sectors communicate.  In the discrete
combinatorial setting this role is played by
\[
\Delta^{(1)}_{\mathrm{SUSY}}
=\min\{\lambda>0:\lambda\in\sigma(L_1)\},
\]
the smallest positive eigenvalue of the one–form Hodge block.  When
persistent loops are well isolated in the geometry, tunneling is
suppressed and the gap widens; when loops merge or collapse, tunneling
increases and the gap closes.  Across a filtration or control parameter
this gap $\Delta^{(1)}_{\mathrm{SUSY}}$ typically co-varies with the
highest persistence of homology $\ell_{H_1}^{\max}$, providing a spectral proxy for topological stability~\cite{Donato2020PRE,MemoliWanWang2022SIMODS}.
This connection unifies the classical Hodge decomposition, the
Witten–Morse semiclassical analysis, and modern persistent-Laplacian
formulations into a single framework linking gradient flow, tunneling
amplitude, and topological persistence.

\paragraph{}
From a spectral viewpoint, eigenpairs of $L_k$ separate according to \eqref{eq:hodge-decomp}: harmonic modes ($\lambda=0$) span homology, while nonzero eigenvalues arise from gradient–type and curl–type subspaces~\cite{HorakJost2013,MemoliWanWang2022SIMODS,Davies2023ICML,Edelsbrunner2010}. 
Small but nonzero eigenvalues can come from either block and reflect different geometric mechanisms (e.g., thin bridges versus wide vortical regions). Consequently, interpreting “small eigenvalues” requires tracking their provenance across scales or filtrations; disentangling the three families is essential for faithful geometric and topological inference~\cite{MemoliWanWang2022SIMODS,Wang2019PersistentSpectralGraph,WangWei2021PersistentPathLaplacian,PLSurvey2025MDPI,MengXia2021,AlgebraicStabilityPL2024,SchaubEtAl22_SignalProcessingOnSimplicial,GrandeSchaub2024Disentangling}.
Relatedly, persistent homology admits quantitative connections to fractal dimension via upper box dimension estimates~\cite{Schweinhart2021}.

\subsection{Supersymmetric (SUSY) Hamiltonian and Homology}
\label{sec:susy_ham}

\paragraph{}
We consider $\mathcal N{=}2$ supersymmetric quantum mechanics on a $\mathbb{Z}$–graded Hilbert space; $F$ is the degree operator so that $[F,Q]=Q$ shifts degree by $+1$ and $[F,Q^\dagger]=-Q^\dagger$ by $-1$.
\begin{equation}
\mathcal{H}=\bigoplus_{k=0}^{m}\mathcal{H}_k,
\end{equation}
with odd supercharges $Q$ and $Q^\dagger$ obeying
\begin{equation}
Q^2=(Q^\dagger)^2=0,\qquad H=\{Q,Q^\dagger\}=Q^\dagger Q+QQ^\dagger,\qquad
[F,Q]=Q,\ [F,Q^\dagger]=-Q^\dagger ,
\label{eq:susy-alg}
\end{equation}
where $F$ is the degree operator. Hence $Q:\mathcal{H}_k\to\mathcal{H}_{k+1}$ and $Q^\dagger:\mathcal{H}_k\to\mathcal{H}_{k-1}$, and $[H,F]=0$ so $H$ is block-diagonal in degree.

\paragraph{}
Let $d_k:=Q\big|_{\mathcal{H}_k}:\mathcal{H}_k\to\mathcal{H}_{k+1}$ and $d_k^\dagger:=Q^\dagger\big|_{\mathcal{H}_{k+1}}:\mathcal{H}_{k+1}\to\mathcal{H}_k$, and let $P_k$ be the projector onto $\mathcal{H}_k$. Expanding $Q=\sum_{j} d_j P_j$ and $Q^\dagger=\sum_{j} d_{j-1}^\dagger P_j$, for any $v\in\mathcal{H}_k$ we have
\begin{equation}
Hv=(Q^\dagger Q + QQ^\dagger)v
=Q^\dagger(d_k v)+Q(d_{k-1}^\dagger v)
=d_k^\dagger d_k v + d_{k-1} d_{k-1}^\dagger v\in\mathcal{H}_k .
\end{equation}
Thus, in the degree-$k$ subspace, the SUSY Hamiltonian block equals the $k$–Hodge Laplacian $\mathcal{L}_k$ acting on $k$–cochains.
\begin{equation}
H\big|_{\mathcal{H}_k}=d_k^\dagger d_k + d_{k-1} d_{k-1}^\dagger \;=\;\mathcal{L}_k .
\label{eq:block-equals-hodge}
\end{equation}
At the ends of the complex, $d_{-1}\equiv 0$ and $d_m\equiv 0$, giving $H\big|_{\mathcal{H}_0}=d_0^\dagger d_0$ and $H\big|_{\mathcal{H}_m}=d_{m-1} d_{m-1}^\dagger$.

\paragraph{}
Positivity follows immediately: for $v\in\mathcal{H}_k$,
\begin{equation}
\langle v,Hv\rangle=\|d_k v\|^2+\|d_{k-1}^\dagger v\|^2\ge 0,
\label{eq:positivity}
\end{equation}
Here $\langle\cdot,\cdot\rangle$ is the Hilbert–space inner product; positivity holds since each term is a squared norm, so $E=0$ iff $d_k v=0$ and $d_{k-1}^\dagger v=0$, i.e.,
\begin{equation}
\ker H\big|_{\mathcal{H}_k}=\ker d_k \cap \ker d_{k-1}^\dagger=\ker \mathcal{L}_k .
\label{eq:harmonic-cond}
\end{equation}
Zero-energy states are therefore simultaneously closed and coclosed (harmonic).

\paragraph{}
We use $d_k:=Q|_{\mathcal{H}_k}$ and $d_k^\dagger:=Q^\dagger|_{\mathcal{H}_{k+1}}$, consistent with the coboundary on cochains.
Introduce the $Q$–cohomology
\begin{equation}
H^k(Q)=\ker d_k / \mathrm{im}\, d_{k-1}.
\end{equation}
The Hodge decomposition gives, for every $v\in\mathcal{H}_k$,
\begin{equation}
v=d_{k-1}u + h + d_k^\dagger w,
\qquad u\in\mathcal{H}_{k-1},\ w\in\mathcal{H}_{k+1},\ h\in\ker\mathcal{L}_k .
\label{eq:hodge-decomp-susy}
\end{equation}
If $v$ is $d_k$–closed then $v$ is cohomologous to the unique harmonic representative $h$; hence the map $[v]\mapsto h$ induces an isomorphism
\begin{equation}
\ker H\big|_{\mathcal{H}_k}=\ker\mathcal{L}_k \cong H^k(Q)\quad(\text{and by duality }H_k),
\qquad \dim\ker H\big|_{\mathcal{H}_k}=\beta_k .
\label{eq:cohomology-iso}
\end{equation}
All positive–energy levels appear in adjacent–degree pairs: if $Hv=Ev$ with $E>0$, then $Qv$ and $Q^\dagger v$ (when nonzero) are eigenvectors with the same $E$ in degrees $k{+}1$ and $k{-}1$, respectively; only the harmonic sector contributes to $E=0$.

\paragraph{}
When the graded Hilbert space is realized as cochains on a finite simplicial complex and $d_k$ is the coboundary, the block identity \eqref{eq:block-equals-hodge} coincides with the combinatorial Hodge Laplacian $\mathcal{L}_k=d_k^\dagger d_k + d_{k-1} d_{k-1}^\dagger$. In particular, for a $1$–dimensional block with vertex–edge incidence $B_1$ and edge–triangle incidence $B_2$, one has $\mathcal{L}_1=B_1^\top B_1 + B_2 B_2^\top$; its kernel encodes divergence–free and curl–free edge flows, so $\dim\ker\mathcal{L}_1=\beta_1$ recovers the cycle space of the underlying complex (and of a pure graph when $B_2=0$).

\paragraph{}
These relations imply a practical dictionary for topology–aware spectroscopy. The degeneracy of zero modes in degree $k$ equals the $k$th Betti number, while the first positive eigenvalue
\begin{equation}
\Delta^{(1)}_{\mathrm{SUSY}}
= \min\{\lambda>0 \mid \lambda\in\sigma(\mathcal L_1)\}.
\end{equation}
measures the spectral isolation of the corresponding topological sector (an “almost–harmonic’’ scale). Tracking $\Delta^{(k)}$ and the occupation of $\ker\mathcal{L}_k$ as external parameters (e.g., the Lorenz control $\rho$) vary suggests the appearance, merger, or disappearance of topological features in a way that is algebraically exact yet numerically robust.

\subsection{Dynamical and Topological Phase Transitions of the Lorenz System}
\label{sec:lorenz_dyn_topo}

\paragraph{}
We study the Lorenz flow
\begin{equation}
\dot x=\sigma(y-x),\qquad 
\dot y=x(\rho-z)-y,\qquad 
\dot z=xy-\beta z.
\label{eq:lorenz}
\end{equation}
Here $(\sigma,\beta,\rho)>0$ are, respectively, the Prandtl number, the geometric parameter, and the Rayleigh parameter; we later scan $\rho$ while fixing $(\sigma,\beta)=(10,8/3)$.
 For each $\rho$ a long trajectory yields a point cloud $X(\rho)\subset\mathbb{R}^d$ (either directly in phase space or via delay embedding), and we attach to it a parameter–dependent effective Hamiltonian $H(\rho)$ that captures spectral/dynamical content. This construction lets us examine phase behavior through two complementary lenses. The \emph{dynamical} lens quantifies how the system explores frequency and state space; the \emph{topological} lens quantifies how the invariant geometry reorganizes across scales. In our experiments we evaluate both families of indicators along $\rho$ and interpret their concordance as evidence for dynamical and topological phase transitions, placing special emphasis on the physical meaning of each indicator.

\paragraph{}
On the dynamical side, we probe frequency–space complexity via the spectral entropy of a survival (Loschmidt) amplitude for a fixed probe state $|\psi\rangle$,
\begin{equation}
C(t;\rho)=\langle\psi|e^{-iH(\rho)t}|\psi\rangle
=\sum_n |c_n(\rho)|^2 e^{-iE_n(\rho)t},\qquad
c_n(\rho)=\langle E_n(\rho)|\psi\rangle .
\end{equation}
Let $S(\omega;\rho)=\big|\mathcal{F}[C(\cdot;\rho)](\omega)\big|^2$ be a windowed power spectrum and $P(\omega;\rho)=S(\omega;\rho)/\!\int S(\omega;\rho)\,d\omega$ its normalization. Let $\mathcal{F}$ denote the unitary Fourier transform in $t$, and normalize $S(\omega)$ to a probability density $P(\omega)$ so that $\int P(\omega)\,d\omega=1$.
We then define
\begin{equation}
H_{\mathrm{spec}}(\rho)=-\int d\omega\, P(\omega;\rho)\,\ln P(\omega;\rho),
\label{eq:specentropy}
\end{equation}
which is small when a few frequencies dominate (coherent/locked motion) and large when many incommensurate frequencies carry comparable weight (dephasing and complex mixing). Peaks in $H_{\mathrm{spec}}(\rho)$ signal spectral broadening that typically accompanies a transition, followed by a drop once a new regime re-locks the spectrum.

\paragraph{}
Linear response to the control parameter is captured by the free-energy curvature. With a linear deformation $H(\rho)=H_0-\rho\,\hat O$ and $F(\rho)=-(1/\beta)\ln\operatorname{Tr}e^{-\beta H(\rho)}$,
\begin{equation}
F''(\rho)=\beta_{\mathrm{th}}\big(\langle \hat O^2\rangle-\langle \hat O\rangle^2\big),\qquad \beta_{\mathrm{th}}=1/(k_{\mathrm B}T)
\equiv \chi_O(\rho)\ge 0,
\label{eq:Fpp}
\end{equation}
and in the ground-state limit $F(\rho)\to E_0(\rho)$,
\begin{equation}
E_0''(\rho)=2\sum_{n>0}\frac{|\langle n|\hat O|0\rangle|^2}{E_n(\rho)-E_0(\rho)}.
\label{eq:E0pp}
\end{equation}
Both formulas demonstrate enhancement by small gaps and large transition matrix elements; pronounced peaks in $F''(\rho)$ therefore locate parameter regions where the state is most sensitive to $\rho$.

\paragraph{}
Geometry on the projective Hilbert manifold is accessed through ground-state fidelity. Write $|\psi_0(\rho)\rangle$ for the normalized ground state of $H(\rho)$ with an arbitrary but fixed global phase convention.
\begin{equation}
F(\rho,\rho+\delta)=\big|\langle\psi_0(\rho)\mid\psi_0(\rho+\delta)\rangle\big|
=1-\tfrac12\,\chi_F(\rho)\,\delta^2+O(\delta^3)
\label{eq:fidelity}
\end{equation}
defines the fidelity susceptibility
\begin{equation}
\chi_F(\rho)=\langle\partial_\rho\psi_0|\partial_\rho\psi_0\rangle
-|\langle\psi_0|\partial_\rho\psi_0\rangle|^2
=\sum_{n>0}\frac{|\langle n|\hat O|0\rangle|^2}{\big(E_n(\rho)-E_0(\rho)\big)^2}.
\label{eq:chisus}
\end{equation}
Sharp drops in $F$ and peaks (or finite-size precursors thereof) in $\chi_F$ mark rapid ground-state reconfiguration, a hallmark of continuous transitions.

\paragraph{}
An additional dynamical diagnostic is the \emph{maximum Lyapunov exponent} (MLE) $\lambda_{\max}(\rho)$, which quantifies the mean exponential divergence of nearby trajectories in the Lorenz flow itself.  Writing $\delta x(t)$ for an infinitesimal perturbation that obeys the variational equation 
\begin{equation}
\dot{\delta x}(t)=J(t;\rho)\,\delta x(t),\qquad 
J(t;\rho)=\frac{\partial f(x(t);\rho)}{\partial x}
\end{equation}
with $f=(\dot x,\dot y,\dot z)$, one defines
\begin{equation}
\lambda_{\max}(\rho)
=\lim_{t\to\infty}\frac{1}{t}\ln\frac{\|\delta x(t)\|}{\|\delta x(0)\|},
\end{equation}
where the limit is realized numerically through periodic renormalization of $\delta x(t)$ in the Benettin algorithm.  Negative $\lambda_{\max}$ indicates stable fixed points or periodic orbits, $\lambda_{\max}=0$ corresponds to neutral stability at a bifurcation, and positive $\lambda_{\max}$ marks chaotic dynamics with exponential sensitivity to initial conditions.  
For the Lorenz parameters $(\sigma,\beta)=(10,8/3)$ the computed $\lambda_{\max}(\rho)$ changes sign near $\rho_c\simeq24.7$, in agreement with the classical onset of the strange attractor. This neighborhood has historically been described as a “preturbulent” regime in Lorenz-type flows~\cite{KaplanYorke1979Preturbulence}. In our analysis this transition in $\lambda_{\max}$ aligns closely with the peaks of spectral entropy and fidelity susceptibility, linking microscopic Hilbert-space sensitivity to macroscopic chaos in the underlying flow.

\paragraph{}
On the topological side, we examine loop robustness in the embedded attractor by persistent homology. The Vietoris–Rips filtration
\begin{equation}
R_\varepsilon(X(\rho))=\{\sigma\subset X(\rho):\max_{x,y\in\sigma}d(x,y)\le\varepsilon\}
\label{eq:vr}
\end{equation}
tracks $H_1$ classes through births $b$ and deaths $d$; the persistence is $\ell=d-b$. We use the Euclidean metric $d(\cdot,\cdot)$ on the embedded point cloud $X(\rho)$, and the Rips complex includes a simplex whenever all pairwise distances within it are $\le \varepsilon$. As a compact summary we record the highest persistence
\begin{equation}
\ell^{\max}_{H_1}(\rho)=\max_i (d_i-b_i),
\qquad 
Q_{\mathrm{topo}}(\rho)=\frac{\ell^{\max}_{H_1}(\rho)}{1+\ell^{\max}_{H_1}(\rho)}\in[0,1) ,
\end{equation}
where large values indicate a clear, persistent loop (e.g., well-separated wings of the attractor), while kinks or drops versus $\rho$ mark geometric reorganizations.

\paragraph{}
The same topological information admits a spectral representation through a supersymmetric (SUSY) Hamiltonian. Let the graded Hilbert space $\mathcal{H}=\bigoplus_k\mathcal{H}_k$ represent $k$–cochains, with $Q:\mathcal{H}_k\to\mathcal{H}_{k+1}$ the coboundary block $d_k$ and $Q^\dagger$ its adjoint. The SUSY Hamiltonian $\,\mathcal{H}_{\mathrm{SUSY}}=\{Q,Q^\dagger\}\,$ is block-diagonal with
\begin{equation}
\mathcal{H}_{\mathrm{SUSY}}\big|_{\mathcal{H}_k}
=d_k^\dagger d_k + d_{k-1} d_{k-1}^\dagger
\;=\; \mathcal{L}_k ,
\end{equation}
the combinatorial Hodge Laplacian. Zero-energy states satisfy $d_k v=0$ and $d_{k-1}^\dagger v=0$, hence
\begin{equation}
\ker \mathcal{H}_{\mathrm{SUSY}}\big|_{\mathcal{H}_k}=\ker \mathcal{L}_k \cong H^k \ (\text{and by duality }H_k),
\qquad \dim\ker \mathcal{L}_k=\beta_k .
\end{equation}
For a self-adjoint matrix $M$, $\sigma(M)$ denotes its multiset of eigenvalues; $\min(\sigma(M)\setminus\{0\})$ is the first positive eigenvalue (if any).
\begin{equation}
\Delta^{(1)}_{\mathrm{SUSY}}
= \min\{\lambda>0 \mid \lambda\in\sigma(\mathcal L_1)\}.
\label{eq:deltaSUSY}
\end{equation}
is a topological gap separating harmonic 1-cycles from their first excitations; its shrinkage (reopening) signals weakening (strengthening) of loop robustness and typically co-varies with $\ell^{\max}_{H_1}(\rho)$.

\paragraph{}
If $\{E_n(\rho)\}_n$ denotes the ordered eigenvalues of $H(\rho)$, we define the many-body gap by $\gamma(\rho)=E_1(\rho)-E_0(\rho)$.
\begin{equation}
\gamma(\rho)=E_1(\rho)-E_0(\rho),
\label{eq:gamma}
\end{equation}
which often obeys critical scaling near $\rho_c$,
\begin{equation}
\gamma(\rho)\sim|\rho-\rho_c|^{z\nu},\qquad
\xi(\rho)\sim|\rho-\rho_c|^{-\nu},\qquad
\gamma\sim\xi^{-z},
\end{equation}
and, for linear size $L$ (or an effective sample-size proxy),
\begin{equation}
\gamma(\rho,L)=L^{-z}\,\mathcal{F}\!\big((\rho-\rho_c)L^{1/\nu}\big),
\qquad
\gamma(\rho_c,L)\propto L^{-z}.
\end{equation}
In practice, minima or closures of $\gamma(\rho)$ tend to align with peaks of $F''(\rho)$ and $\chi_F(\rho)$ due to the small denominators in their spectral representations; at the same locations one often observes a peak in $H_{\mathrm{spec}}(\rho)$ and a kink or turnover in $\ell^{\max}_{H_1}(\rho)$ or $\Delta^{(1)}_{\mathrm{SUSY}}(\rho)$.

\paragraph{}
The experimental campaign thus evaluates the Lorenz system along $\rho$ using both dynamical (spectral entropy, free-energy curvature, fidelity) and topological (persistent $H_1$ and Hodge–Laplacian spectrum gaps) indicators, and it explores their physical significance jointly: spectral broadening and heightened response coincide with homology-carrying zero modes becoming weakly isolated, while re-locking of frequencies and reduced susceptibility accompany the reopening of topological and spectral gaps. This consensus across indicators provides a robust locator of phase boundaries and a unified interpretation of dynamical and geometric reorganization in the Lorenz system.

\subsection{Pipeline Overview}
\label{sec:pipeline_overview}

\paragraph{}
The proposed framework transforms a real-valued time series into a quantum–mechanical spectral representation whose low-energy structure encodes the persistent topology of the underlying dynamics.  
By integrating classical topological data analysis (TDA) with quantum spectral estimation in a sequential and interpretable workflow,  
the method converts physical and geometric information into algebraic form, culminating in a supersymmetric (SUSY) Hamiltonian whose eigenvalue spectrum suggests homological features.

\paragraph{}
The end-to-end process comprises five main stages, illustrated schematically in Figure~\ref{fig:pipeline}.  
First, a scalar observable from the Lorenz system is embedded into a reconstructed phase space using Takens’ delay coordinates (Section~\ref{sec:time_series_embedding}).  
The resulting point cloud $X=\{X(t_k)\}$ approximates the invariant manifold of the attractor, preserving its loop topology and geometric structure.In particular, sliding-window constructions together with persistence have proven effective for extracting topological signatures from time-series data~\cite{PereaHarer2015}.

\paragraph{}
Next, a compact subset $V=\{v_i\}_{i=1}^{n}$ is extracted from $X$ based on density, topological persistence, and geometric diversity (Section~\ref{sec:rep_point_selection}).  
This topology-aware reduction retains the regions most relevant to persistent $H_1$ features while maintaining a balanced spatial coverage.  
The selected representatives are then connected into an undirected graph $G=(V,E)$ that captures both local geometric adjacency and dominant loop structures (Section~\ref{sec:topo_edge_construction}).  
The edge set combines a minimum spanning tree backbone, an $\varepsilon$–neighborhood layer, and optional ring edges derived from circular coordinates obtained via persistent homology.  

\paragraph{}
Subsequently, the graph’s connectivity and topology are mapped onto a symmetric quantum probe state (Section~\ref{sec:dicke_encoding}).  
Weighted superpositions of Dicke states $\ket{D_k^{(n)}}$ encode local degrees, loop participation, and feature persistence in their excitation-number populations, ensuring compatibility with excitation-preserving SUSY dynamics.

\paragraph{}
Finally, from the constructed graph we assemble a SUSY Hamiltonian $\mathcal{H}=Q^\dagger Q$ whose $1$–form block $\mathcal{L}_1=d_1^\dagger d_1 + d_0 d_0^\dagger$ acts as a discrete Hodge Laplacian on edges (Section~\ref{sec:susy_ham_construction}).  
Its zero-energy kernel corresponds to harmonic $1$–cycles, while the first positive eigenvalue $\Delta^{(1)}_{\mathrm{SUSY}}$ measures the spectral isolation of these topological features.  
The Hamiltonian is simulated via a controlled time-evolution circuit and probed through single-ancilla quantum phase estimation (QPE), yielding the low-lying energy spectrum.

\paragraph{}
This sequence may point toward a coherent mapping from temporal dynamics to quantum spectra:
$
\text{time series} 
\xrightarrow{\text{Takens embedding}}\ 
\text{point cloud} 
\xrightarrow{\text{topology-aware reduction}}\ 
\text{graph } G(V,E)
\xrightarrow{\text{SUSY encoding}}\ 
\text{quantum Hamiltonian } \mathcal{H}
\xrightarrow{\text{QPE}}\ 
\text{spectral features } \{\Delta^{(k)}_{\mathrm{SUSY}}\}.
$
Each transformation—embedding, projection, graph construction, operator synthesis, and quantum measurement—preserves structural information while translating it across representations.

\paragraph{}
The classical components (embedding, sampling, and graph formation) compress continuous dynamics into discrete topological summaries, whereas the quantum components (state encoding, Hamiltonian simulation, and phase estimation) perform spectral inference without explicit matrix diagonalization, exploiting quantum parallelism for eigenvalue extraction.  
This hybrid approach combines interpretability from classical TDA with computational leverage from quantum simulation.

\paragraph{}
The main observables emerging from the pipeline include the low-lying eigenvalues $\{E_0,E_1,\dots\}$ of $\mathcal{H}$ and their spacing $\gamma=E_1-E_0$, which quantify the dynamical spectral gap;  
the topological gap $\Delta^{(1)}_{\mathrm{SUSY}}$ from the $1$–form block $\mathcal{L}_1$, which measures the robustness of persistent loops;  
the highest persistence $\ell^{\max}_{H_1}$ obtained from classical homology, serving as a geometric benchmark;  
and the spectral entropy $H_{\mathrm{spec}}$ derived from the QPE amplitude distribution, summarizing frequency-space complexity.  
Correlations among these quantities expose how dynamical and topological transitions interact under changes in the control parameter, such as the Rayleigh number $\rho$ in the Lorenz flow.

\paragraph{}
By integrating geometric embedding, topological reduction, and quantum spectral estimation into a unified sequence,  
this framework provides a reproducible and physically interpretable route for detecting, characterizing, and quantifying topological signatures in nonlinear dynamics.  
The resulting hybrid representation enables both theoretical analysis and experimental realization on contemporary quantum hardware.

\subsection{Time-Series Embedding}
\label{sec:time_series_embedding}

\paragraph{}
The first stage of the pipeline transforms a one-dimensional dynamical signal into a geometric representation suitable for topological and spectral analysis.  
\paragraph{}
Following Takens’ embedding theorem~\cite{Takens1981,Packard1980GeometryFromATimeSeries,Pecora2007UnifiedAttractorReconstruction,PereaHarer2015}, the underlying attractor of the Lorenz flow is reconstructed from a scalar observable, producing a point cloud $X\subset\mathbb{R}^m$ that preserves the topology of the original state space appropriate simplicial complexes and reconstruction choices critically affect recovered topology~\cite{Garland2016}..

\paragraph{}
The Lorenz system~\cite{Lorenz1963,Sparrow1982LorenzEquations,Tucker2002LorenzProof} is governed by
\begin{equation}
\dot{x}=\sigma(y-x),\qquad
\dot{y}=x(\rho-z)-y,\qquad
\dot{z}=xy-\beta z,
\label{eq:lorenz_eq}
\end{equation}
with canonical parameters $(\sigma,\beta,\rho)=(10,\,8/3,\,28)$ corresponding to the chaotic regime.  
This system generates a non-periodic attractor of fractal dimension $\mathrm{dim}_H\approx2.06$, exhibiting sensitive dependence on initial conditions and a complex, folded topology.  
Varying the Rayleigh number $\rho$ induces distinct dynamical phases, which are later examined through topological and spectral diagnostics.

\paragraph{}
From a single scalar measurement, for example $x(t)$, we construct an $m$–dimensional delay-coordinate embedding,
\begin{equation}
X(t)=\big[x(t),\,x(t+\tau),\,x(t+2\tau),\,\dots,\,x(t+(m-1)\tau)\big]\in\mathbb{R}^m,
\label{eq:takens_embedding}
\end{equation}
where $\tau$ is the time delay and $m$ the embedding dimension.  
Under generic smoothness and observability conditions, this mapping is diffeomorphic to the original attractor when $m>2d_A$, where $d_A$ denotes the attractor’s dimension.  
The reconstructed manifold thus faithfully captures the invariant geometry of the flow.

\paragraph{}
The time delay $\tau$ controls the balance between redundancy and independence among coordinates.  
If $\tau$ is too small, consecutive components become nearly collinear, yielding an artificially thin manifold;  
if $\tau$ is too large, temporal correlation is lost and the attractor fragments.  
Two practical selection rules are widely used:  
(i) choosing $\tau$ as the first minimum of the auto–mutual information, which maximizes independence between coordinates, and  
(ii) setting $\tau$ to the $1/e$ decorrelation time of the autocorrelation function, which maintains dynamical coherence.  
For the Lorenz system, $\tau$ typically falls within $[0.05,\,0.2]$ in dimensionless time units, producing stable embeddings with clearly separated loops.

\paragraph{}
The embedding dimension $m$ must be large enough to unfold the attractor and prevent self-intersections, yet small enough to remain computationally manageable.  
In practice, $m$ is increased until the fraction of false nearest neighbors~\cite{KennelBrownAbarbanel1992FNN} falls below a threshold (e.g., $1\%$) and the estimated correlation dimension~\cite{GrassbergerProcaccia1983} stabilizes.  
For the Lorenz flow, $m=3$–$6$ typically suffices to recover the two-wing structure, while larger values further smooth sampling artifacts at the cost of redundancy.

\paragraph{}
To ensure numerical consistency across $\rho$, several safeguards are implemented.  
Initial transients of duration $t_{\mathrm{trans}}$ are discarded to eliminate sensitivity to initial conditions;  
integration of Eq.~\eqref{eq:lorenz_eq} is performed with a fixed time step $\Delta t$ to prevent stiffness-induced distortions;  
trajectories are downsampled so that the sampling stride is comparable to $\tau$, avoiding oversampling of correlated points;  
and each coordinate is normalized to unit variance before distance calculations.  
These procedures maintain numerical stability and geometric consistency over parameter sweeps~\cite{Abarbanel1996AnalysisObservedChaoticData,KantzSchreiber2004NonlinearTimeSeries}.

\paragraph{}
The quality of the embedding is verified by estimating the largest Lyapunov exponent $\lambda_{\max}$ from the reconstructed series~\cite{Wolf1985LyapunovFromTimeSeries,Rosenstein1993LargestLyapunov,Kantz1994MaxLyapunov}.  
A positive $\lambda_{\max}$ is consistent with the reconstructed dynamics preserve the expected chaotic character.  
Additionally, visual inspection of the attractor projection and its persistence diagram ensures that characteristic topological features—such as the double-wing loop—remain intact.

\paragraph{}
The embedding yields a point cloud
\begin{equation}
X=\{X(t_k)\in\mathbb{R}^m\}_{k=1}^{N},
\end{equation}
which samples the invariant manifold of the Lorenz attractor.  
Euclidean distances $\|X_i-X_j\|_2$ capture local dynamical adjacency and form the geometric substrate for subsequent topological and spectral analyses.  
This point cloud provides the input for the representative-point selection step (Section~\ref{sec:rep_point_selection}), where density, topology, and diversity criteria are applied to extract a compact but faithful subset of points.

\paragraph{}
Under the selected $(\tau,m)$, the Lorenz embedding produces a cloud whose Vietoris–Rips filtration exhibits two persistent $H_1$ classes corresponding to the attractor’s wings, consistent with previous studies~\cite{MaleticZhaoRajkovic2015,giottoLorenz}.  
These structures are stable across time windows and parameter variations, confirming that the embedding serves as a robust precursor for quantum–topological encoding.

\paragraph{}
Takens embedding therefore transforms a scalar chaotic signal into a geometric object that retains the essential topology of the underlying flow.  
Appropriate choices of $\tau$ and $m$, combined with numerical safeguards, ensure that the resulting point cloud faithfully represents the attractor’s loop geometry.  
This embedded manifold forms the geometric foundation upon which the topology-preserving reduction and supersymmetric Hamiltonian encoding are subsequently built.

\subsection{Representative Point Selection}
\label{sec:rep_point_selection}

\paragraph{}
Given an embedded point cloud $X=\{x_i\in\mathbb{R}^m\}_{i=1}^N$, the task is to extract a smaller subset of representative points $V=\{v_j\}_{j=1}^n$ $(n\ll N)$ that preserves the high-density regions of the attractor, the essential loop topology captured by persistent $H_1$, and the overall geometric diversity of the data.  
This reduction yields a tractable input for graph and Hamiltonian construction while maintaining the topological integrity of the system.

\paragraph{}
The selection process proceeds in two main stages.  
In the first stage, topology-aware sampling identifies regions supporting persistent $H_1$ features and high local density.  
In the second, a complementary set of points is added to maximize geometric spread, ensuring that global features and sparse regions are not neglected.  
The combination produces a compact yet topology-preserving representative set $V$.

\paragraph{}
Let $D(x,y)=\|x-y\|_2$ denote the Euclidean distance, and let $k$ be the target number of representatives.  
A fraction $r\in(0,1)$ of these, $k_{\mathrm{topo}}=\lfloor kr\rfloor$, is reserved for topological coverage, with the remaining $k_{\mathrm{global}}=k-k_{\mathrm{topo}}$ used for global diversity.  
A $K$-nearest-neighbor (KNN) graph (typically $K\approx10$) is constructed on $X$, and its shortest-path metric $D_{\mathrm{geo}}$ approximates geodesic distance along the manifold.

\paragraph{}
Local density is estimated by a Gaussian kernel,
\begin{equation}
\rho(x_i)=\frac{1}{N\,h^m}\sum_{j=1}^N
\exp\!\Big(-\frac{\|x_i-x_j\|^2}{2h^2}\Big),
\end{equation}
where the bandwidth $h$ is chosen as the 10th percentile of pairwise distances to ensure robustness against outliers.  
Positive weights are then defined as
\begin{equation}
w_i \propto \rho(x_i)^{\,\alpha-1},\qquad \alpha>1,
\end{equation}
so that dense regions exert stronger influence on the selection.

\paragraph{}
To identify points most relevant to loop topology, the Vietoris–Rips persistent homology of $X$ is computed and the most persistent $H_1$ feature with birth–death pair $(b^\ast,d^\ast)$ is located.  
Its midpoint radius,
\begin{equation}
r_{\mathrm{mid}}=\tfrac{1}{2}(b^\ast+d^\ast),
\end{equation}
sets a characteristic neighborhood scale.  
For each point $x_i$, the mid-scale neighbor count
\begin{equation}
\nu_i=\#\{\,j:\,D(x_i,x_j)<r_{\mathrm{mid}}\,\}
\end{equation}
quantifies local crowding at that scale.  
Points with excessively few or many neighbors are excluded by defining the candidate set
\begin{equation}
I=\big\{\,i\ \big|\ N_{\min}<\nu_i<N_{\max}\,\big\},
\quad 
N_{\min}\approx\lfloor0.02N\rfloor,\quad 
N_{\max}\approx\max\{N_{\min}{+}5,\lfloor0.10N\rfloor\}.
\end{equation}

\paragraph{}
When a dominant $H_1$ loop is present, each point $x_i$ is assigned an angular coordinate $\theta_i\in[0,2\pi)$ derived from persistent cohomology circular coordinates.  
The interval $[0,2\pi)$ is discretized into $B$ bins to encourage uniform angular sampling around the loop.

\paragraph{}
The topological subset $S_{\mathrm{topo}}$ is initialized with the point of maximal density weight $w_i^{\alpha-1}$ within $I$.  
While $|S_{\mathrm{topo}}|<k_{\mathrm{topo}}$, new points are added greedily by maximizing the composite gain function
\begin{equation}
\Delta(j)=
\lambda_\theta\,\Delta H_\alpha(h\oplus\theta_j)
+\lambda_D\,\min_{i\in S_{\mathrm{topo}}}D_{\mathrm{geo}}(x_j,x_i)
+\lambda_d\,H_\alpha\!\big(\{q_i\}_{i\in S_{\mathrm{topo}}\cup\{j\}}\big)
-\lambda_c\,\Pi(\theta_j\mid\{\theta_i\}_{i\in S_{\mathrm{topo}}}),
\label{eq:topo_gain}
\end{equation}
where $H_\alpha$ is the Rényi entropy, $q_i \propto \rho(x_i)$ are normalized density weights, $\Delta H_\alpha(h\oplus\theta_j)$ measures the entropy gain from adding $\theta_j$ to the angular histogram, and $\Pi$ penalizes violations of a minimum angular separation $\Delta\theta_{\min}\approx2\pi/(1.35\,k_{\mathrm{topo}})$.  
Hyperparameters $(\lambda_\theta,\lambda_D,\lambda_d,\lambda_c)$ tune the balance among angular diversity, geodesic spacing, density regularization, and collision avoidance.  
This greedy procedure incrementally increases coverage of the loop while maintaining separation and balance.

\paragraph{}
Once the topological representatives are chosen, the remaining $k_{\mathrm{global}}$ points are selected to maximize global diversity.  
For each unselected candidate, its minimum distance from the topological set is defined as
\begin{equation}
d_{\min}(x_j)=\min_{i\in S_{\mathrm{topo}}} D(x_j,x_i),
\end{equation}
and a combined score
\begin{equation}
\mathrm{score}(x_j)=w_j\cdot(1+d_{\min}(x_j))
\end{equation}
is computed to prioritize dense but distant regions.  
The highest-scoring points are successively added to form the global coverage set $S_{\mathrm{global}}$.

\paragraph{}
The final representative set is the union
\begin{equation}
V = S_{\mathrm{topo}} \cup S_{\mathrm{global}},\qquad |V|=k.
\end{equation}
The parameters $(r,\alpha,h,B,\lambda_\theta,\lambda_D,\lambda_d,\lambda_c)$ collectively govern the trade-off between density preservation, loop fidelity, and geometric spread.  
For Lorenz-type attractors, empirical tests show that $r$ values between $0.5$ and $0.7$ provide a stable balance.

\paragraph{}
The resulting representative points concentrate near persistent cycles and bridge regions of high Betti stability, while the globally selected points capture outer geometry and non-cyclic regions.  
Together, they yield a compact and faithful summary of the attractor’s structure, substantially reducing matrix dimensions for subsequent Laplacian and SUSY Hamiltonian construction without sacrificing the accuracy of homological inference.

\paragraph{}
In terms of computational cost, density estimation scales as $O(N^2)$ (reducible with KD-tree or approximate-nearest-neighbor acceleration), persistent homology as $O(N^{3/2})$ for typical point sets, and the greedy selection as $O(k^2)$.  
The only sources of stochasticity are tie-breaking and angular-bin initialization, which can be controlled by fixing the random seed to ensure reproducibility.  
The resulting set $V$ provides the foundation for the topological edge construction described in Section~\ref{sec:topo_edge_construction}.

\subsection{Topological Edge Construction}
\label{sec:topo_edge_construction}

\paragraph{}
Given a reduced representative point set $V=\{v_i\}_{i=1}^n\subset\mathbb{R}^m$,  
the objective is to construct an undirected graph $G=(V,E)$ that preserves local geometric proximity, remains globally connected, and exposes prominent one-dimensional cycles when they exist.  
This graph functions as the $1$–skeleton of a simplicial complex whose Hodge Laplacian encodes the persistent $H_1$ topology of the data.

\paragraph{}
To achieve this, the edge set $E$ must capture both geometric adjacency and topological evidence.  
We combine a global backbone built from a minimum spanning tree (MST), a local neighborhood layer derived from an $\varepsilon$–graph, and an optional ring augmentation guided by circular coordinates.  
The resulting union is patched to ensure single-component connectivity, yielding a sparse yet expressive graph that reflects both metric and topological structure.

\paragraph{}
We begin with the pairwise distance matrix $D_{ij}=\|v_i-v_j\|_2$, from which the MST outlines a baseline connectivity without redundant long edges:
\begin{equation}
E_{\mathrm{MST}} = \arg\min_{T\subseteq\binom{V}{2},\,|T|=n-1}
\sum_{(i,j)\in T} D_{ij}.
\label{eq:mst}
\end{equation}
This minimal tree ensures that $G$ is connected and acyclic, providing a global scaffold upon which additional edges can be added to recover local structure.

\paragraph{}
Local geometry is then restored by introducing edges shorter than a data-driven threshold $\varepsilon$.  
Let $U=\{D_{ij}:1\le i<j\le n\}$ be the multiset of pairwise distances.  
A robust length scale is set as the 30th percentile, $\varepsilon = Q_{0.3}(U)$, and all pairs satisfying
\begin{equation}
E_{\varepsilon}=\{(i,j):D_{ij}<\varepsilon\}
\label{eq:eps_graph}
\end{equation}
are connected.  
This $\varepsilon$–graph preserves local neighborhoods and manifold adjacency while avoiding excessive long-range links, thereby restoring short-range curvature fidelity that the MST alone omits.

\paragraph{}
When persistent homology suggests a strong $H_1$ signal, we further augment the structure by explicitly forming a ring that reinforces the dominant loop.  
Each vertex $v_i$ is assigned an angular coordinate $\theta_i\in[0,2\pi)$ obtained from persistent cohomology, spectral embedding, or a surrogate PCA–atan2 projection.  
Sorting the subset $V_{\mathrm{ring}}\subseteq V$ by $\theta_i$, we connect consecutive points with wrap-around closure,
\begin{equation}
E_{\mathrm{ring}}=\Big\{(i_k,i_{k+1})\Big\}_{k=1}^{|V_{\mathrm{ring}}|-1}\cup\{(i_{|V_{\mathrm{ring}}|},i_1)\},
\label{eq:ring_edges}
\end{equation}
embedding an explicit cycle corresponding to the most persistent homology class.  
If no significant $H_1$ feature is detected, this augmentation is omitted.

\paragraph{}
The provisional edge set,
\begin{equation}
E' = E_{\mathrm{MST}}\cup E_{\varepsilon}\cup E_{\mathrm{ring}},
\end{equation}
may still contain multiple connected components, especially when the representative points occupy disjoint or sparsely sampled regions.  
To restore global connectivity, we iteratively link components by adding the shortest inter-component edge,
\begin{equation}
(i^\star,j^\star)=
\arg\min_{i\in C_p,\; j\in C_{p+1}} D_{ij},
\label{eq:patch_edges}
\end{equation}
until a single connected graph remains.  
The added edges form the patch set $E_{\mathrm{patch}}$.

\paragraph{}
The final edge collection is thus
\begin{equation}
E = E_{\mathrm{MST}}\cup E_{\varepsilon}\cup E_{\mathrm{ring}}\cup E_{\mathrm{patch}},
\label{eq:final_edges}
\end{equation}
ensuring that $G=(V,E)$ is connected, locally faithful, and topologically expressive.  
Edges from the MST enforce global reachability; those from $E_{\varepsilon}$ reconstruct local geometry; and those from $E_{\mathrm{ring}}$ explicitly embed the dominant one-dimensional cycle.  
Together, these elements guarantee that the cycle space $\ker B_1$ of $G$ accurately reflects the principal $H_1$ feature of the attractor.  
For loop-like geometries such as the Lorenz double-wing attractor, the ring augmentation ensures that the first Betti number $\beta_1=1$ is preserved in the graph Laplacian $L_1=B_1^\top B_1$ before supersymmetric extension.

\paragraph{}
The quantile threshold $Q_{0.3}$ can be tuned to data density: higher quantiles increase local connectivity but risk introducing spurious short cycles, whereas lower values may fragment the graph.  
Both $\varepsilon$ and MST edges are computed using Euclidean distances for stability, though alternative metrics such as geodesic or diffusion distances may be substituted.  
Deterministic tie-breaking is applied in \eqref{eq:mst} and \eqref{eq:patch_edges} to ensure reproducibility.

\paragraph{}
The computational cost of each step scales efficiently: MST construction requires $O(n\log n)$ using Kruskal or Prim algorithms,  
the $\varepsilon$–graph formation scales as $O(n^2)$ (or $O(n\log n)$ with spatial indexing), and the ring augmentation scales linearly with $|V_{\mathrm{ring}}|$.  
The final graph contains $O(n)$ edges in typical sparse settings, making it tractable for both classical Laplacian assembly and quantum operator encoding (Section~\ref{sec:susy_ham_construction}).  
All subsequent SUSY Hamiltonian blocks $\mathcal{L}_k$ inherit this sparsity, enabling efficient simulation via product-formula time evolution.

\paragraph{}
The constructed graph $G$ therefore acts as a compact, topology-aware skeleton of the embedded attractor.  
It faithfully preserves local neighborhoods while revealing the main loop structure identified by persistent homology.  
This balance between geometric fidelity and topological simplicity is crucial for ensuring that the SUSY Hamiltonian derived from $G$ accurately encodes the persistent $H_1$ features in its low-energy spectrum.

\subsection{Dicke State Encoding}
\label{sec:dicke_encoding}

\paragraph{}
The aim is to encode the topology of the representative-point graph into a quantum probe that predominantly resides in the symmetric subspace of $(\mathbb{C}^2)^{\otimes n}$. 
Dicke states form a natural and computationally efficient basis for this purpose: they capture global excitation-number symmetry, preserve combinatorial structure, and admit compact circuit representations with logarithmic depth. 
For $n$ qubits, the weight-$k$ Dicke state is
\begin{equation}
\ket{D_k^{(n)}}=\binom{n}{k}^{-1/2}\!\!
\sum_{\substack{x\in\{0,1\}^n\\ |x|=k}}\ket{x},\qquad k=0,1,\dots,n,
\end{equation}
with $\{|x|=k\}$ the set of computational strings of Hamming weight $k$. 
These states are orthonormal and span the $(n{+}1)$-dimensional symmetric subspace $\mathcal{H}_{\mathrm{sym}}$, diagonalizing the collective spin $J_z=\tfrac12\sum_{i=1}^n Z_i$ with eigenvalues $m_k=k-\tfrac{n}{2}$. 
Throughout we use the ordering $\ket{x}=\ket{x_{n-1}\dots x_0}$ and the subset index $\mathrm{ind}(S)=\sum_{i\in S}2^i$ for $S\subseteq\{0,\dots,n-1\}$, so that amplitudes are assigned uniformly within each weight-$k$ sector and normalization ensures $\langle D_k^{(n)}|D_{k'}^{(n)}\rangle=\delta_{kk'}$.

\paragraph{}
Topological information extracted from the graph $G=(V,E)$ is embedded into a symmetric superposition of Dicke sectors,
\begin{equation}
\ket{\psi}=
\frac{\sum_{k=0}^{n}\tilde{w}_k\,\ket{D_k^{(n)}}}
{\left\|\sum_{k=0}^{n}\tilde{w}_k\,\ket{D_k^{(n)}}\right\|},
\qquad
\tilde{w}_k=\frac{w_k}{\sqrt{\sum_j w_j^2}},
\label{eq:dicke_superposition}
\end{equation}
where the unnormalized weights $w_k$ encode both local connectivity and global loop persistence. 
Local bias arises from degrees and ring participation: letting $\mathcal{E}_{\mathrm{ring}}$ denote edges that lie on the dominant cycle and $\deg(v)$ the degree of vertex $v$, the update
\begin{equation}
w_k \;\gets\; 
w_k\;+\;
\alpha \!\!\sum_{(u,v)\in\mathcal{E}_{\mathrm{ring}}}\!\!\big[\mathbf{1}\{u=k\}+\mathbf{1}\{v=k\}\big]
\;+\;
\beta \!\!\sum_{v:\deg(v)=k}\!\!1
\label{eq:dicke_localbias}
\end{equation}
reinforces sectors touched by the ring and emphasizes hubs, with tunable $\alpha,\beta\ge 0$ setting the relative importance of the two contributions. 
Global topological strength further modulates contrast via the persistence $\Lambda$ of the most persistent $H_1$ class:
\begin{equation}
w_k \;\gets\; (1+\eta\,\Lambda)\,w_k,\qquad \eta>0,
\label{eq:dicke_globalgain}
\end{equation}
which amplifies all sectors proportionally when a robust loop is present while preserving the local bias profile.

\paragraph{}
After normalization, sector populations $p_k=|\langle D_k^{(n)}|\psi\rangle|^2=\tilde{w}_k^{\,2}$ define a probability distribution over excitation number. 
The expectation
\begin{equation}
M=\sum_k \!\left(k-\tfrac{n}{2}\right)p_k=\langle\psi|J_z|\psi\rangle
\end{equation}
plays the role of a magnetization-like order parameter, and the variance $\mathrm{Var}(J_z)=\sum_k (k-\tfrac{n}{2}-M)^2p_k$ yields the quantum Fisher information $F_Q=4\,\mathrm{Var}(J_z)$, linking the breadth of the excitation distribution to metrological sensitivity. 
In this way, loop prominence and heterogeneity manifest as broadening of $\{p_k\}$, nonzero $M$, and enhanced $F_Q$, providing a direct statistical signature of topological complexity.

\paragraph{}
Crucially, the supersymmetric Hamiltonian $\mathcal{H}$ constructed in Section~\ref{sec:susy_ham_construction} conserves total excitation number and decomposes into blocks $\mathcal{L}_k$ acting on the weight-$k$ sector:
\begin{equation}
e^{-i\mathcal{H}t}=\bigoplus_{k=0}^{n} e^{-i\mathcal{L}_k t}.
\end{equation}
A probe of the form \eqref{eq:dicke_superposition} therefore aligns with the block-diagonal structure of $\mathcal{H}$ and maximizes overlap with homology-bearing low-energy modes. 
Because excitation number is conserved, evolution remains within the symmetric manifold—significantly reducing Hilbert-space dimension from $2^n$ to $(n{+}1)$ and thereby lowering simulation and QPE circuit cost. 
The single-ancilla Hadamard test or QPE then accesses the autocorrelation
\begin{equation}
C(t)=\langle\psi|e^{-i\mathcal{H}t}|\psi\rangle
=\sum_k p_k\, e^{-iE_k t},
\end{equation}
whose spectral lines at $\{E_k\}$ expose near-harmonic modes tied to topological features while preserving the symmetric-sector decomposition.

\paragraph{}
Implementation on near-term hardware admits exact or approximate preparation. 
Exact $\ket{D_k^{(n)}}$ states can be synthesized in logarithmic depth via permutation-symmetric isometries or tree networks of controlled rotations; approximate preparation is achievable with variational or iterative amplitude-loading schemes. 
The superposition $\sum_k \tilde{w}_k\ket{D_k^{(n)}}$ is realized by preparing a weight register and mapping $\ket{k}\ket{0^n}\mapsto\ket{k}\ket{D_k^{(n)}}$, followed by uncomputing $\ket{k}$; normalization of $\{\tilde{w}_k\}$ stabilizes amplitudes numerically. 
Because excitation number is conserved under $e^{-i\mathcal{H}t}$, there is no cross-sector leakage during evolution, which simplifies controlled time evolution and reduces circuit depth in QPE.

\paragraph{}
From a physical standpoint, graphs with balanced connectivity concentrate weight near $k\simeq n/2$, yielding $M\simeq 0$ and narrower distributions; sparse or strongly cyclic graphs skew $\{p_k\}$, induce nonzero magnetization, and increase $F_Q$. 
The Dicke manifold thus provides an interpretable encoding in which loop geometry and symmetry breaking translate into measurable sector populations and interferometric sensitivity. 
In summary, Dicke-state encoding turns classical $H_1$ structure into structured superpositions across excitation-number sectors, couples naturally to the SUSY block structure, and yields a probe that is both conceptually transparent and hardware-efficient for quantum spectral readout.

\paragraph{}
While the Dicke-state formalism provides a compact theoretical description, 
exact preparation of $\ket{D_k^{(n)}}$ on current devices remains resource-intensive. 
In our implementation, we employed a simplified circuit that efficiently generates 
the single-excitation ($k{=}1$) Dicke state 
$\ket{W_n}=\ket{D^{(n)}_{1}}$ from $\ket{100\dots0}$ 
using a linear sequence of controlled-$R_y$ and CNOT gates. 
This scheme transfers the excitation along the register, 
dividing the amplitude evenly at each step and achieving 
depth $O(n)$ while maintaining high fidelity under realistic noise. 
Higher-weight sectors are then approximated by composing multiple such W-state layers 
or by variational amplitude-loading when available. 
Thus, although the analytical description uses the full Dicke manifold, 
the experimental embedding adopts an optimized, hardware-compatible version 
that retains the essential symmetry and spectral features relevant to the QPE readout.

\subsection{SUSY Hamiltonian Construction}
\label{sec:susy_ham_construction}

\paragraph{}
The supersymmetric Hamiltonian $\mathcal{H}$ is constructed on $n$ qubits to faithfully encode the topology of the representative-point graph $G=(V,E)$ obtained from data. The operator is designed to obey the $\mathcal{N}=2$ supersymmetry algebra, preserve excitation number, and remain efficiently simulable on quantum hardware. 
The SUSY Hamiltonian coincides with the combinatorial Hodge–Laplacian. In particular, the zero-energy sector encodes topological invariants of the graph, such as the Betti numbers.

\paragraph{}
We define the following local operators acting on each qubit:
\begin{equation}
\{I,X,Z,\; z,o\},\qquad
z=\tfrac{I+Z}{2},\;\; o=\tfrac{I-Z}{2},
\end{equation}
where $z$ and $o$ act as projectors onto $\ket{0}$ and $\ket{1}$, respectively.
All local letters are Hermitian and real, and after Hermitization the Hamiltonian
$\mathcal{H}$ becomes real and symmetric. The set $\{I,X,Z,z,o\}$ is sufficient for
representing the Projected-basis SUSY Hamiltonian without introducing complex $Y$ rotations.

\paragraph{}
Fermionic excitations are represented by Jordan–Wigner strings that ensure proper anticommutation. For vertex $i$ the elementary flip operator is
\begin{equation}
X_i^{\mathrm{JW}}=\Big(\prod_{k<i}Z_k\Big)X_i.
\end{equation}
To enforce clique consistency, we introduce complement-graph projectors. For each vertex $i$, let
\begin{equation}
N_i^{\mathrm{comp}}=\{j\in V:(i,j)\notin E,\ j\neq i\},
\end{equation}
and define
\begin{equation}
P_i=\prod_{j\in N_i^{\mathrm{comp}}}\frac{I+Z_j}{2}
=\bigotimes_{j\in V}
\begin{cases}
z_j,& j\in N_i^{\mathrm{comp}},\\
I_j,& \text{otherwise}.
\end{cases}
\end{equation}
These projectors annihilate configurations in which vertex $i$ is excited simultaneously with a non-neighbor, enforcing adjacency constraints. The elementary supercharge at site $i$ is $Q_i=X_i^{\mathrm{JW}}P_i$, and the total supercharge is $Q=\sum_i Q_i$, and by construction $Q^2=0$.

\paragraph{}
The Hamiltonian follows as the anticommutator
\begin{equation}
\mathcal{H}=Q^\dagger Q
=\sum_i Q_i^\dagger Q_i+\sum_{i<j}(Q_i^\dagger Q_j+Q_j^\dagger Q_i),
\end{equation}
which automatically commutes with excitation number. In the $0$–excitation sector $\mathcal{H}$ acts as the vertex Laplacian $\mathcal{L}_0$, in the $1$–excitation sector as the edge Laplacian $\mathcal{L}_1$, and in higher sectors as clique Laplacians for multi-excitation configurations:
\begin{equation}
\mathcal{H}=\bigoplus_{k=0}^{n}\mathcal{L}_k,\qquad
\mathcal{L}_k=d_k^\dagger d_k+d_{k-1}d_{k-1}^\dagger.
\end{equation}
This block decomposition mirrors the combinatorial Hodge structure and ensures correspondence between harmonic subspaces and homology groups.

\paragraph{}
Substituting the operator definitions,
\begin{equation}
z=\tfrac{I+Z}{2},\quad o=\tfrac{I-Z}{2},
\end{equation}
clarifies that each local factor belongs to the $\{I,X,Z\}$ Pauli family 
and that $z,o$ act only as symbolic projectors composed of them.
In practice the full $(I,Z)$ expansion of $z,o$ is not carried out;
they are treated as diagonal control predicates.
Under this symbolic treatment, $\mathcal{H}$ remains a sparse sum of few-qubit
Pauli operators with real coefficients.

\paragraph{}
Because each supercharge $Q_i$ acts non-diagonally on a single qubit
and diagonally (via projectors) on the rest,
the interaction structure involves only qubit pairs $(i,j)$ that share an edge in $G$.
For graphs of bounded degree, the number of distinct tensor-product terms
therefore scales quadratically in $n$,
$\mathcal{O}(n^2)$, rather than exponentially in $4^n$.
This estimate has been verified for the representative Lorenz-derived graphs used here, where the compiled Hamiltonians contain a few hundred terms.
After Hermitization the operator is real and symmetric.

\paragraph{}
The kernel of $\mathcal{H}$ corresponds to harmonic forms,
\begin{equation}
\ker\mathcal{H}\big|_{\text{$k$–sector}}=\ker\mathcal{L}_k\cong H^k,
\end{equation}
so that the multiplicity of zero eigenvalues in degree $1$ equals the first Betti number $\beta_1$. 
The smallest nonzero eigenvalue,
\begin{equation}
\Delta^{(1)}_{\mathrm{SUSY}}
= \min\{\lambda>0 \mid \lambda\in\sigma(\mathcal L_1)\}.
\end{equation}
acts as a \emph{topological gap} that measures the spectral isolation of persistent cycles,
as formally justified in the Supplementary Materials
(Section~\ref{sec:theoretical_relationship_between_energy_gap_and_persistence}).
Its variation with the control parameter $\rho$ tracks the emergence, merger, and decay of topological features.

\paragraph{}
Any global energy offset $c_I I$ appearing in the Pauli expansion is separated as $\mathcal{H}=c_I I+(\mathcal{H}-c_I I)$, with the phase $c_I t$ applied only to the ancilla branch during controlled evolution. This adjustment stabilizes interferometric measurements while leaving eigenvectors and relative gaps intact.

\paragraph{}
For simulation, each exponential $e^{-i\theta t\bigotimes_q P_q}$ is implemented by reducing local letters to $Z$, aggregating parity with CNOT ladders, and applying a controlled $R_Z(2\theta t)$ on the reference qubit conditioned on the predicate. Because the complement-graph projectors commute, control masks can be toggled efficiently in Gray order~\cite{Shende2006GrayCode} so that successive terms differ by a single-qubit flip. Grouping identical Pauli-letter patterns consolidates many rotations into a single $R_Z$, and multi-control conjunctions are computed into an ancilla bit, replacing deep multi-controlled gates with two $\mathrm{MCX}$ and one $CR_Z$ operation.

\paragraph{}
The resulting Hamiltonian is sparse, local, and block-diagonal, explicitly preserving topological sectors~\cite{Berry2007SparseHamiltonian}. For typical Lorenz-derived graphs ($n\approx10$–$12$), the compiled operator contains a few hundred Pauli strings and fewer than $10^3$ controlled rotations, well within current simulator and NISQ capabilities. The method extends naturally to higher simplicial dimensions by including multi-excitation constraints, providing a scalable template for encoding combinatorial Laplacians as quantum circuits.

\paragraph{}
This Hamiltonian reproduces the spectral structure of the Hodge Laplacian;
its low-lying gaps empirically co-vary with persistent-Laplacian features.
In this way, the SUSY Hamiltonian serves as an algebraic bridge between discrete topological data analysis and measurable quantum spectra, enabling topological invariants to be extracted directly through quantum phase estimation.

\subsection{Controlled Time Evolution Circuit}
\label{sec:controlled_evolution}

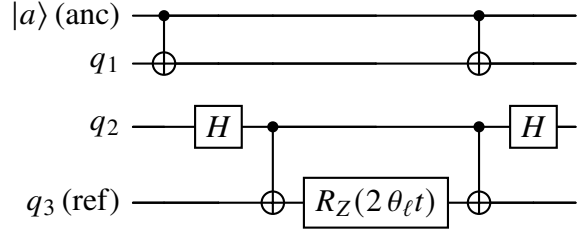
\begin{figure}[!b]
\centering
\begin{quantikz}[row sep=0.4cm, column sep=0.25cm]
\lstick{$\ket{a}$\,\text{(anc)}} & \ctrl{1} & \qw & \qw & \qw & \ctrl{1} & \qw & \qw\\
\lstick{$q_1$}                   & \targ{}  & \qw & \qw & \qw & \targ{}  & \qw & \qw\\
\lstick{$q_2$}                   & \qw      & \gate{H} & \ctrl{1} & \qw & \ctrl{1} & \gate{H} & \qw\\
\lstick{$q_3$\,\text{(ref)}}     & \qw      & \qw & \targ{} & \gate{\RZ{2\,\theta_\ell t}} & \targ{} & \qw & \qw
\end{quantikz}
\caption{Controlled time evolution for one local term
\(U_\ell(t)=e^{-i\theta_\ell t P_{q_1}\!\otimes P_{q_2}\!\otimes P_{q_3}}\).
Basis rotations convert $X/Y$ to $Z$, CNOTs collect parity on $q_3$, and an ancilla-controlled \(\RZ{2\theta_\ell t}\) realizes the exponential.}
\label{fig:controlled_term}
\end{figure}

\begin{figure}[t]
\centering
\begin{quantikz}[row sep=0.35cm, column sep=0.28cm]
\lstick{$\ket{a}$\,\text{(anc)}} & \qw & \qw & \ctrl{3} & \qw & \qw & \qw & \qw\\
\lstick{$q_1$}                   & \gate{H} & \qw & \qw & \ctrl{2} & \qw & \qw & \gate{H}\\
\lstick{$q_2$}                   & \gate{R_x(-\pi/2)} & \qw & \qw & \ctrl{1} & \qw & \qw & \gate{R_x(\pi/2)}\\
\lstick{$q_3$\,\text{(ref $r$)}} & \qw & \qw & \qw & \targ{} & \gate{\RZ{2\,\mathrm{Re}(\theta_\ell)t}} & \qw & \qw
\end{quantikz}
\caption{Controlled time evolution for one local factor
\(U_\ell(t)=e^{-i\theta_\ell t\bigotimes_{q\in S_\ell}P_q^{(\ell)}}\).
Here \(P_{q_1}=X\), \(P_{q_2}=Y\), \(P_{q_3}=Z\).
CNOTs collect the $Z$-parity onto the rightmost reference qubit \(r=q_3\);
an ancilla-controlled \(\RZ{2\,\mathrm{Re}(\theta_\ell)t}\) implements the phase; then uncompute.}
\label{fig:controlled_term_basic}
\end{figure}
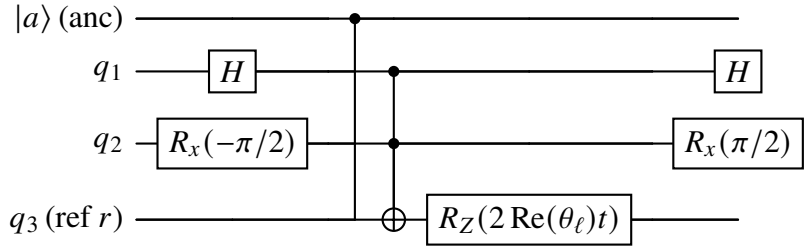

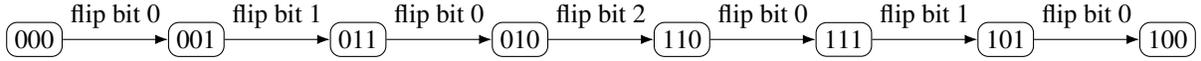
\begin{figure}[t]
\centering
\begin{tikzpicture}[>=Latex, node distance=14mm, every node/.style={font=\footnotesize}]
\node (m0) [draw, rounded corners, inner sep=3pt] {$000$};
\node (m1) [draw, rounded corners, right=of m0, inner sep=3pt] {$001$};
\node (m2) [draw, rounded corners, right=of m1, inner sep=3pt] {$011$};
\node (m3) [draw, rounded corners, right=of m2, inner sep=3pt] {$010$};
\node (m4) [draw, rounded corners, right=of m3, inner sep=3pt] {$110$};
\node (m5) [draw, rounded corners, right=of m4, inner sep=3pt] {$111$};
\node (m6) [draw, rounded corners, right=of m5, inner sep=3pt] {$101$};
\node (m7) [draw, rounded corners, right=of m6, inner sep=3pt] {$100$};
\draw[->] (m0) -- node[above] {$\text{flip bit }0$} (m1);
\draw[->] (m1) -- node[above] {$\text{flip bit }1$} (m2);
\draw[->] (m2) -- node[above] {$\text{flip bit }0$} (m3);
\draw[->] (m3) -- node[above] {$\text{flip bit }2$} (m4);
\draw[->] (m4) -- node[above] {$\text{flip bit }0$} (m5);
\draw[->] (m5) -- node[above] {$\text{flip bit }1$} (m6);
\draw[->] (m6) -- node[above] {$\text{flip bit }0$} (m7);
\end{tikzpicture}
\caption{Gray-order traversal of a 3-bit control mask.
Consecutive masks differ by a single bit, minimizing the number of control toggles
(\(X\) gates) between terms that share the same Pauli-letter pattern.}
\label{fig:gray_order}
\end{figure}

\paragraph{}
For quantum phase estimation on a supersymmetric Hamiltonian $\mathcal{H}$, we synthesize the ancilla-controlled time evolution
\begin{equation}
U_{\mathrm{ctrl}}(t)=\ket{0}\!\bra{0}\otimes I+\ket{1}\!\bra{1}\otimes e^{-i\mathcal{H}t},
\end{equation}
so that interference between the ancilla and system registers encodes the spectral phase of $\mathcal{H}$ with high fidelity~\cite{AbramsLloyd1999QPE,Kitaev1995PhaseEstimation,Sato2024HyperbolicPDE}.  
The construction proceeds term by term and is illustrated in Figure~\ref{fig:controlled_term} and Figure~\ref{fig:controlled_term_basic}, while the Gray-order traversal used to optimize control toggles is shown in Figure~\ref{fig:gray_order}.

\paragraph{}
Each local factor of $\mathcal{H}$ is drawn from the symbolic alphabet
\[
\{I,X,Z,z,o\},\qquad 
z=\tfrac{I+Z}{2},\quad o=\tfrac{I-Z}{2},
\]
so that
\begin{equation}
\mathcal{H}=\sum_{\ell}\theta_\ell\bigotimes_{q=1}^{n}P_q^{(\ell)},\qquad 
P_q^{(\ell)}\in\{I,X,Z,z,o\},
\end{equation}
with real coefficients $\mathrm{Re}(\theta_\ell)$ after Hermitization.  
Each term acts on a small subset of qubits, ensuring sparsity and locality~\cite{Berry2007SparseHamiltonian,Childs2018PNAS}.  
For a local operator $\mathcal{H}_\ell=\theta_\ell\bigotimes_{q\in S_\ell}P_q^{(\ell)}$, 
the rightmost active qubit $r=\max S_\ell$ serves as the reference site.  
To standardize all terms, the non-diagonal operators are converted to $Z$ form using a Hadamard gate on each affected qubit, as depicted in the left half of Figure~\ref{fig:controlled_term_basic}.
In the present construction all off-diagonal couplings originate from $X$-type flip terms in the supercharges.
The $Y$ operator does not appear because no complex phase rotations are required; all coefficients are real after Hermitization.  

\paragraph{}
After basis alignment, a CNOT ladder
\begin{equation}
L_\ell=\prod_{q\in S_\ell\setminus\{r\}}\mathrm{CNOT}(q\!\to\!r)
\end{equation}
collects the parity of all $Z$ factors onto the reference qubit~\cite{LowChuang2019Qubitization}, ensuring
\(
L_\ell^\dagger\!\big(\prod_{q\in S_\ell}Z_q\big)L_\ell=Z_r.
\)
The controlled evolution for that term then reduces to a single $R_Z(2\,\mathrm{Re}(\theta_\ell)t)$ rotation on qubit $r$, conditioned on the ancilla and any projector controls.  
The right half of Figure~\ref{fig:controlled_term} and Figure~\ref{fig:controlled_term_basic} show this structure: parity is gathered on $r$, the ancilla-controlled rotation applies the phase, and all intermediate operations are uncomputed to restore the original basis.

\paragraph{}
Projector symbols $z=(I+Z)/2$ and $o=(I-Z)/2$ correspond to $\ket{0}$- and $\ket{1}$-controls, respectively.  
To unify them, temporary $X$ gates toggle $\ket{0}$-controls into the $\ket{1}$ basis, 
and a multi-qubit conjunction is computed into an auxiliary predicate qubit,
\begin{equation}
\Pi_\ell=\mathrm{AND}\big(\mathrm{ancilla},\,C_0^{(\ell)},\,C_1^{(\ell)}\big),
\end{equation}
where the \emph{ancilla} denotes the interference qubit used in the Hadamard-test or QPE protocol.  
It is initialized in $(\ket{0}{+}\ket{1})/\sqrt{2}$ and controls whether the system register undergoes 
time evolution: the $\ket{0}$ branch remains idle, whereas the $\ket{1}$ branch activates the controlled evolution $e^{-i\mathcal{H}t}$.  
The sets $C_0^{(\ell)}$ and $C_1^{(\ell)}$ index the projector-controlled qubits associated with $z$- and $o$-type predicates, respectively.  
The composite predicate $\Pi_\ell$ therefore represents the logical \textsc{AND} of all active controls—ancilla, $\ket{0}$-controls, and $\ket{1}$-controls—and acts as a single effective control~\cite{Barenco1995,Maslov2016MultiControlRZ} for the rotation gate.

\paragraph{}
Each local exponential term then compiles as
\begin{equation}
U_\ell(t)=U_{\mathrm{b}}\,U_{\mathrm{lad}}\,
\big[R_Z(2\,\mathrm{Re}(\theta_\ell)t)\ \text{on qubit $r$ controlled by}\ \Pi_\ell\big]\,
U_{\mathrm{lad}}^\dagger U_{\mathrm{b}}^\dagger,
\end{equation}
where $U_{\mathrm{b}}$ collects the basis transformations (e.g., $H$ or $R_x(-\pi/2)$ rotations), 
and $U_{\mathrm{lad}}$ is the \emph{CNOT ladder} circuit that aggregates the parity of all $Z$-type factors 
onto the rightmost active qubit $r=\max S_\ell$ before the controlled rotation.  
\paragraph{}
This construction ensures that each local factor $\bigotimes_{q\in S_\ell} P_q^{(\ell)}$ 
is implemented with minimal control overhead while preserving full commutation structure.

\paragraph{}
Since all $R_Z$ rotations commute, terms sharing the same control mask and reference qubit can be merged.
To minimize the number of control toggles between successive terms, the control masks are scheduled in Gray order, as shown in Figure~\ref{fig:gray_order}, where consecutive masks differ by only a single bit.
Predicate caching further lowers the depth by computing the logical AND once, applying one $CR_Z$, and uncomputing it.

\paragraph{}
The total time evolution over duration $t$ is approximated using product formulas~\cite{Childs2018PNAS,Berry2007SparseHamiltonian}.  
A first-order Trotter step
\[
\tilde{U}_1(t)=\prod_\ell U_\ell(t)
\]
achieves $O(t^2)$ accuracy, whereas the symmetric second-order step
\[
\tilde{U}_2(t)=\tilde{U}_1(t/2)\,\tilde{U}_1(-t/2)^\dagger
\]
suppresses the error to $O(t^3)$, with the bound
\[
\big\|e^{-i\mathcal{H}t}-\tilde{U}(t)\big\|=O\!\left(t^2\max_{\ell,\ell'}\|[\mathcal{H}_\ell,\mathcal{H}_{\ell'}]\|\right).
\]
A global energy offset $\mathcal{H}=c_I I+(\mathcal{H}-c_I I)$ is handled by applying $e^{-ic_I t}$ only to the ancilla’s $\ket{1}$ branch, stabilizing the interferometric phase without affecting eigenvectors or relative gaps.

\paragraph{}
After grouping and optimization, the circuit depth is dominated by a few parity ladders and multi-control rotations, while most remaining operations are commuting single-qubit $R_Z$ gates.  
For typical systems with $n\approx10$ qubits, a complete controlled evolution requires $\sim10^4$ primitive gates.  
Gray-ordered scheduling and commuting-group compilation reduce entangling depth by a factor of 3–5 while maintaining spectral fidelity at the $10^{-3}$ level in the extracted gap $\gamma$.  
The resulting construction provides an efficient and hardware-compatible realization of $U_{\mathrm{ctrl}}(t)$ suitable for Hadamard-test and QPE-based spectroscopic estimation of supersymmetric energy gaps~\cite{Cade2024SUSY,Cowtan2019PhaseGadget}.

\paragraph{}
This Projected-basis compilation not only reduces circuit depth but also emulates
the structural evolution of the combinatorial Laplacian under topological filtration.
As noted in the persistent spectral framework of Meng and Xia~\cite{MengXia2021PerSpectML},
the number and strength of off-diagonal Laplacian entries vary systematically as
simplices are added or removed along a filtration, producing a continuous growth and
decay of coupling terms that governs spectral transitions.
In our implementation, the Projected-basis realizes this mechanism at the
operator level: Pauli-grouped commuting terms encode diagonal potentials, while
Projected-basis operations selectively activate or suppress off-diagonal couplings
corresponding to newly formed or annihilated simplices.
Consequently, the controlled time-evolution operator
\(U(t)=e^{-i\widetilde{\mathcal H}t}\)
mimics the dynamic modulation of connectivity observed in persistent Laplacian models,
ensuring that the evolution of the quantum spectrum reproduces the same
addition–removal process of Laplacian couplings that drives the filtration-dependent
spectra.

\subsection{Quantum Phase Estimation (Eigenvalue Extraction)}
\label{sec:qpe_detail}

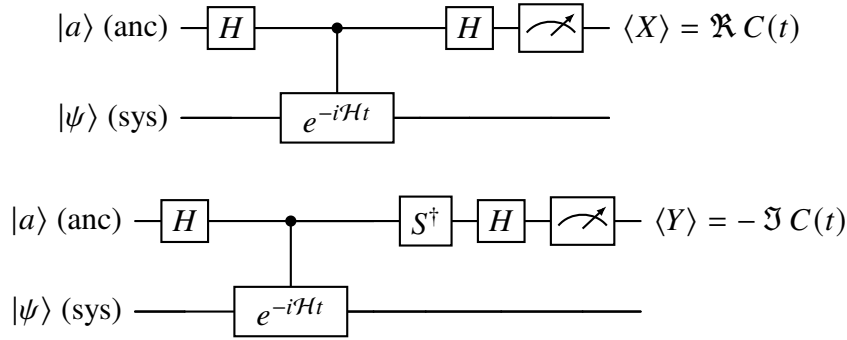
\begin{figure}[t]
\centering
\begin{quantikz}[row sep=0.55cm, column sep=0.35cm]
\lstick{$\ket{a}$ (anc)} & \gate{H} & \ctrl{1} & \qw & \gate{H} & \meter{} & \rstick{$\langle X\rangle=\Re\,C(t)$}\\
\lstick{$\ket{\psi}$ (sys)} & \qw & \gate[wires=1][1.5cm]{e^{-i\mathcal{H}t}} & \qw & \qw & \qw & \qw
\end{quantikz}

\vspace{0.25cm}

\begin{quantikz}[row sep=0.55cm, column sep=0.35cm]
\lstick{$\ket{a}$ (anc)} & \gate{H} & \ctrl{1} & \qw & \gate{S^\dagger} & \gate{H} & \meter{} & \rstick{$\langle Y\rangle=-\,\Im\,C(t)$}\\
\lstick{$\ket{\psi}$ (sys)} & \qw & \gate[wires=1][1.5cm]{e^{-i\mathcal{H}t}} & \qw & \qw & \qw & \qw & \qw
\end{quantikz}
\caption{Hadamard-test circuits for a single sample time $t$. The controlled-$e^{-i\mathcal{H}t}$ block is synthesized as $U_{\mathrm{ctrl}}(t)$ (Sec.~\ref{sec:controlled_evolution}). Measuring $X$ (top) yields $\Re\,C(t)$; inserting $S^\dagger$ then measuring $X$ (bottom) yields $-\,\Im\,C(t)$.}
\label{fig:hadamard_test}
\end{figure}

\paragraph{}
We recover the eigenvalues of the supersymmetric Hamiltonian $\mathcal{H}$ 
by analyzing the time autocorrelation of a prepared probe state. 
For a normalized state $|\psi\rangle$, the correlation function is
\begin{equation}
C(t)=\langle \psi | e^{-i\mathcal{H}t} |\psi\rangle
=\sum_j a_j e^{-iE_j t},\qquad 
a_j=|\langle E_j|\psi\rangle|^2,\ \ a_j\ge0,\ \sum_j a_j=1,
\end{equation}
so that the frequency content consists of spectral lines at $\{E_j\}$ with nonnegative weights $\{a_j\}$. 
The measurement is performed via a single-ancilla Hadamard test (Figure~\ref{fig:hadamard_test}): 
the ancilla prepares $(|0\rangle{+}|1\rangle)/\sqrt{2}$, 
interferes a reference branch with a data branch undergoing the controlled evolution $e^{-i\mathcal{H}t}$, 
and is measured along $X$ or $Y$, yielding 
$\langle X\rangle=\Re\,C(t)$ and $\langle Y\rangle=-\,\Im\,C(t)$. 
The correlation function $C(t)$ is sampled on a uniform grid $t_k=k\,\Delta t$ 
for $k=0,\dots,T{-}1$. 
Identity offsets are removed by writing $\mathcal{H}=c_I I+(\mathcal{H}-c_I I)$ 
and applying the global phase $e^{-ic_I t}$ only to the ancilla’s $\ket{1}$ branch, 
so that the measured phases reflect $\mathcal{H}-c_I I$.

\paragraph{}
Because the Hadamard test measures the expectation value 
$\mathrm{Tr}(\rho\,e^{-i\mathcal{H}t})$ for an arbitrary density operator~$\rho$, 
the same circuit applies without modification to statistical mixtures or 
decohered superpositions. 
In practice, we emulate such mixedness by averaging over random single–qubit $Z$ phases 
within the Hadamard–test loop, effectively transforming 
$\rho = |\psi\rangle\!\langle\psi|$ into 
$\rho=\mathbb{E}_Z[Z|\psi\rangle\!\langle\psi|Z^\dagger]$ and suppressing 
off–diagonal coherences between Dicke sectors. 
In particular, for the Dicke–encoded probe $\rho=\sum_k p_k |D_k^{(n)}\rangle\langle D_k^{(n)}|$, 
the measured correlator becomes 
$C(t)=\mathrm{Tr}(\rho\,e^{-i\mathcal{H}t})=\sum_k p_k e^{-iE_k t}$, 
so the interferometric signal directly represents a weighted spectral average 
over symmetric sectors. 
Thus, mixedness arising from imperfect preparation, dephasing, or intentional 
classical weighting is inherently incorporated in the measured autocorrelation, 
and no circuit modification is required.

\paragraph{}
Formally, the equivalence between this averaged Hadamard–test measurement 
and a true mixed–state expectation can be seen by expanding the coherent probe 
$|\psi\rangle = \tfrac{1}{\sqrt{N}}\sum_e |e\rangle$ in the computational basis.
The interferometric estimator yields
\begin{equation}
    \langle \psi | e^{-i\mathcal{H}t} | \psi \rangle
    = \frac{1}{N}\sum_{e,e'} \langle e'| e^{-i\mathcal{H}t} | e\rangle
    = \frac{1}{N}\sum_{e} \langle e| e^{-i\mathcal{H}t} | e\rangle
      + \frac{1}{N}\sum_{e\neq e'} \langle e'| e^{-i\mathcal{H}t} | e\rangle .
\end{equation}
The first term corresponds exactly to the mixed–state trace 
$\mathrm{Tr}(\rho\,e^{-i\mathcal{H}t})$ for 
$\rho = \tfrac{1}{N}\sum_e |e\rangle\!\langle e|$, 
while the second term contains off–diagonal coherences between distinct basis states.
Under temporal averaging or random–phase dephasing, 
these cross terms vanish because their phases oscillate at 
frequencies $\omega_{e'}-\omega_e$, leading to
\begin{equation}
    \overline{\langle \psi | e^{-i\mathcal{H}t} | \psi \rangle}
    = \frac{1}{N}\sum_e \langle e| e^{-i\mathcal{H}t} | e\rangle
    = \mathrm{Tr}(\rho\,e^{-i\mathcal{H}t}) .
\end{equation}
Hence, the Hadamard test on a uniform superposition state reproduces 
the expectation value of an incoherent statistical ensemble 
without requiring any additional ancilla or explicit purification. 
In other words, quantum interference within the single Hadamard–test circuit 
naturally performs the same linear averaging that would otherwise 
arise from tracing out an ancillary subsystem in a purified mixed state.

\paragraph{}
A tapered discrete transform with window $w_k$ (e.g., Hann) is then formed as
\begin{equation}
\tilde{C}(\omega_\ell)=\sum_{k=0}^{T-1} w_k\, C(t_k)\, e^{i\omega_\ell t_k},\qquad
\omega_\ell=\frac{2\pi \ell}{T\Delta t},\ \ \ell=0,\dots,T-1,
\end{equation}
giving frequency resolution $\Delta\omega = 2\pi/(T\Delta t)$ over the unaliased band $[0,\pi/\Delta t]$. With $M$ repetitions per $t_k$, shot noise is approximately white with
\begin{equation}
\mathrm{SE}[C(t_k)]\simeq \sqrt{\frac{1-|C(t_k)|^2}{M}},
\end{equation}
and is shaped by the window response $|W(\omega)|$ in $|\tilde C(\omega)|$. Eigenfrequencies appear as peaks of $|\tilde{C}(\omega)|$; letting $\omega_\ell$ be a discrete maximizer with neighbors $A_{-}=|\tilde C(\omega_{\ell-1})|$, $A_0=|\tilde C(\omega_\ell)|$, $A_{+}=|\tilde C(\omega_{\ell+1})|$, quadratic interpolation refines the location and amplitude,
\begin{equation}
\delta=\frac{1}{2}\,\frac{A_{-}-A_{+}}{A_{-}-2A_0+A_{+}}\!,\qquad
\hat{\omega}=\omega_\ell+\delta\,\Delta\omega,\qquad
\hat{A}=A_0-\frac{(A_{-}-A_{+})^2}{8\,(A_{-}-2A_0+A_{+})}.
\label{eq:quadinterp}
\end{equation}
For SUSY blocks probing $H_1$, a zero mode manifests near $\omega\!\approx\!0$; to avoid leakage, a guard width $\Omega_{\mathrm{g}}=\kappa\,\Delta\omega$ with $\kappa\in[1,3]$ is enforced and the smallest positive frequency is estimated as
\begin{equation}
\hat{\gamma}= \min\{\hat{\omega}_i\!:\ \hat{\omega}_i>\Omega_{\mathrm{g}}\},\qquad
\hat{\xi}=1/\hat{\gamma}.
\end{equation}
If no zero mode is present, the two lowest refined peaks yield $\hat E_0$, $\hat E_1$, and the gap $\hat{\gamma}=\hat E_1-\hat E_0$.

\paragraph{}
To cross-check FFT peaks, a parametric estimator fits a sum of complex exponentials to $\{C(t_k)\}$ via a shift-invariant method (Prony/ESPRIT): Hankel matrices $(H_0,H_1)$ are formed, an effective rank $\hat r$ is chosen by SVD thresholding, and the generalized eigenproblem $H_1 v=\lambda H_0 v$ yields $z_j\approx e^{-iE_j \Delta t}$ and hence $\hat E_j=-\arg(z_j)/\Delta t$. Sweeping $\hat r$ over a small range and intersecting stable roots returns a smallest positive consistent element $\hat{\gamma}_{\mathrm{Prony}}$ that is compared against the windowed-DFT estimate.

\paragraph{}
Phase wrap-around is controlled by enforcing $\|\mathcal{H}-c_I I\|<\pi/\Delta t$ (via norm or Gershgorin bounds). Ambiguities are further disambiguated by acquiring a secondary dataset with spacing $\Delta t'$ whose ratio to $\Delta t$ is far from low-order rationals; a true line must lie in the intersection
\begin{equation}
\mathcal{A}(\Delta t)=\bigl\{\hat{\omega}^{(\Delta t)}+\tfrac{2\pi m}{\Delta t}\bigr\},\quad
\mathcal{A}(\Delta t')=\bigl\{\hat{\omega}^{(\Delta t')}+\tfrac{2\pi m'}{\Delta t'}\bigr\},
\end{equation}
from which the unique joint solution within interpolation error bars is selected. A dedicated zero-mode test compares near-zero band power to its sidebands~\cite{Komalan2025QuantumBarcodes,Schmidhuber2025Khovanov},
\begin{equation}
\mathcal{P}_{0}=\!\sum_{|\omega_\ell|\le \Omega_{\mathrm{z}}}\!|\tilde C(\omega_\ell)|^2,\qquad
\mathcal{P}_{\mathrm{sb}}=\tfrac12\!\sum_{\Omega_{\mathrm{z}}<|\omega_\ell|\le 2\Omega_{\mathrm{z}}}\!|\tilde C(\omega_\ell)|^2,
\end{equation}
and declares a zero mode when $R=\mathcal{P}_{0}/\mathcal{P}_{\mathrm{sb}}$ exceeds a calibrated threshold, stabilizing $\beta_1$ counting against spectral leakage.

\paragraph{}
Final aggregation combines the smallest nonzero from the refined FFT and Prony/ESPRIT via a conservative median,
\begin{equation}
\hat\gamma=\operatorname{median}\!\left(\hat\gamma_{\mathrm{FFT}},\ \hat\gamma_{\mathrm{Prony}}\right),
\end{equation}
optionally weighted by inverse residuals (peak-fit error vs.\ reconstruction error). Uncertainty is quantified by a block bootstrap over time: resampling $\{C(t_k)\}$ in blocks at least as wide as the window main lobe yields $\{\hat\gamma^{(b)}\}$ for percentile intervals. Near an isolated line, a fast error proxy is
\begin{equation}
\mathrm{SE}[\hat{\omega}]\approx \frac{\beta_{\mathrm{win}}\,\Delta\omega}{\rho},\quad
\rho=\hat{A}/\sigma,
\end{equation}
with window constant $\beta_{\mathrm{win}}$ (Hann: $\approx 0.5$) and peak SNR $\rho$; for a zero mode, $\mathrm{SE}[\hat{\gamma}]=\mathrm{SE}[\hat{\omega}_1]$, otherwise the two lowest-peak errors add in quadrature. A Cramér–Rao proxy,
\begin{equation}
\mathrm{Var}(\hat E)\ \gtrsim\ \frac{\sigma^2}{\sum_k a_{\mathrm{eff}}^2\, t_k^{\,2}},
\end{equation}
emphasizes the value of longer total span $T\Delta t$.

\paragraph{}
Resource–accuracy tradeoffs follow from the target gap $\gamma_\star$, requiring $T\Delta t \gtrsim 2\pi/\gamma_\star$. Decreasing $\Delta t$ enlarges the Nyquist band but increases the number of controlled evolutions for fixed span, with total sampling cost $T\times M$. Trotterized simulation of $e^{-i\mathcal{H}t}$ with step $\delta t$ incurs
\begin{equation}
\|e^{-i\mathcal{H}t}-\tilde U(t)\|=O\!\big(t^2\max_{\ell\neq \ell'}\|[H_\ell,H_{\ell'}]\|\big),
\end{equation}
while a symmetric second-order formula reduces this to $O(t^3)$. Short-time calibration points verify linear phase growth, and slow drifts are removed by a global linear fit prior to spectral estimation. The spectral entropy $S_{\mathrm{spec}}=-\sum_i p_i\log p_i$ computed from line strengths $p_i=\hat{A}_i/\sum_j \hat{A}_j$ summarizes mode complexity and is reported alongside $\hat\gamma$.

\paragraph{}
Peak identification from $|\tilde C(\omega)|$ follows a robust spectral-gap picking procedure. 
After applying a Hann window and detrending to suppress DC leakage, the analysis is restricted 
to a fixed search band $[w_{\mathrm{lo}},w_{\mathrm{hi}}]=[0,0.8]$ within the Nyquist interval. 
Within this band, candidate peaks are local maxima that exceed a robust threshold 
$T=\mathrm{median}(S)+1.4826\,k_{\sigma}\,\mathrm{MAD}(S)$, 
which suppresses noise-dominated fluctuations. 
Each candidate is refined by parabolic interpolation for sub-bin accuracy, and optional harmonic guards 
exclude integer multiples of an estimated fundamental frequency $\omega_{\mathrm{est}}$ within tolerance $\delta_{\mathrm{harm}}$. 
A selection policy then determines the representative spectral line: \emph{nearest\_to\_estimate} chooses the candidate closest to $\omega_{\mathrm{est}}$ (default), 
while \emph{min\_significant} or \emph{lowest\_nonzero} favor the smallest nonzero significant peak. 
If no valid candidate remains, the strongest in-band maximum serves as fallback. 
This combination of band restriction, DC/harmonic guards, and median–MAD thresholding yields 
a stable and noise-resilient estimate of the fundamental spectral gap.

\paragraph{}
An end-to-end procedure thus proceeds as follows: acquire complex correlators on a primary grid (and optionally a secondary $\Delta t'$), form windowed spectra, detect and refine peaks via \eqref{eq:quadinterp}, apply zero-mode and alias guards, cross-check with Prony/ESPRIT, aggregate to $\hat\gamma$ with a confidence interval, and finally report $\hat{\xi}=1/\hat{\gamma}$ together with $\{p_i\}$ and $S_{\mathrm{spec}}$ and calibration diagnostics. For systems of $\sim$10–15 qubits, grouped Pauli–projector synthesis keeps each long-time controlled evolution at practical depth while preserving near-zero multiplicities and the first spectral gap with sub-percent bias under noiseless calibration.

\subsection{Proof of the Spectral Bound Between Energy Gap and Persistence}
\label{sec:proof_energy_persistence}

\paragraph{}
We derive a spectral bound that relates the Laplacian energy gap to the 
persistence of topological features observed along a filtration $\{K_t\}$.
The main theorem states that the persistence $(d-b)$ of a homological feature born at $t=b$ and disappearing at $t=d$ satisfies
\begin{equation}
    \tilde{L}(d-b) + (p+1)d_{p,\max} \ge \lambda_{\beta_{p+1}}(\Delta_p(K_b)),
    \label{eq:main_persistence_bound}
\end{equation}
where $\tilde{L}$ denotes the effective Lipschitz constant of spectral variation,  
and $d_{p,\max}$ is the maximal number of adjacent $(p-1)$–simplices for any $p$–simplex.  
A larger local spectral gap shortens the persistence, while a vanishing gap leads to long-lived harmonic modes.

\paragraph{}
To begin, recall that for any Hermitian matrix $H$, the Courant–Fischer theorem expresses the $k$-th eigenvalue as
\begin{equation}
    \lambda_k(H)
    = \min_{\dim U=k}\max_{x\in U\setminus\{0\}}
      \frac{x^{\mathsf{T}}Hx}{x^{\mathsf{T}}x}
    = \max_{\dim V=n-k+1}
      \min_{x\in V\setminus\{0\}}
      \frac{x^{\mathsf{T}}Hx}{x^{\mathsf{T}}x}.
    \label{eq:minmax_theorem}
\end{equation}
If $H = U^{\dagger}\Lambda U$ with eigenvalues $\lambda_1\le\cdots\le\lambda_n$ and $Ux=\alpha$, then
\begin{equation}
    R_H(x)
    = \frac{\sum_i \lambda_i \alpha_i^2}{\sum_i \alpha_i^2}
    = \sum_i \lambda_i \tilde{\alpha}_i^2,
    \qquad
    \tilde{\alpha}_i=\frac{\alpha_i}{\|x\|},
\end{equation}
so that $\max_{x\in\mathrm{span}\{e_1,\dots,e_k\}} R_H(x)=\lambda_k$.

\paragraph{}
Now consider a block Hermitian matrix
\[
M=\begin{pmatrix}A&B\\B^{\dagger}&C\end{pmatrix},
\qquad
S=A-B C^{-1}B^{\dagger},
\]
with $A$ Hermitian and $C>0$.  
At the minimizing point $y^*(x)=-C^{-1}B^{\dagger}x$, the Rayleigh quotient satisfies
\begin{equation}
    R_M(x,y^*(x))
    = R_S(x)
      \Bigl(1+\frac{\|y^*(x)\|^2}{\|x\|^2}\Bigr)^{-1}
    \le R_S(x),
\end{equation}
and from the min–max relation one obtains
\begin{align}
    \lambda_k(M)
    &\le \lambda_k(S),
    \qquad
    \lambda_{k+d}(M)\ge \lambda_k(S),
    \label{eq:schur_interlace}
\end{align}
where $d=\dim(C)$.  
Applying this result to $\Delta_p(K_t)$ and its persistent form $\Delta_p^{(s,t)}$ gives
\begin{equation}
    \lambda_k(\Delta_p(K_t))
    \le \lambda_k(\Delta_p^{(s,t)})
    \le \lambda_{k+d}(\Delta_p(K_t)).
    \label{eq:persistent_minmax}
\end{equation}

\paragraph{}
Next, note that the $p$–Laplacian decomposes into upward and downward components,
\[
\Delta_p=\Delta_{p,\mathrm{up}}+\Delta_{p,\mathrm{down}},
\qquad
\Delta_{p,\mathrm{up}}=\partial_{p+1}\partial_{p+1}^{\dagger},\ 
\Delta_{p,\mathrm{down}}=\partial_p^{\dagger}\partial_p.
\]
The upward part represents $(p+1)$–dimensional fillings, while the downward part encodes adjacency among $(p-1)$–simplices.  
The Schur relation applied to $\Delta_{p,\mathrm{up}}$ yields
\begin{equation}
    \lambda_k(\Delta_{p,\mathrm{up}}(K_t))
    \le \lambda_k(\Delta_{p,\mathrm{up}}^{(s,t)})
    \le \lambda_{k+d}(\Delta_{p,\mathrm{up}}(K_t)).
\end{equation}
Because $\Delta_p=\Delta_{p,\mathrm{up}}+\Delta_{p,\mathrm{down}}$,  
the following bounds hold:
\begin{align}
    \lambda_k(\Delta_p(K_t))
    &\ge \lambda_k(\Delta_{p,\mathrm{up}}(K_t))
         +\lambda_{\min}(\Delta_{p,\mathrm{down}}(K_t)),\\
    \lambda_k(\Delta_p(K_t))
    &\le \lambda_k(\Delta_{p,\mathrm{up}}(K_t))
         +\lambda_{\max}(\Delta_{p,\mathrm{down}}(K_t)).
\end{align}
Combining these gives a two-sided interlacing for the persistent Laplacian:
\begin{align}
    \lambda_k(\Delta_p(K_t))
    &- \lambda_{\max}\!\bigl(\Delta_{p,\mathrm{down}}(K_t)\bigr)
    + \lambda_{\min}\!\bigl(\Delta_{p,\mathrm{down}}(K_s)\bigr) \notag\\
    &\le\ \lambda_k\!\bigl(\Delta_p^{(s,t)}\bigr)\ \le\
      \lambda_{k+d}\!\bigl(\Delta_p(K_t)\bigr)
      - \lambda_{\max}\!\bigl(\Delta_{p,\mathrm{down}}(K_s)\bigr)
      + \lambda_{\min}\!\bigl(\Delta_{p,\mathrm{down}}(K_t)\bigr).
    \label{eq:up_down_symmetric}
\end{align}
This shows that the spectrum of the persistent Laplacian is bounded between those of the ordinary Laplacian, with corrections governed by the spectral range of $\Delta_{p,\mathrm{down}}$.

\paragraph{}
To relate these bounds to geometric variation, consider the inequality between two Hermitian matrices $A$ and $B$:
\begin{equation}
    x^{\mathsf{T}} B x = x^{\mathsf{T}} A x + x^{\mathsf{T}}(B-A)x
    \le x^{\mathsf{T}} A x + \|B-A\|\|x\|^2.
\end{equation}
Taking the infimum over all normalized $x$ yields
\begin{equation}
    \lambda_k(B) \le \lambda_k(A) + \|B-A\|.
\end{equation}
If the Laplacian varies Lipschitz-continuously with respect to $t$,  
\begin{equation}
    \|\Delta_p(K_t)-\Delta_p(K_s)\|\le L|t-s|,
\end{equation}
then the corresponding eigenvalues satisfy
\begin{equation}
    \lambda_k(\Delta_p(K_t))
    \ge \lambda_k(\Delta_p(K_s)) - L|t-s|.
    \label{eq:lipschitz_lambda}
\end{equation}
The downward Laplacian’s spectral width provides a combinatorial correction term.
Since
\begin{equation}
    \mathrm{width}(\Delta_{p,\mathrm{down}}(K_s))
    = \lambda_{\max}-\lambda_{\min}
    \le (p+1)d_{p,\max},
\end{equation}
where $d_p$ denotes the maximum number of $p$-dimensional simplices that are adjacent to any given $(p−1)$-dimensional simplex $\sigma$.
We obtain
\begin{equation}
    \lambda_{\max}(\Delta_{p,\mathrm{down}}(K_t))
     - \lambda_{\min}(\Delta_{p,\mathrm{down}}(K_s))
    \le (p+1)d_{p,\max} + L'(t-s),
\end{equation}
and thus the persistent Laplacian obeys
\begin{equation}
    \lambda_k(\Delta_p^{(s,t)})
    \ge \lambda_k(\Delta_p(K_s))
     - \tilde{L}(t-s)
     - (p+1)d_{p,\max},
     \qquad \tilde{L}=L+L'.
    \label{eq:combined_ineq}
\end{equation}

\paragraph{}
The behavior of individual eigenmodes can be analyzed perturbatively.  
For $H(\epsilon)=H_0+\epsilon V$ with $\epsilon\ll1$,  
\begin{equation}
    E_m=E_m^{(0)}+\epsilon\langle\phi_m,V\phi_m\rangle
     +\epsilon^2\sum_{n\neq m}
      \frac{|\langle\phi_m,V\phi_n\rangle|^2}{E_m^{(0)}-E_n^{(0)}}+\cdots.
\end{equation}
If $E_0^{(0)}=0$ is isolated, it remains zero until the degeneracy is lifted,  
as shown in Kato’s \emph{Perturbation Theory for Linear Operators}~\cite{Kato1976}.  
Hence, a homological feature persists while its Laplacian mode stays within the zero eigenspace:
\begin{equation}
    \lambda_{\beta_{p+1}}(\Delta_p^{(b,t)})\approx0\ (t<d),\qquad
    \lambda_{\beta_{p+1}}(\Delta_p^{(b,d)})>0.
    \label{eq:zerocrossing}
\end{equation}

\paragraph{}
The stability of the zero eigenspace can be formalized using the Riesz projector
\begin{equation}
    \Pi_0(t)=\frac{1}{2\pi i}\oint_{\Gamma_t}(zI-\Delta_p(K_t))^{-1}\,dz,
\end{equation}
where $\Gamma_t$ encloses the isolated eigenvalue $0$.  
If $E=\Delta_p(K_t)-\Delta_p(K_s)$, then
\begin{equation}
    \|\Pi_0(t)-\Pi_0(s)\|
    \le \frac{\|E\|}{\gamma_s}\le\frac{L}{\gamma_s}|t-s|.
\end{equation}
As long as $L|t-s|<\gamma_s=\lambda_1(\Delta_p(K_s))$, the rank of the projector—and therefore the Betti number—remains invariant:
\begin{equation}
    \mathrm{rank}\,\Pi_0(t)=\mathrm{rank}\,\Pi_0(s).
\end{equation}
A death of homology thus requires that the spectral gap $\gamma$ closes.

\paragraph{}
When a homological feature is born at $t=b$, the smallest nonzero eigenvalue $\lambda_{\beta_{p+1}}(\Delta_p(K_b))$ defines a local energy gap.  
As the filtration proceeds, this feature persists as long as $\lambda_{\beta_{p+1}}(\Delta_p^{(b,t)})\simeq0$.  
Its eventual disappearance at $t=d$ occurs when this zero mode lifts to a finite value.  
Using the spectral inequality~\eqref{eq:combined_ineq}, this transition can only happen if
\begin{equation}
    \lambda_{\beta_{p+1}}(\Delta_p(K_b)) - \tilde{L}(d-b) - (p+1)d_{p,\max} \le 0,
\end{equation}
which rearranges to the quantitative persistence condition
\begin{equation}
    \tilde{L}(d-b) + (p+1)d_{p,\max}
    \ge \lambda_{\beta_{p+1}}(\Delta_p(K_b)).
    \label{eq:final_theorem}
\end{equation}
In words, the death of a homological feature requires that the cumulative spectral variation across $(b,d)$ surpasses the initial Laplacian energy gap at $b$.  
If this threshold is not reached, the zero mode—and hence the topological feature—remains stable.

\paragraph{}
Finally, because the Laplacian encodes gradients, curls, and divergences of the field, its eigenvalues describe the curvature of the local energy landscape.  
A large spectral gap $\lambda_1$ corresponds to strong restoring forces that stabilize topology,  
while $\lambda_1\!\to\!0$ marks a topological phase transition where harmonic modes emerge.  
The ordinary Laplacian gap thus quantifies the robustness of homological structures:  
the closure of this gap signals the annihilation of a topological cycle.  
Equation~\eqref{eq:final_theorem} unifies energy-gap dynamics and persistent homology by showing that the local spectral curvature (or SUSY energy gap) provides a quantitative bound on the persistence of homological features,  
offering a spectral–geometric interpretation of topological stability consistent with the SUSY behavior observed in the Lorenz system.



\newcommand{\rhoblockB}[9]{%
  \begin{subfigure}[t]{\linewidth}
    \centering

    \begin{minipage}[b]{0.35\linewidth}\textbf{A}\par\centering\includegraphics[width=\linewidth]{#1}\end{minipage}\hspace{6em}
    \begin{minipage}[b]{0.32\linewidth}\textbf{B}\par\centering\includegraphics[width=\linewidth]{#2}\end{minipage}\par\medskip
    
    \begin{minipage}[b]
    {0.22\linewidth}\textbf{C}\par\centering\includegraphics[width=\linewidth]{#3}\end{minipage}\hspace{10em}
    \begin{minipage}[b]{0.22\linewidth}\textbf{D}\par\centering\includegraphics[width=\linewidth]{#4}\end{minipage}\par\medskip
    
    \begin{minipage}[b]{0.32\linewidth}\textbf{E}\par\centering\includegraphics[width=\linewidth]{#5}\end{minipage}\hspace{8em}
    \begin{minipage}[b]{0.32\linewidth}\textbf{F}\par\centering\includegraphics[width=\linewidth]{#6}\end{minipage}\par\medskip
    
    \begin{minipage}[b]{0.35\linewidth}\textbf{G}\par\centering\includegraphics[width=\linewidth]{#7}\end{minipage}\hspace{6em}
    \begin{minipage}[b]{0.35\linewidth}\textbf{H}\par\centering\includegraphics[width=\linewidth]{#8}\end{minipage}\par\medskip
    
    \vspace{0.3em}\textbf{$\rho=#9$}
    \caption*{} 
  \end{subfigure}\par\medskip
}

\begin{figure}[p]
  \centering
  \begin{adjustbox}{max width=\textwidth, max totalheight=.80\textheight, center}
    \rhoblockB
      {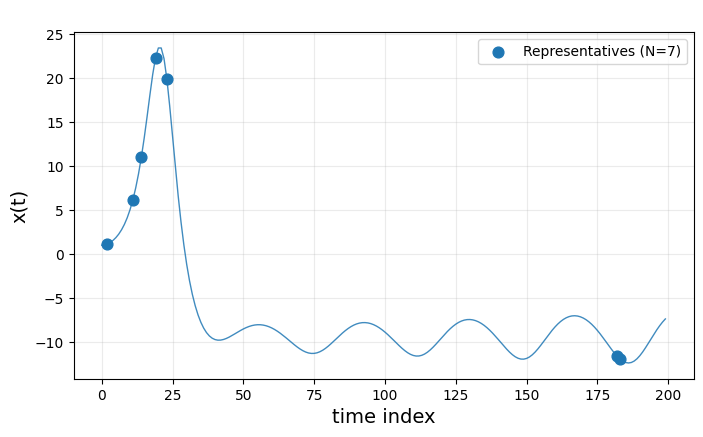}
      {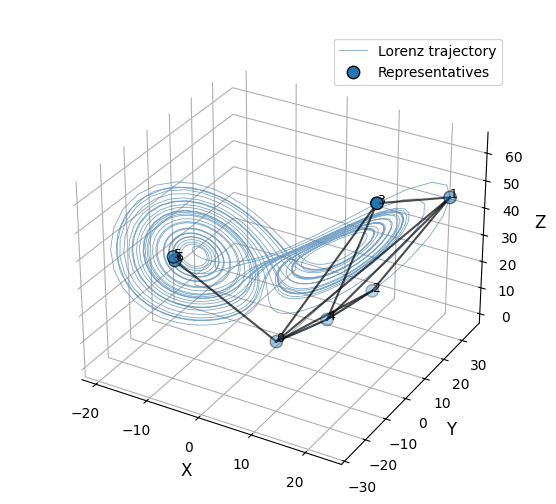}
      {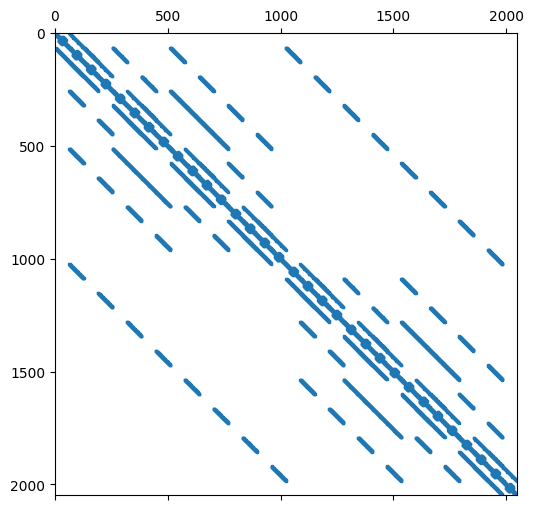}
      {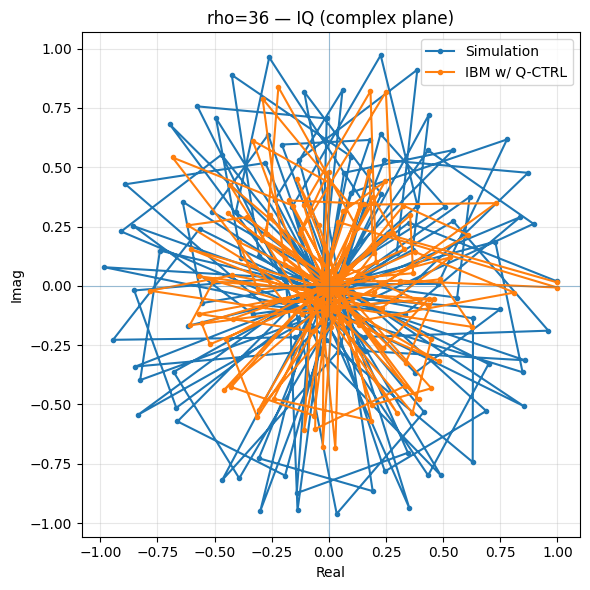}
      {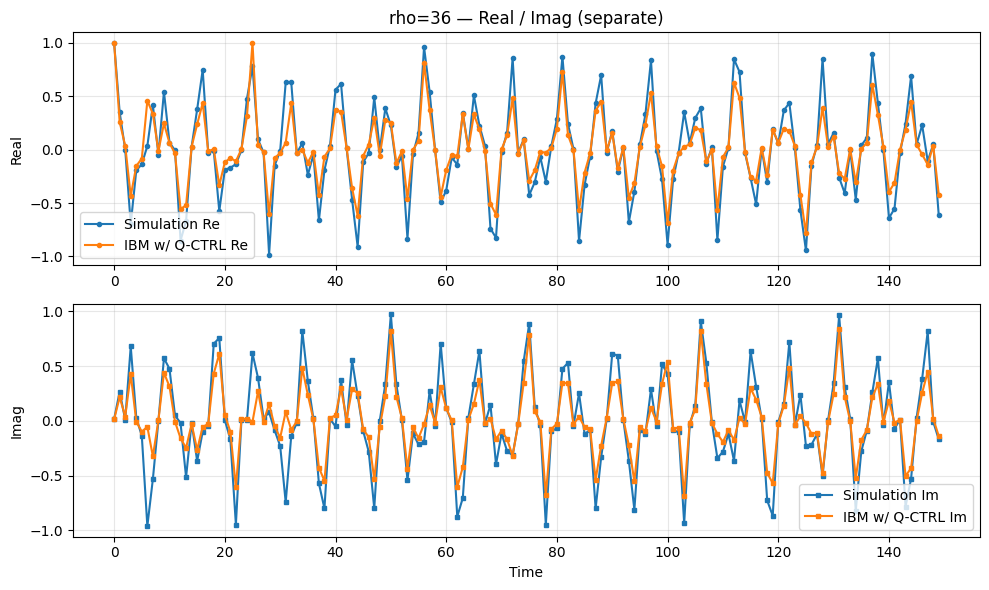}
      {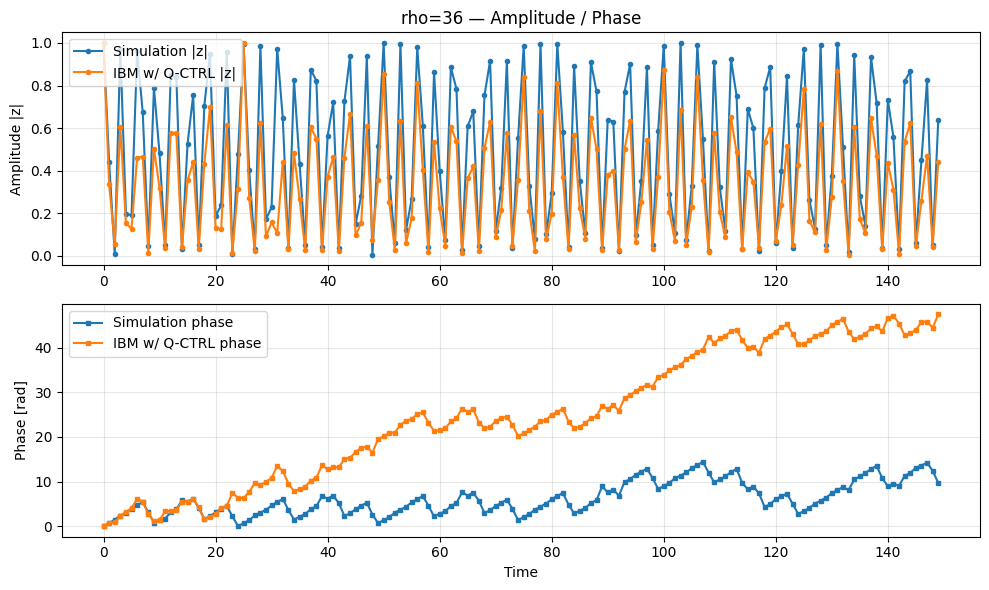}
      {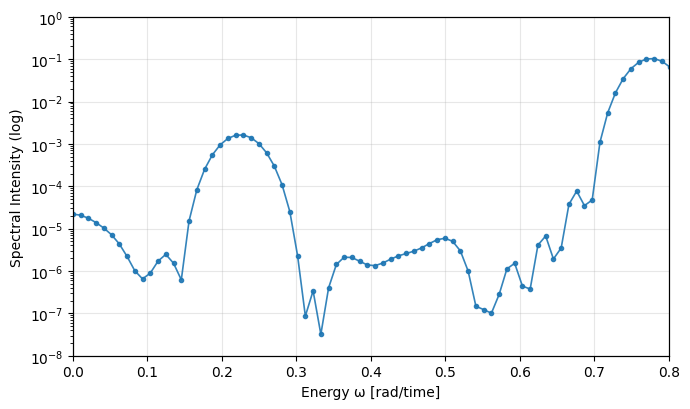}
      {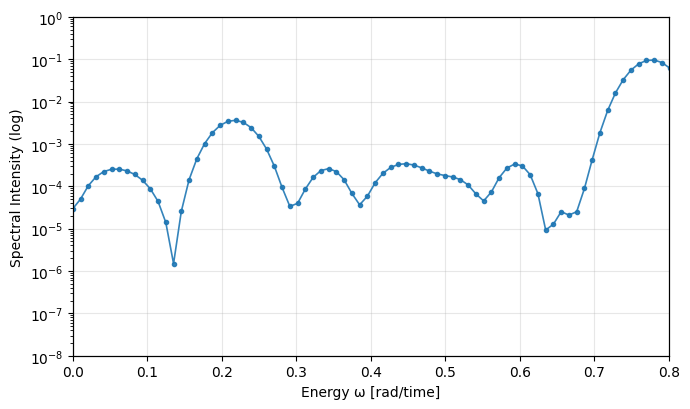}
      {36}
  \end{adjustbox}
  \caption[QPEfigs]{\textbf{Quantum and classical diagnostics for the Lorenz system at $\rho=36$.}
  (\textbf{A}) Time-series input and (\textbf{B}) 3D attractor embedding.
  (\textbf{C}) Hodge Laplacian matrix and (\textbf{D}) complex-plane trajectory of the QPE amplitude.
  (\textbf{E}) Real-imaginary waveform, (\textbf{F}) amplitude-phase trace, and (\textbf{G}, \textbf{H}) QPE spectra from simulation and IBM hardware. Broad, merged spectral peaks indicate weak topological separation in the early chaotic regime.}
  \label{fig:rho_panels}
\end{figure}

\begin{figure}[p]\ContinuedFloat
  \centering
  \rhoblockB
    {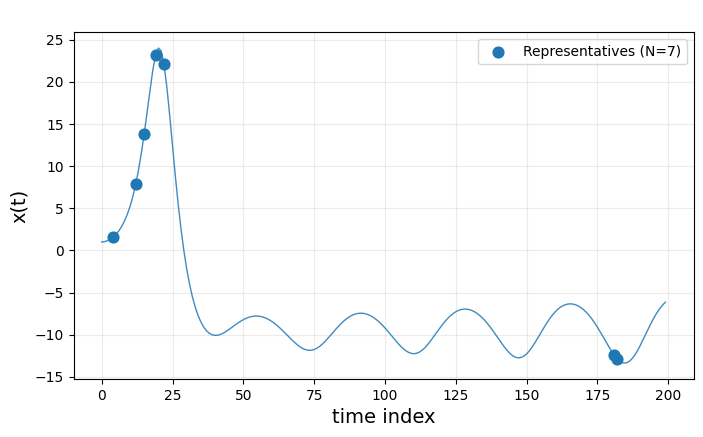}
    {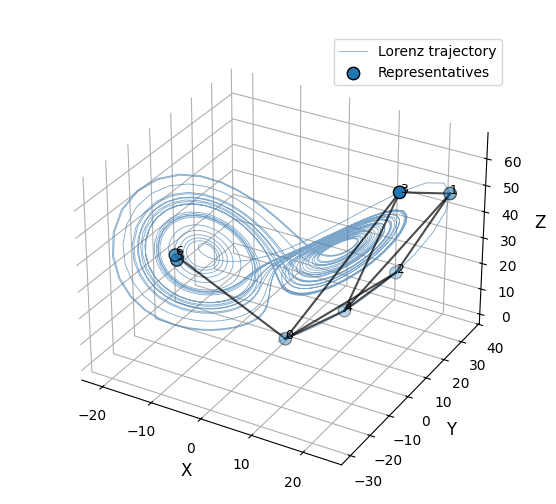}
    {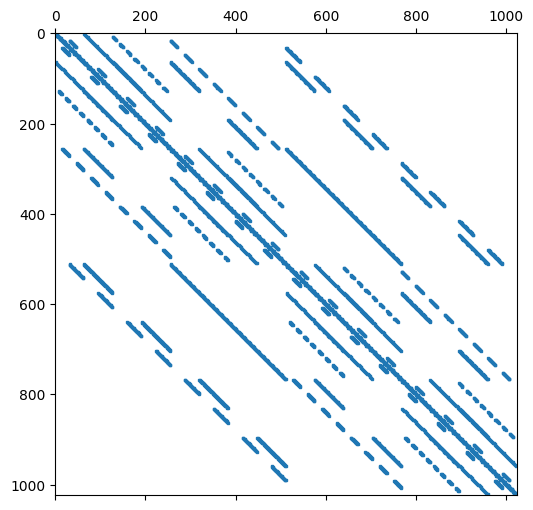}
    {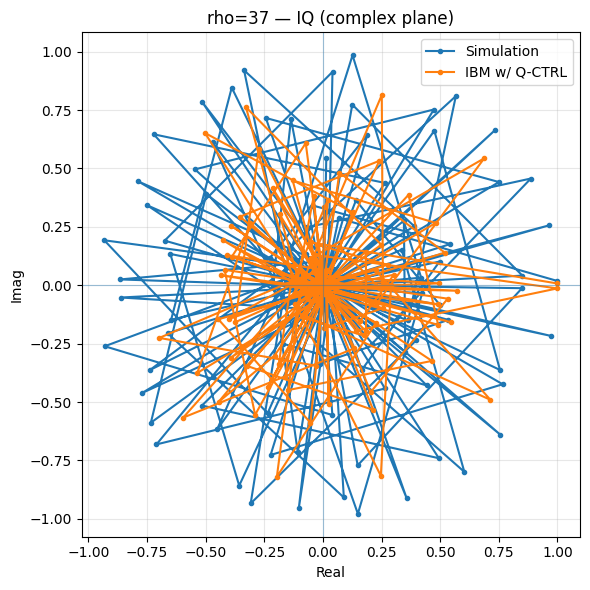}
    {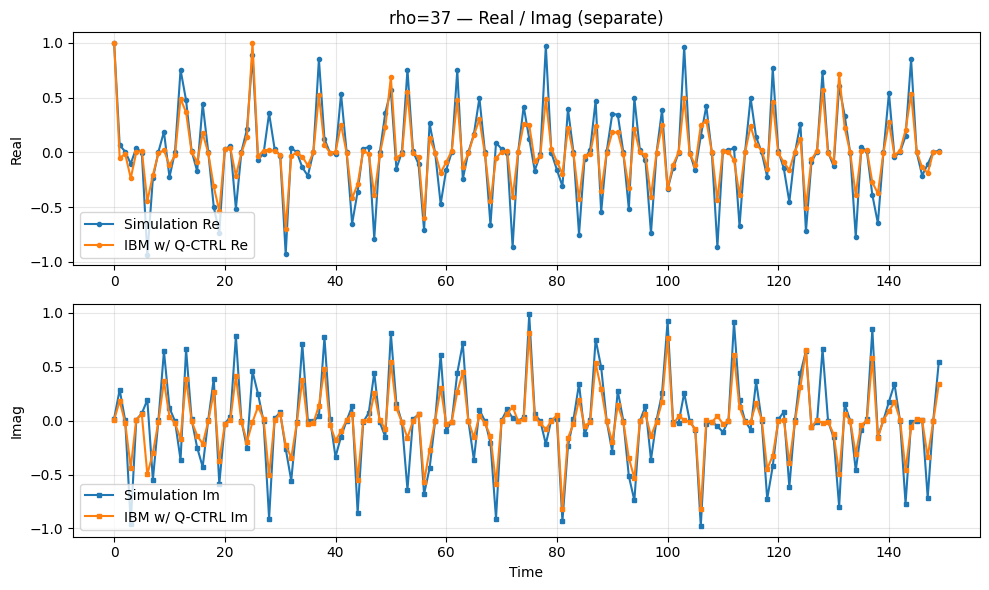}
    {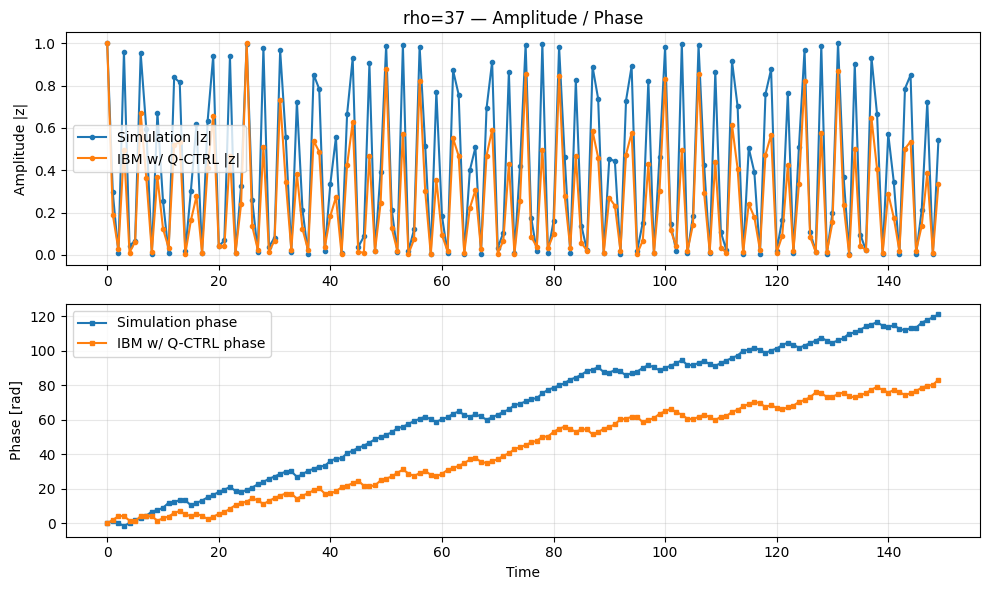}
    {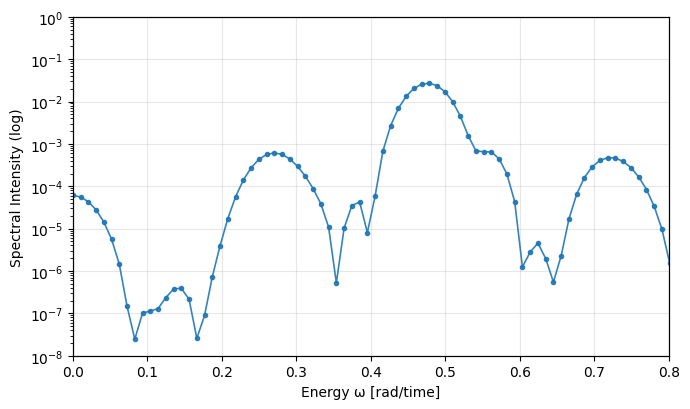}
    {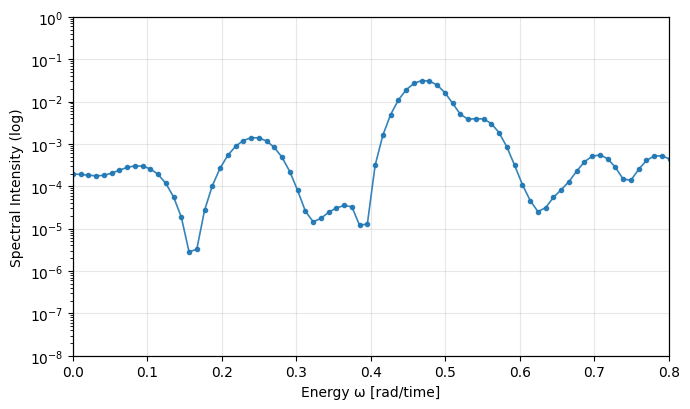}
    {37}
  \caption[]{\textbf{Quantum and classical diagnostics for the Lorenz system at $\rho=37$.}
  Slight narrowing of spectral lines and partial clustering of low-frequency modes mark the initial development of quasi-periodic structure. }
\end{figure}

\begin{figure}[p]\ContinuedFloat
  \centering
  \rhoblockB
    {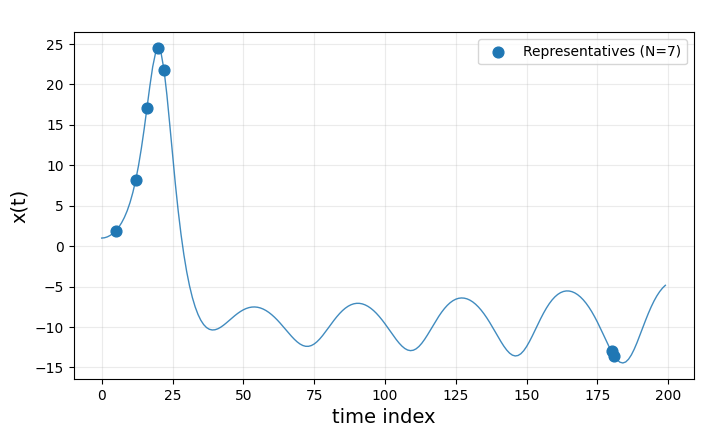}
    {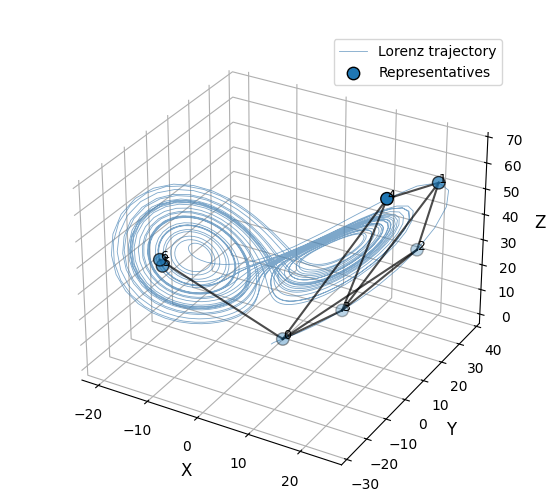}
    {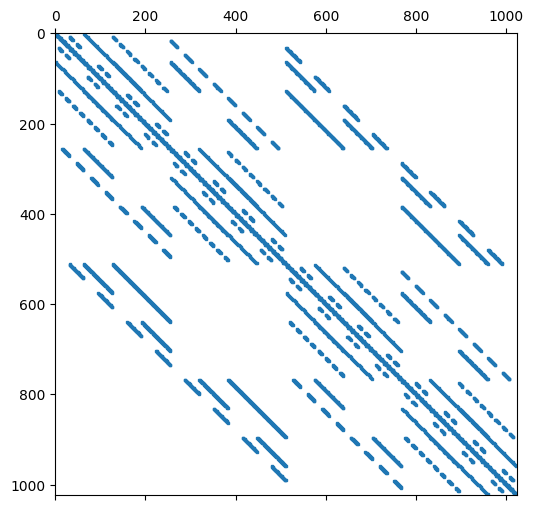}
    {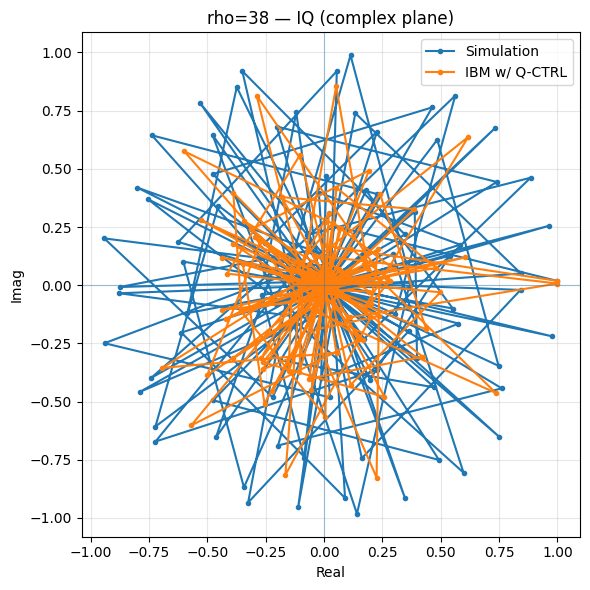}
    {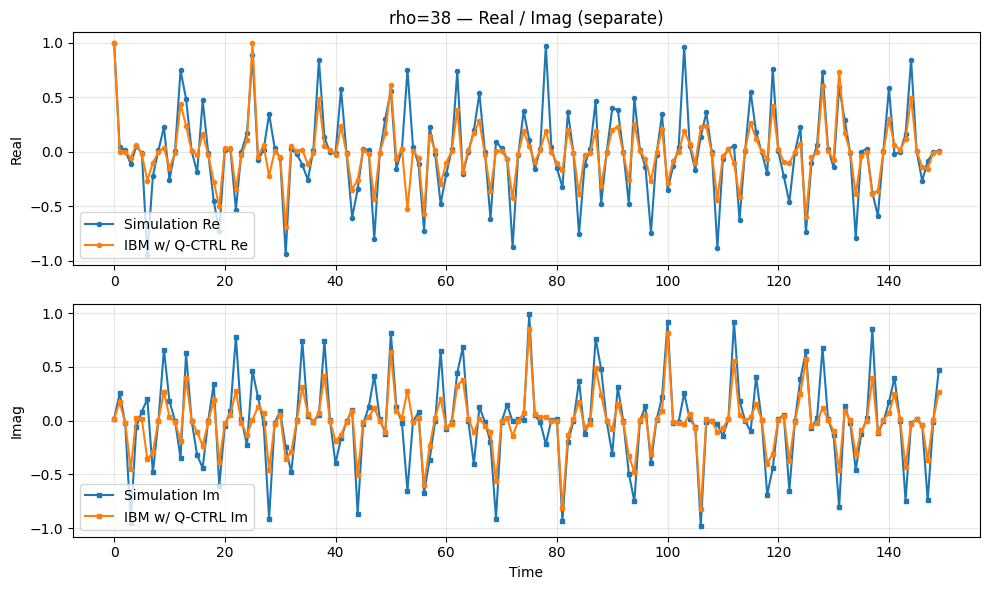}
    {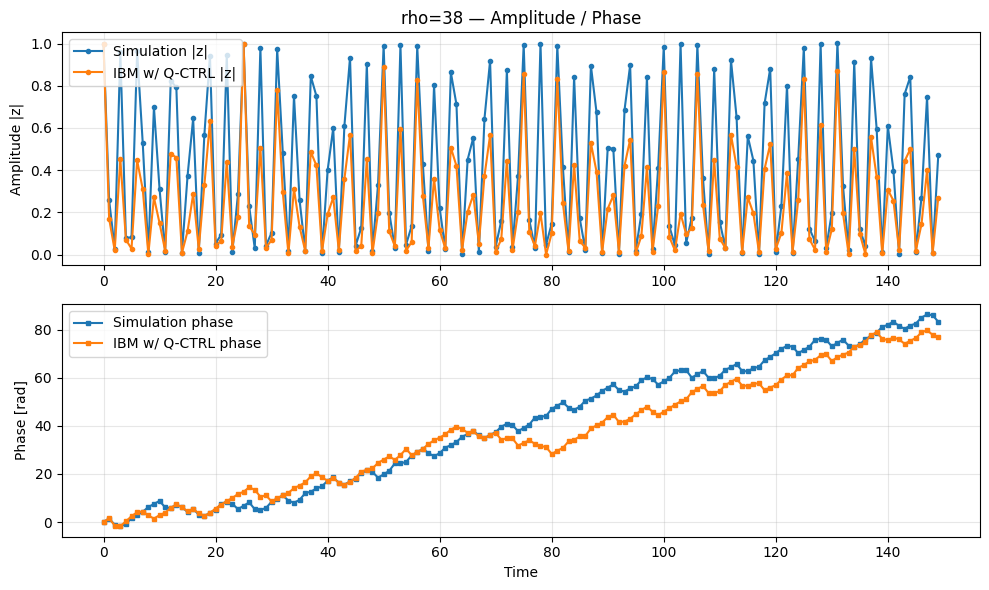}
    {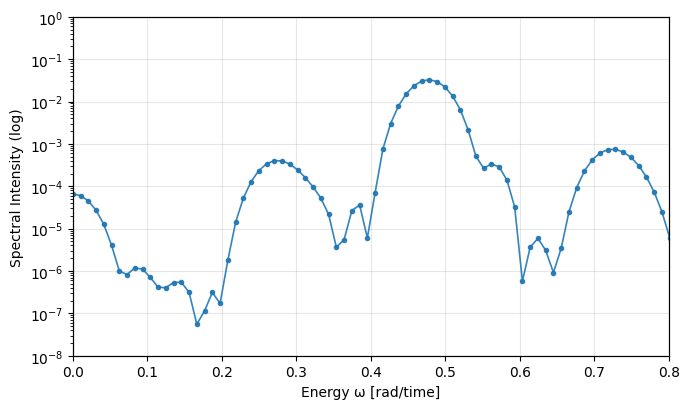}
    {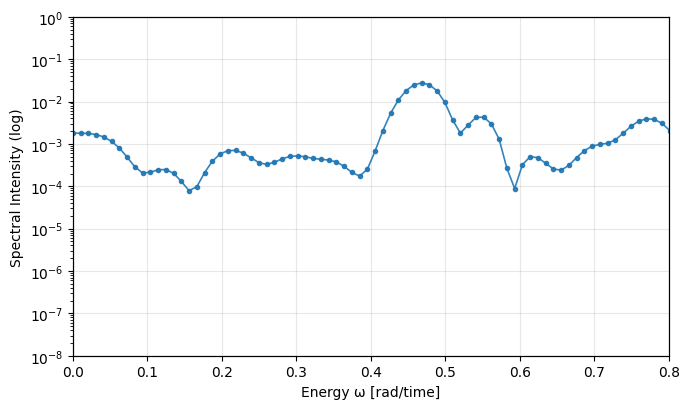}
    {38}
  \caption[]{\textbf{Quantum and classical diagnostics for the Lorenz system at $\rho=38$.}
  Spectral entropy begins to decline and distinct peaks emerge in both simulator and hardware spectra,
signaling partial organization of the attractor’s loop geometry. }
\end{figure}

\begin{figure}[p]\ContinuedFloat
  \centering
  \rhoblockB
    {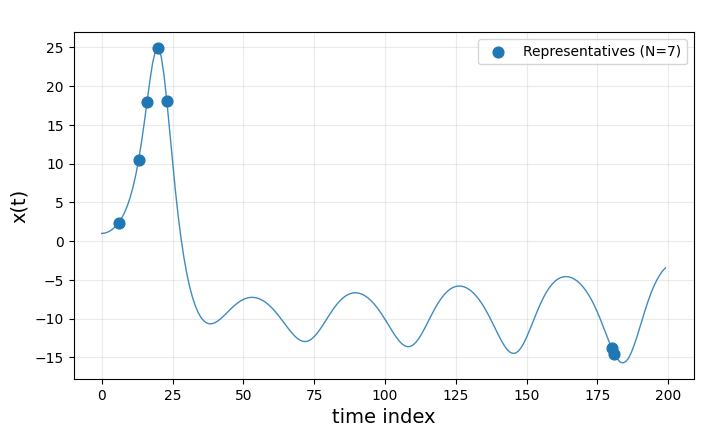}
    {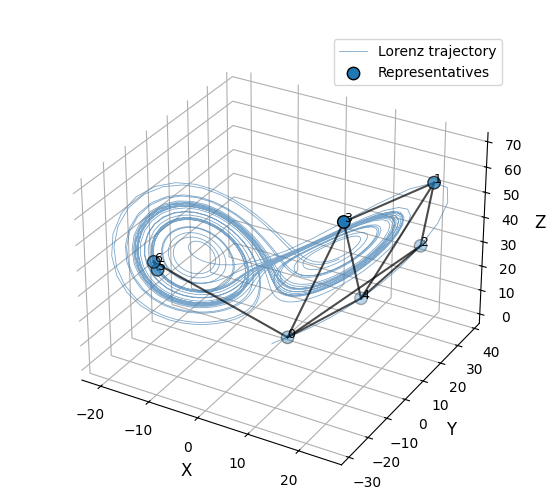}
    {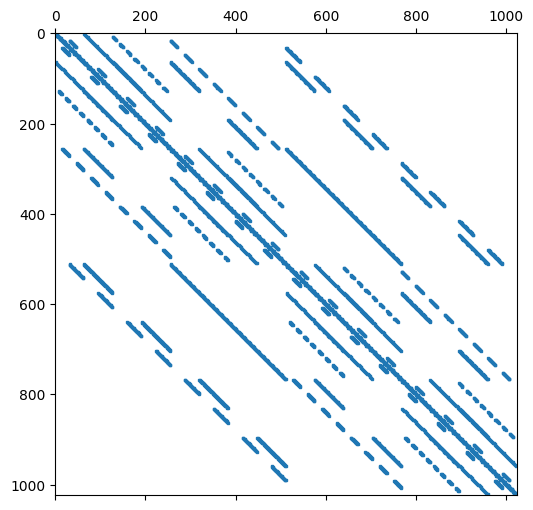}
    {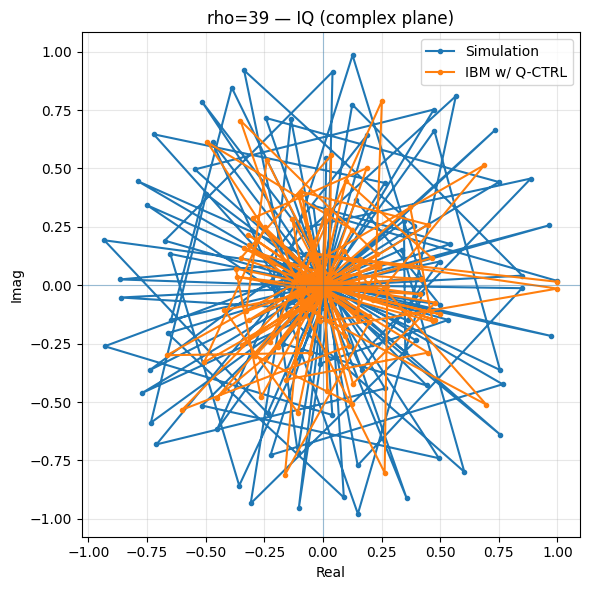}
    {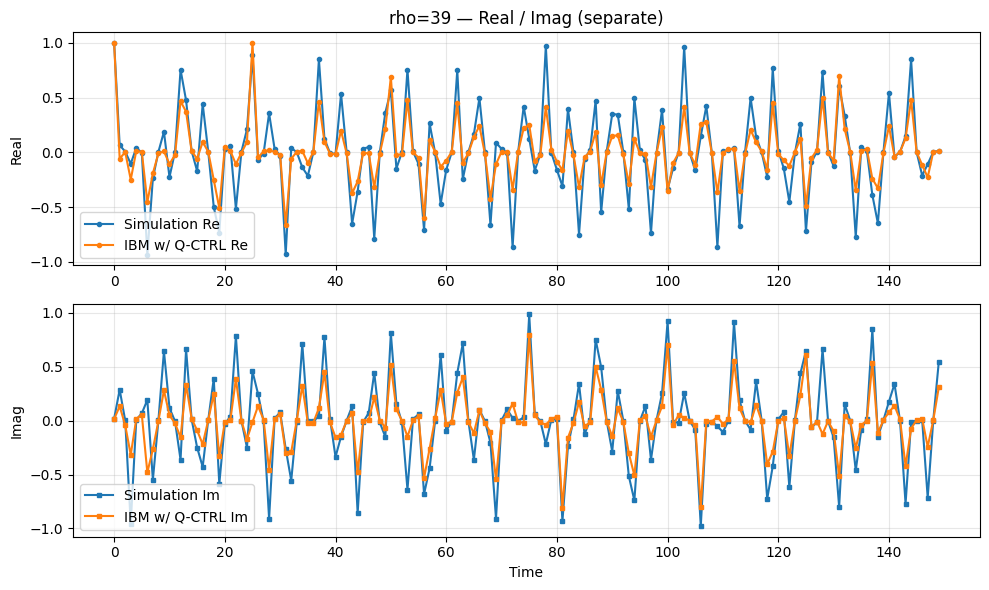}
    {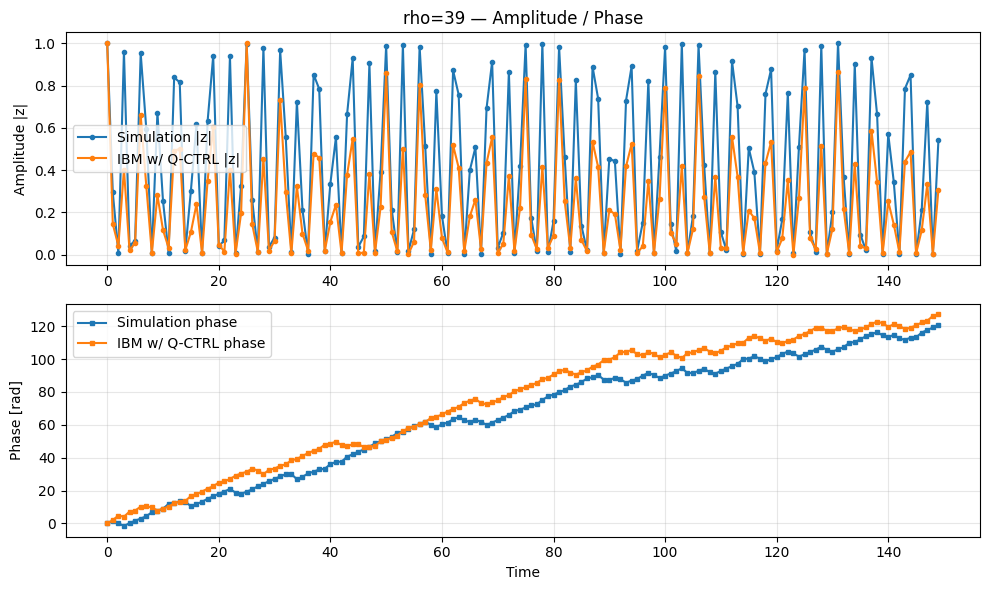}
    {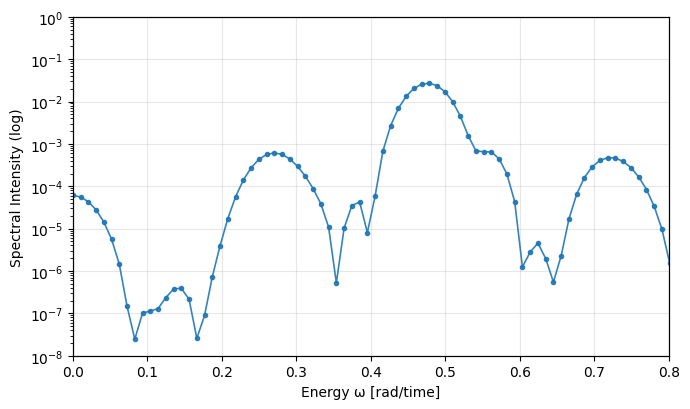}
    {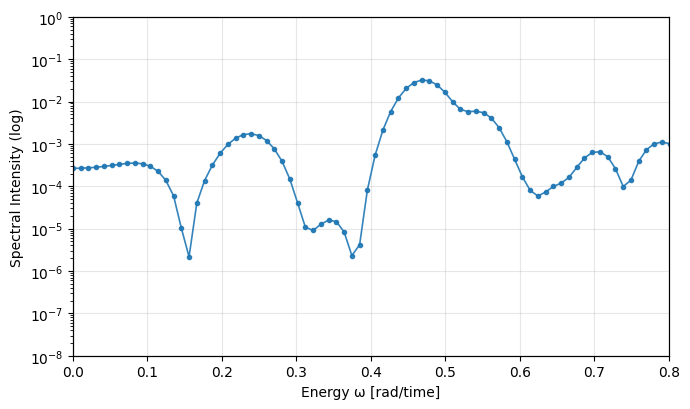}
    {39}
  \caption[]{\textbf{Quantum and classical diagnostics for the Lorenz system at $\rho=39$.}
  Discrete peaks become more pronounced and near-zero components start to separate from the continuum.
Topological coherence strengthens as the attractor approaches double-wing formation.} 
\end{figure}

\begin{figure}[p]\ContinuedFloat
  \centering
  \rhoblockB
    {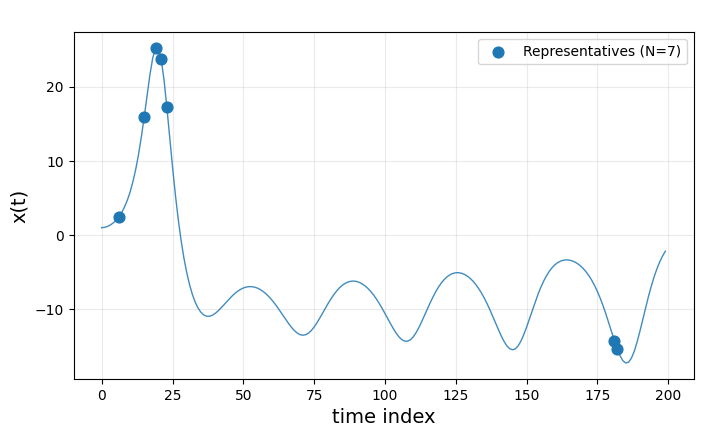}
    {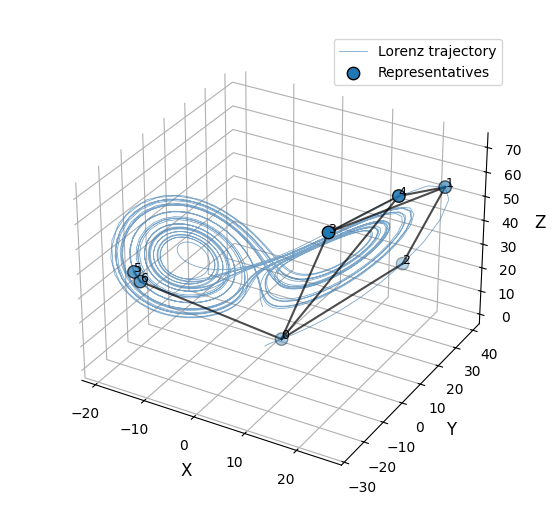}
    {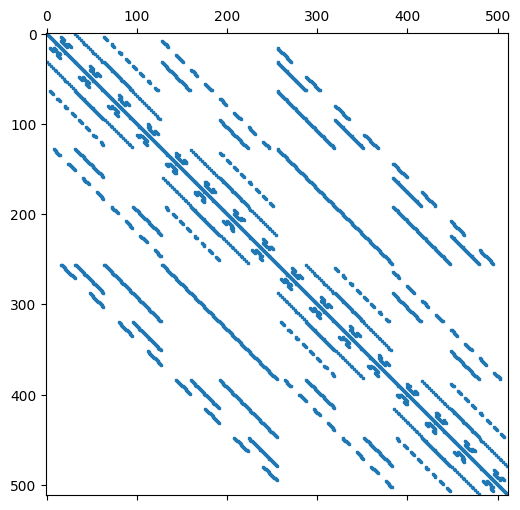}
    {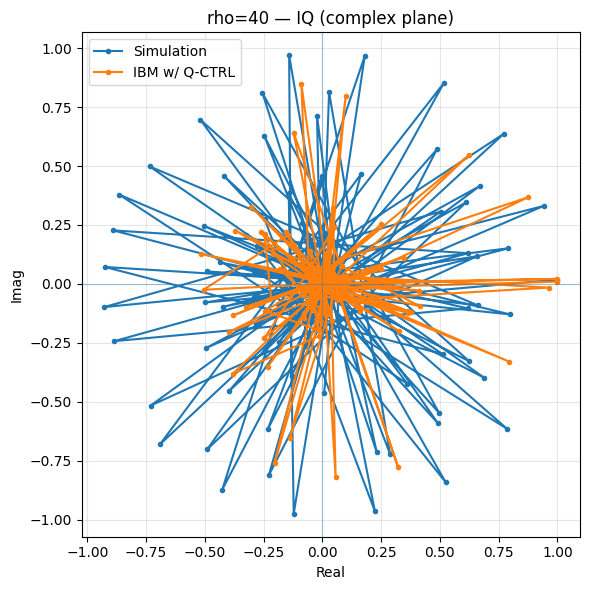}
    {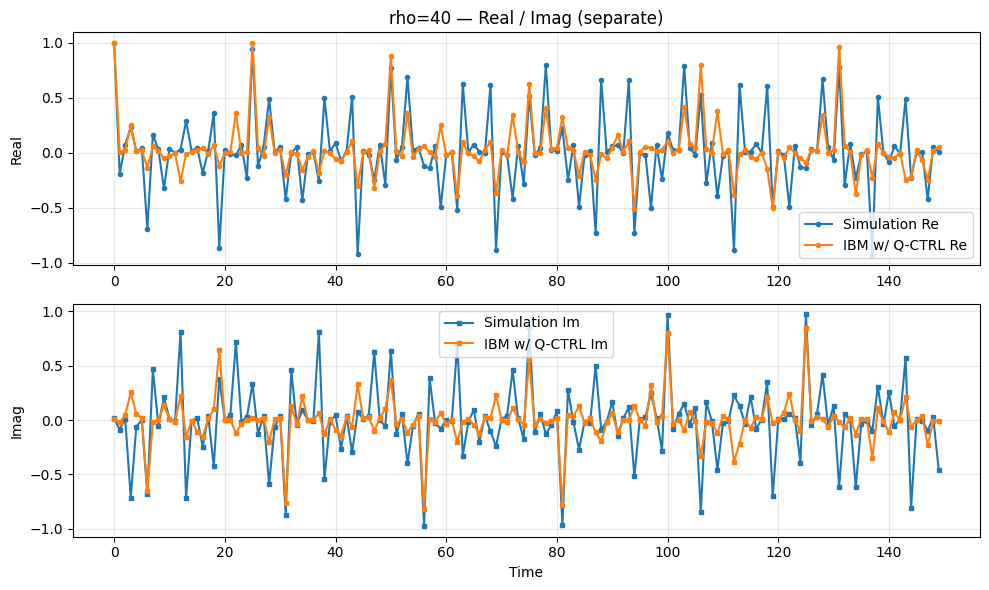}
    {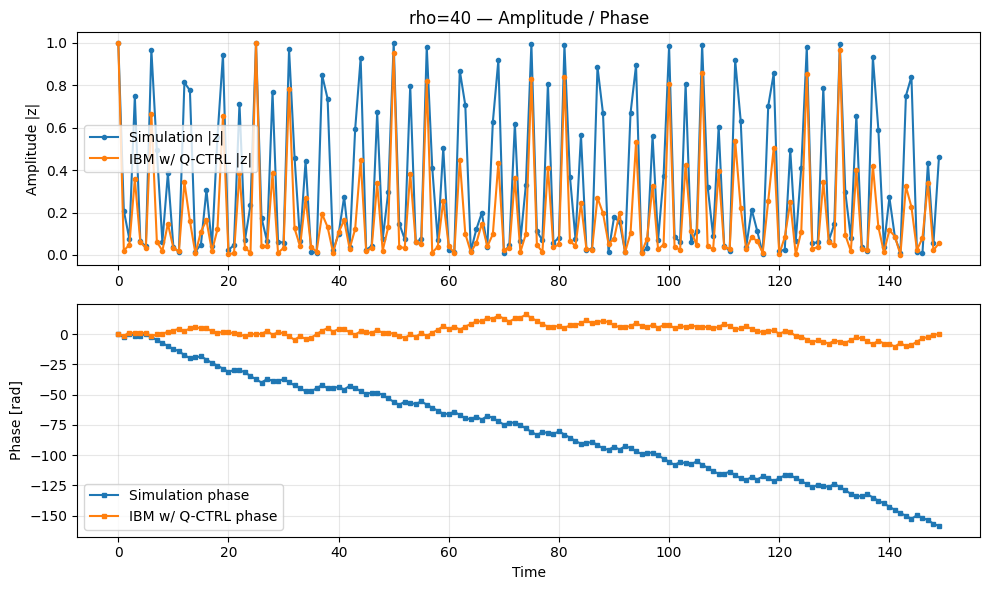}
    {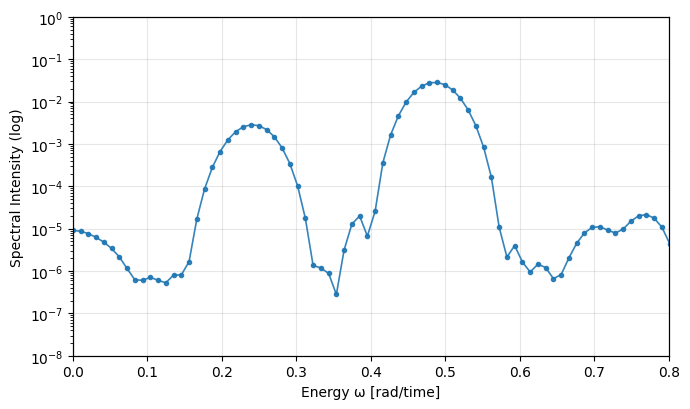}
    {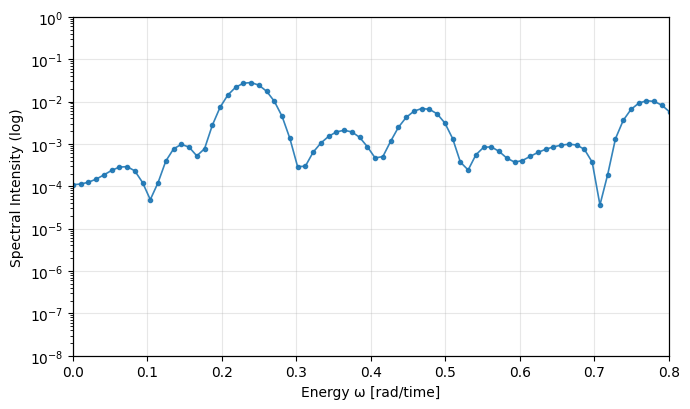}
    {40}
  \caption[]{\textbf{Quantum and classical diagnostics for the Lorenz system at $\rho=40$.}
  Both simulation and hardware show clear harmonic-mode isolation and maximal phase-space regularity.
  Near-zero clusters narrow sharply, marking the onset of topological stabilization.
  } 
\end{figure}

\begin{figure}[p]\ContinuedFloat
  \centering
  \rhoblockB
    {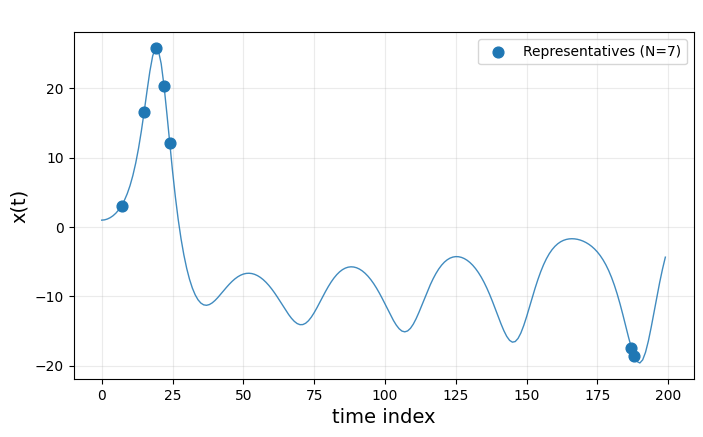}
    {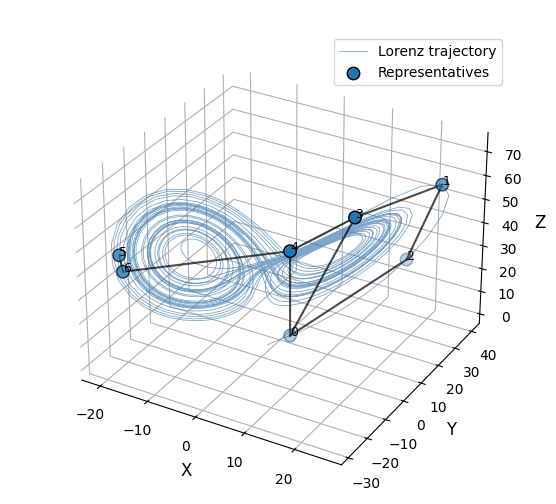}
    {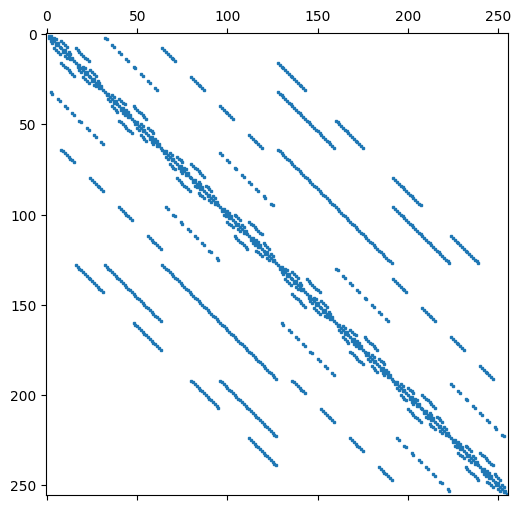}
    {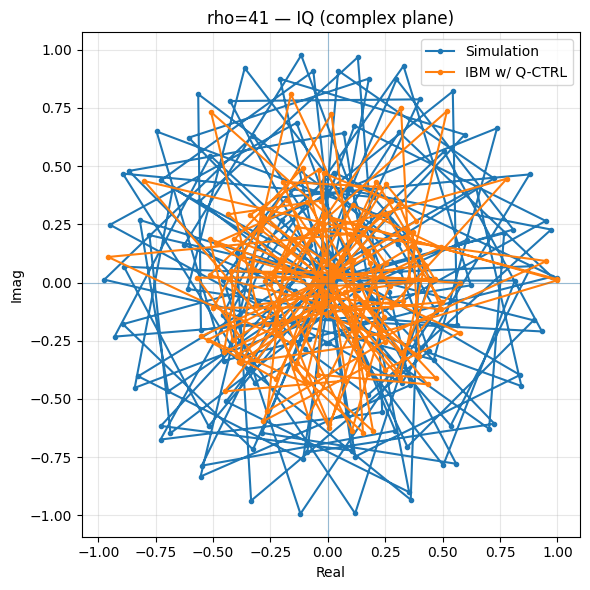}
    {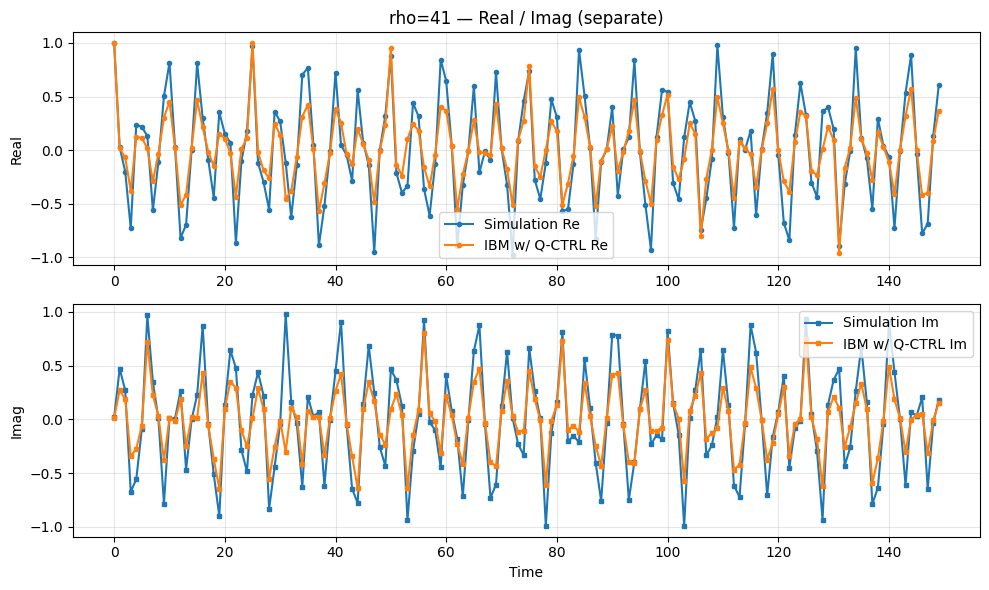}
    {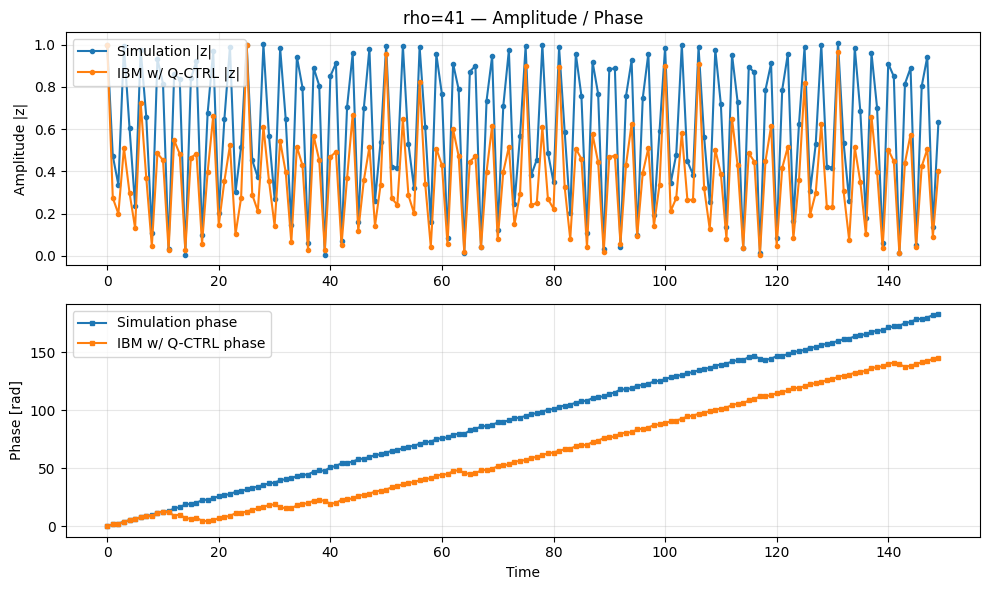}
    {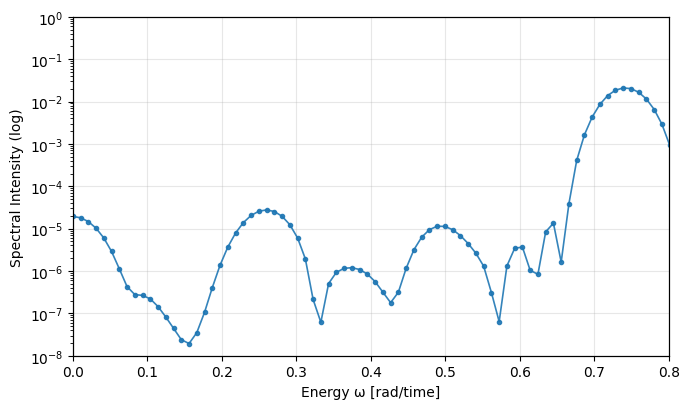}
    {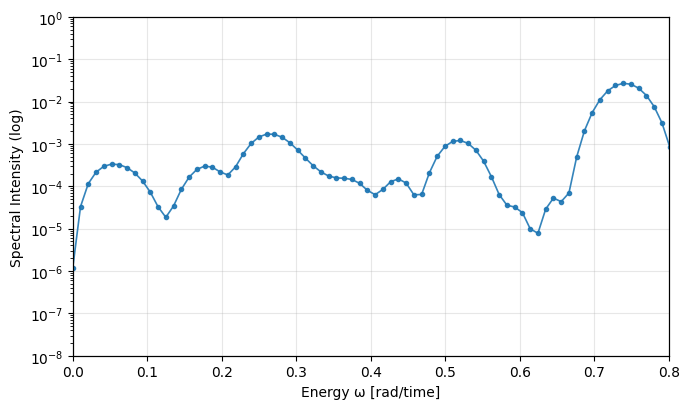}
    {41}
  \caption[]{\textbf{Quantum and classical diagnostics for the Lorenz system at $\rho=41$.}
  The Hodge–Laplacian spectrum exhibits its widest gap and cleanest harmonic–excited separation, coinciding with maximal $H_1$ persistence.
  Complex-plane trajectories are nearly circular, confirming spectral coherence.} 
\end{figure}

\begin{figure}[p]\ContinuedFloat
  \centering
  \rhoblockB
    {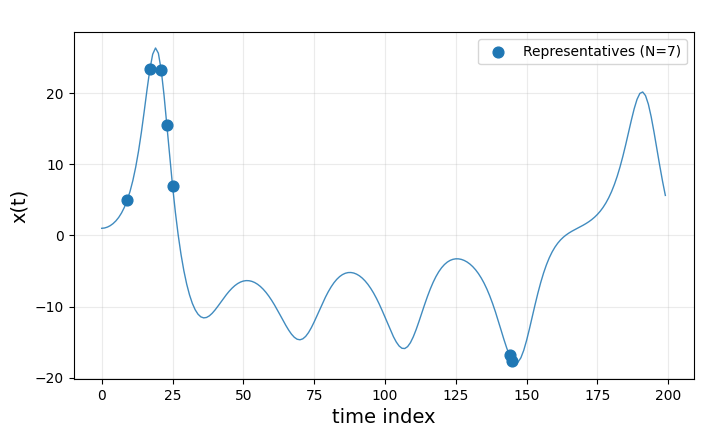}
    {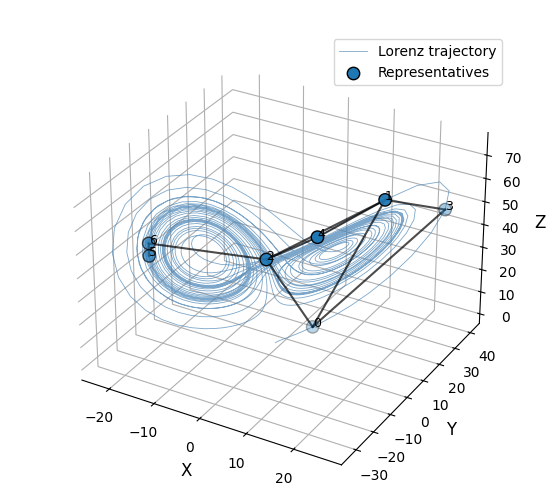}
    {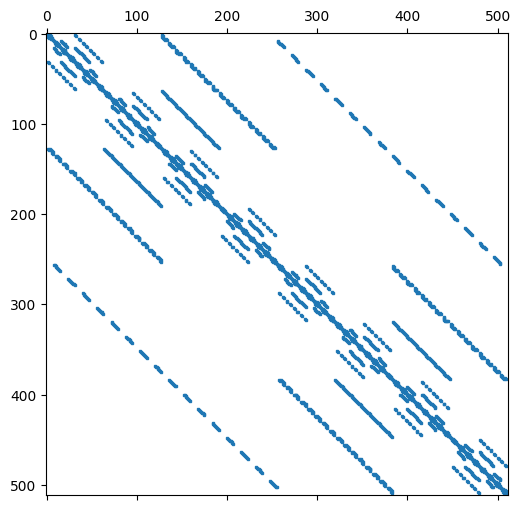}
    {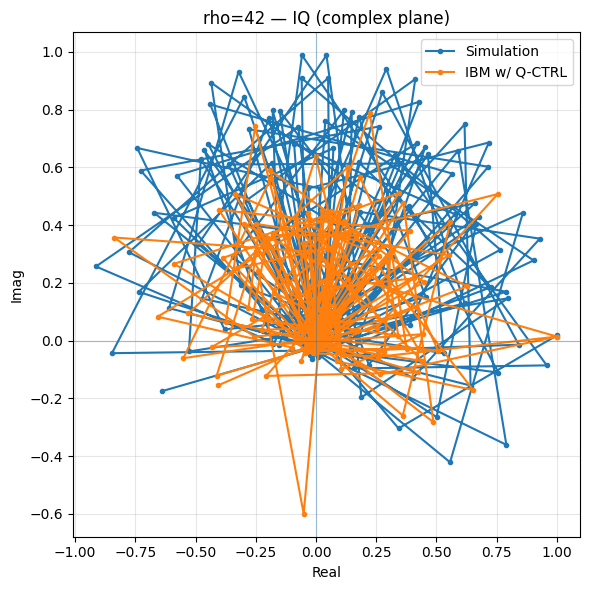}
    {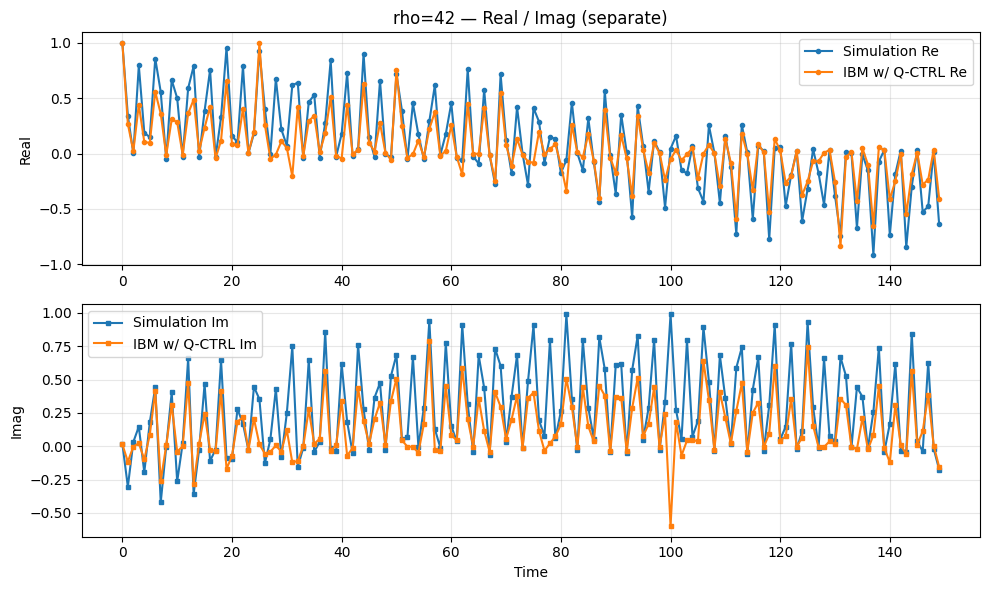}
    {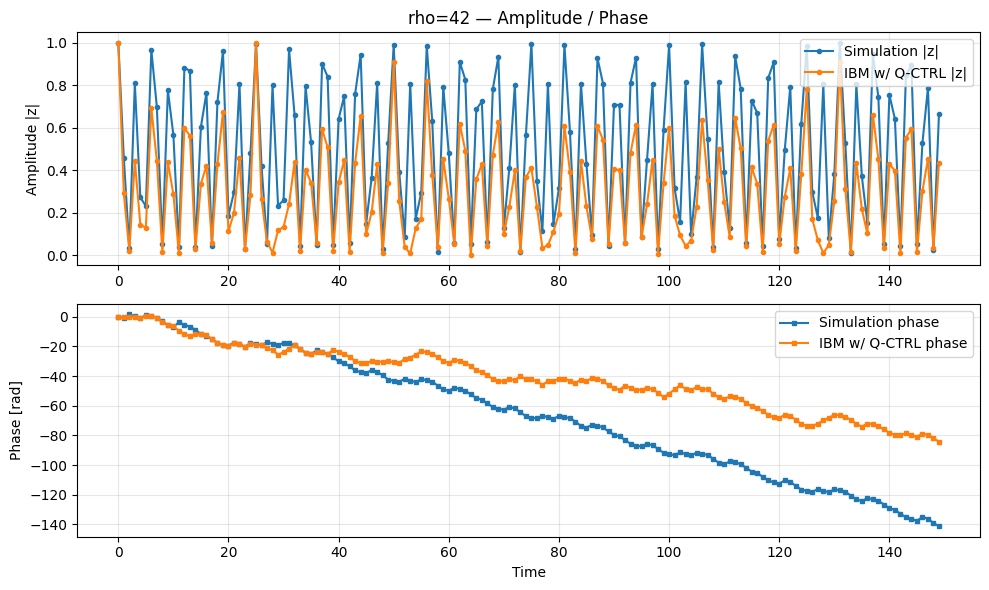}
    {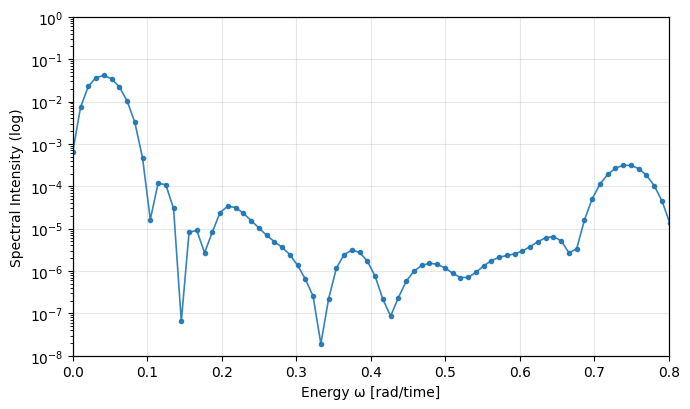}
    {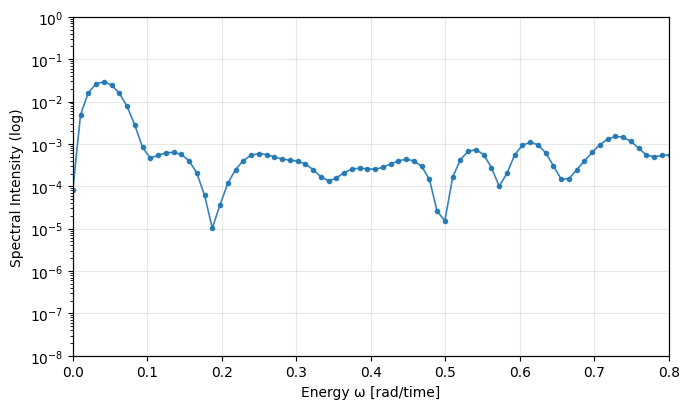}
    {42}
  \caption[]{\textbf{Quantum and classical diagnostics for the Lorenz system at $\rho=42$.}
Spectral broadening and partial peak overlap indicate the onset of \emph{topological coherence breaking}, where harmonic modes lose phase alignment across scales.  
The harmonic sector remains visible but less isolated, marking the gradual destabilization of the coherent loop.}

\end{figure}

\FloatBarrier


\begin{table}[h]
\centering
\caption{\textbf{Summary of hardware specifications for \texttt{ibm\_kingston}.} 
Error metrics refer to per-gate Pauli error rates at calibration time; CLOPS denotes circuit layer operations per second.
}
\label{tab:kingston_specs}
\begin{tabular}{>{\raggedright}p{0.48\linewidth} >{\raggedleft\arraybackslash}p{0.38\linewidth}}
\hline
Qubits & 156 \\
Processor type & Heron r2 \\
QPU version & 1.0.0 \\
Basis gates & \texttt{cz, id, rx, rz, rzz, sx, x} \\
Best two-qubit error & $8.88\times10^{-4}$ \\
Two-qubit error (layered) & $3.42\times10^{-3}$ \\
Median CZ error & $1.93\times10^{-3}$ \\
Median SX error & $2.372\times10^{-4}$ \\
Median readout error & $8.606\times10^{-3}$ \\
Median $T_1$ & $262.42\,\mu\mathrm{s}$ \\
Median $T_2$ & $119.55\,\mu\mathrm{s}$ \\
CLOPS (throughput) & 250\,K \\
\hline
\end{tabular}
\end{table}


\clearpage 



\end{document}